\documentstyle[prd,aps,amsfonts,eqsecnum,preprint,tighten]{revtex}
\begin{document}
\preprint{\vbox to 50 pt
{\hbox{IHES/P/95/57}\hbox{CPT-95/P.3167}\vfil}}
\draft
\title{Testing gravity to second post-Newtonian order~:\\
a field-theory approach}
\author{Thibault Damour}
\address{Institut des Hautes Etudes Scientifiques, F--91440
Bures-sur-Yvette, France\\
and D\'epartement d'Astrophysique Relativiste et de Cosmologie,
Observatoire de Paris,\\
Centre National de la Recherche Scientifique, F--92195 Meudon, France}
\author{Gilles Esposito-Far\`ese}
\address{Centre de Physique Th\'eorique\cite{UPR},
Centre National de la Recherche Scientifique,\\
Luminy, Case 907, F--13288 Marseille Cedex 9, France}
\date{June 27, 1995}
\maketitle
\begin{abstract}
A new, field-theory-based framework for discussing and interpreting
experimental tests of relativistic gravity, notably at the second
post-Newtonian (2PN) level, is introduced. Contrary to previous
frameworks which attempted at parametrizing any conceivable
phenomenological deviation from general relativity, we focus on the
most general class of gravity models of the type suggested by unified
theories: namely models in which gravity is mediated by a tensor field
together with one or several scalar fields. The 2PN approximation of
these ``tensor--multi-scalar'' theories is obtained thanks to a
diagrammatic expansion which allows us to compute the Lagrangian
describing the motion of $N$ bodies. In contrast with previous studies
which had to introduce many phenomenological parameters, we find that
the 2PN deviations from general relativity can be fully described by
introducing only two new 2PN parameters, $\varepsilon$ and $\zeta$,
beyond the usual (Eddington) 1PN parameters
$\overline\beta\equiv\beta-1$ and $\overline\gamma\equiv\gamma-1$. It
follows from the basic tenets of field theory (notably the absence of
negative-energy excitations), that $\overline\beta$, $\varepsilon$ and
$\zeta$ (as well as all the further parameters entering higher
post-Newtonian orders) must tend to zero with $\overline\gamma$. It is
also found that $\varepsilon$ and $\zeta$ do not enter the 2PN
equations of motion of light. Therefore, second-order light-deflection
or time-delay experiments cannot probe any theoretically motivated 2PN
deviation from general relativity. On the other hand, these experiments
can give a clean access to $\overline\gamma$, which is of greatest
significance as it measures the basic coupling strength of matter to
the scalar fields. Because of the importance of self-gravity effects in
neutron stars, binary-pulsar experiments are found to constitute a
unique testing ground for the 2PN structure of relativistic gravity. A
simplified analysis of current data on four binary-pulsar systems
already leads to significant constraints on the two 2PN parameters:
$|\varepsilon|<7\times 10^{-2}$, $|\zeta|<6\times10^{-3}$.
\end{abstract}
\narrowtext
\section{Introduction}
The last three decades have been the golden era of experimental gravity:
from Pound and Rebka to Hulse and Taylor, many complementary aspects of
general relativity have been successfully tested. In particular,
solar-system experiments allowed one to map out fairly completely
weak-field gravity at the first post-Newtonian (1PN) approximation,
{\it i.e.}, to put stringent numerical constraints on a large class of
possible deviations from general relativity at order $1/c^2$. Let us
recall the useful role played in this respect by the first-order
parametrized post-Newtonian (PPN) formalism
\cite{E23,S60,R62,B67,N68,W71,WN72,W81} which introduced, in its
extended versions, about 10 independent phenomenological parameters to
describe possible non-Einsteinian 1PN effects. Improved experiments are
now planned to reach the second post-Newtonian (2PN) level --order
$1/c^4$--, such as microsecond level light deflection experiments. Let
us also mention that a 2PN treatment of the periastron advance is
already significant for the binary pulsar PSR 1913+16 \cite{DS88}. It
is therefore timely to undertake a systematic theoretical study of
gravitational theories at this approximation.

The 2PN limit of general relativity has already been studied in depth
\cite{OOKH,H76,ES80,FF80,DD,RM82,RM83,DS85,S87,DS88,OK89}, but we also
need to know what can be the possible deviations from these results in
alternative theories of gravity. An ambitious program developed by
Nordtvedt and Benacquista in \cite{BN88,B92,N93} tries to extend
directly the PPN formalism at order $1/c^4$, {\it i.e.}, it aims at
introducing a large number of parameters describing any possible
relativistic theory at this order. Although it has only been partially
implemented at the present time, this approach allowed one to derive
some relations between these 2PN parameters by imposing the concept of
``extended Lorentz invariance'' ({\it i.e.}, by requiring that the
gravitational physics of subsystems, influenced by external masses,
exhibit Lorentz-invariance) \cite{N85,N93}. In spite of its
partial achievements, the ability of such a general phenomenological
approach to delineate the physically most important structures at the
2PN level is unclear. For instance, it was claimed in \cite{B92} that
10 ``parameters'' are required to map Lorentz-invariant theories of
gravity at the 2PN level; however, a careful reading of this article
shows that several of these ``parameters'' are in fact {\it
functions\/} of the distances between massive bodies, and could
depend {\it a priori\/} on an infinite number of real parameters.

In the present paper, we shall follow an entirely different methodology
by developing a ``theory-dependent'' approach initiated in \cite{DEF1}:
Instead of considering any conceivable phenomenological deviation from
general relativity, we focus on the simplest and best motivated class
of non-Einsteinian theories, in which gravity is mediated by a tensor
field ($g^*_{\mu\nu}$) together with one or several scalar fields
($\varphi^a$). These ``tensor--multi-scalar'' theories arise naturally
in theoretical attempts at quantizing gravity or at unifying it with
other interactions (Kaluza--Klein and superstrings theories). Moreover,
they are the only consistent field theories, containing only fields of
infinite range, able to satisfy the weak equivalence principle
(universality of free-fall of laboratory-size objects)
\cite{DHouches2}\footnote{Note, however, that the scalar couplings
coming out naturally from unifying theories violate the equivalence
principle \cite{DP}.}. Indeed, massless gravitational theories
incorporating, besides the metric $g^*_{\mu\nu}$, vector fields, a
second symmetric tensor field or an antisymmetric tensor field are
known to present in general many flaws, such as discontinuities in the
field degrees of freedom, negative-energy modes, causality violations,
ill-posedness of the Cauchy problem, {\it etc.}, not to mention the
lack of theoretical motivations for considering
equivalence-principle-preserving couplings for such fields. By
contrast, tensor--multi-scalar theories are well motivated, consistent
and simple enough to allow their observational predictions to be fully
worked out \cite{DEF1}. Moreover, we believe that these field theories
are the only ones satisfying the extended Lorentz invariance required
in \cite{BN88,B92,N93,N85}. It would be an interesting program to prove
it rigorously. In that case, our field-theoretical approach gives a
much more complete control of their structure than that of
Refs.~\cite{BN88,B92,N93,N85}, as exemplified by our results below.

A detailed study of the 1PN limit of tensor--multi-scalar theories has
been performed in \cite{DEF1}, as well as the generalization of this
approximation to the case of compact bodies (like neutron stars),
called the first post-Keplerian (1PK) limit. We recall some of our
results in section II below. Out of the 10 post-Newtonian parameters
describing conceivable deviations from general relativity at the 1PN
level, only two do not vanish in this class of theories: the parameters
$\overline\beta\equiv\beta-1$ and $\overline\gamma\equiv\gamma-1$,
introduced long ago (on different grounds) by Eddington
\cite{E23}\footnote{The intuitively preferred role played by $\overline
\beta$ and $\overline \gamma$ in the PPN formalism is a further
argument for working in the framework of tensor--scalar theories.}. In
\cite{DN93,DP}, it was shown that the cosmological evolution
generically drives these parameters towards values $\lesssim 10^{-7}$ at
our present epoch. This class of theories gives therefore a natural
explanation (requiring no fine tuning nor the {\it a priori\/} presence
of small parameters) to the bounds $|\overline\gamma|<2\times 10^{-3}$
\cite{R79} and $|\overline\beta|<6\times 10^{-4}$ \cite{Di94} found at
the 1PN level in solar-system experiments, and furnish a motivation for
increasing the precision of these measurements to the $10^{-7}$ level.
Such an increase in accuracy down to a level comparable to 2PN effects
($Gm_\odot/R_\odot c^2 \sim 10^{-6}$) makes it necessary to determine
how 2PN effects can influence such high-precision measurements, {\it
i.e.}, whether there are new and {\it a priori\/} unknown 2PN
parameters which could complicate the interpretation of 1PN
experiments. An example of this is given by higher-order
light-deflection experiments which have been claimed to involve a new
2PN parameter \cite{ES80,FF80,RM82}. We shall, however, prove below
that this claim is incorrect in the framework of tensor--scalar
theories.

The questions we shall address are thus: What are the new degrees of
freedom describing the possible deviations of tensor--multi-scalar
theories from general relativity at the 2PN order, and can the
corresponding effects be separated from those associated with
$\overline\beta$ and $\overline\gamma$~? On the other hand, do
experimental bounds on $\overline\beta$ and $\overline\gamma$ give
constraints on possible 2PN non-Einsteinian effects~?

Before entering into a detailed study of the 2PN limit of tensor--scalar
theories, let us quote one of our main results: Two, and only two, new
parameters arise at the 2PN level. We have denoted them by
$\varepsilon$ and $\zeta$, {\it cf.} Eqs.~(\ref{eq3.30}) below. The
possible 2PN {\it deviations\/} from the general relativistic physical
metric tensor are given by
\begin{mathletters}
\label{eq1.1}
\begin{eqnarray}
\delta g_{00}(x) & = & {\varepsilon\over 3c^6} U^3({\bf x})
+{\varepsilon\over c^6}\int d^3{\bf x}'\,
{G \sigma({\bf x}')U^2({\bf x}')
\over |{\bf x}-{\bf x}'|}
+{2\zeta\over c^6}\int d^3{\bf x}'\, {G\sigma({\bf x}')\over
|{\bf x}-{\bf x}'|}\int d^3{\bf x}''{G\sigma({\bf x}'')U({\bf x}'')\over
|{\bf x}'-{\bf x}''|}
\nonumber \\
& & + {2\zeta\over c^6} U({\bf x})\int d^3{\bf x}'{G\sigma({\bf x}')
U({\bf x}')\over |{\bf x}-{\bf x}'|}
+O\left({\overline\beta\over c^6},{\overline\gamma\over c^6}\right)
+O\left({1\over c^8}\right)
\ ,
\label{eq1.1a} \\
\delta g_{0i}(x) & = &
O\left({\overline\beta\over c^5},{\overline\gamma\over c^5}\right)
+O\left({1\over c^7}\right) \ ,
\label{eq1.1b} \\
\delta g_{ij}(x) & = &
O\left({\overline\beta\over c^4},{\overline\gamma\over c^4}\right)
+O\left({1\over c^6}\right)\ ,
\label{eq1.1c}\end{eqnarray}
\end{mathletters}
where $\sigma({\bf x})$ is the mass density and $U({\bf x}) = \int
d^3{\bf x}'\, G\sigma({\bf x}')/|{\bf x}-{\bf x}'|$ the Newtonian
potential. To increase the readability of Eqs.~(\ref{eq1.1}), we have
suppressed the symbol $\ \widetilde{}\ $ which should decorate all
the quantities appearing in it: $\widetilde g_{00}(\widetilde x),
\widetilde U, \widetilde G, \widetilde \sigma, \ldots$; see below. The
same parameters $\varepsilon$, $\zeta$ will be found to define the 2PN
renormalizations of various Newtonian or 1PN quantities under the
influence of self-gravity or external gravitational fields.

In section II, we recall the action and the equations of motion of
tensor--multi-scalar theories, as well as a few useful results
concerning their 1PN limit. We also recall how the motion of
self-gravitating bodies can be described in these theories. The main
discussion of our paper, in section III, is devoted to the Lagrangian
describing the motion of $N$ massive bodies at the 2PN level. Our main
technical tool is a diagrammatic expansion, which allows us to compute
straightforwardly all the 2PN effects. In section IV we derive the 2PN
metric corresponding to the Lagrangian of section III, and we verify
and complement our results by considering the metric generated by one
static and spherically symmetric body, whose exact solution has been
derived in \cite{DEF1}. In section V, we discuss the impact of our
findings on future relativistic experiments. We summarize our results
and give our conclusions in section VI. To relieve the tedium,
technical details are relegated to various appendices. Appendix~A gives
the explicit diagrammatic calculation of the 2PN Lagrangian. In
Appendix~B, we discuss the renormalizations of the Newtonian and 1PN
coupling parameters due to 2PN effects. Finally, Appendix~C derives the
explicit 2PN formulae for the deflection of light and the perihelion
shift of test masses.

\section{Tensor--multi-scalar theories}

In this section we define our notation for dealing with tensor--scalar
theories, and recall the results of \cite{DEF1} that we need below to
study their 2PN approximation.

\subsection{Action and field equations}
For simplicity, we consider in the present paper only theories
respecting exactly the weak equivalence principle, {\it i.e.}, theories
in which matter is universally coupled to {\it one\/} second rank
symmetric tensor, say $\widetilde g_{\mu\nu}(x^\lambda)$. The action
describing matter can then be written as a functional
\begin{equation}
S_m[\psi_m,\widetilde g_{\mu\nu}]\ ,
\label{eq2.1}\end{equation}
where $\psi_m$ denotes globally all matter fields, including gauge
bosons. Actually, from the perspective of modern unified theories, this
class of models seems rather {\it ad hoc}. For instance, string theory
does suggest the possibility that there exist long-range scalar fields
contributing to the interaction between macroscopic bodies, but all
such scalar fields have composition-dependent couplings. However, a
recent study of a large class of superstring-inspired tensor--scalar
models \cite{DP} has found that (because of deep physical facts) the
composition-dependent effects represent only fractionally small ($\sim
10^{-5}$) corrections to standard post-Newtonian effects.

At a fundamental level, the matter action $S_m$ should be chosen as the
curved-spacetime version of the action of the Standard Model of
electroweak and strong interactions, obtained by replacing the flat
metric $f_{\mu\nu}={\rm diag}(-1,1,1,1)$ by $\widetilde g_{\mu\nu}$ and
partial derivatives by $\widetilde g$-covariant ones. At a
phenomenological level, the action describing a system of $N$ (non
self-gravitating) pointlike particles is
\begin{equation}
S_m = -\sum_{A=1}^{N}\int \widetilde m_A c\, d\widetilde s_A\ ,
\label{eq2.2}\end{equation}
where $d\widetilde s_A\equiv[-\widetilde g_{\mu\nu}(x_A^\lambda)
dx_A^\mu dx_A^\nu]^{1/2}$, and the $\widetilde m_A$'s denote the
(constant) inertial masses of the different particles. The universal
coupling to $\widetilde g_{\mu\nu}$ implies in particular that
laboratory rods and clocks measure this metric, which will therefore be
called the ``physical metric'' [the names Jordan, Fierz, or Pauli metric
are also used in the literature].

The difference with general relativity lies in that the physical metric
$\widetilde g_{\mu\nu}$, instead of being a pure spin-2 field, is in
tensor--scalar theories a mixing of spin-2 and spin-0 degrees of
freedom. More precisely, it can be written as
\begin{equation}
\widetilde g_{\mu\nu} = A^2(\varphi^a) g^*_{\mu\nu}\ ,
\label{eq2.3}\end{equation}
where $A(\varphi^a)$ is a function of $n$ scalar fields (we choose $A>0$
to simplify some equations below). The dynamics of the pure spin-2
field $g^*_{\mu\nu}$, usually called the ``Einstein metric'', is
described by the Einstein-Hilbert action
\begin{equation}
S_{\rm spin 2} = {c^4\over 4\pi G_*}\int {d^4 x\over c}
\sqrt{g_*}\, {R^*\over 4} \ ,
\label{eq2.4}\end{equation}
where $G_*$ is a constant (the bare gravitational constant), $R^*$ is
the scalar curvature of $g^*_{\mu\nu}$ (with the sign conventions of
\cite{MTW}), and $g_* \equiv |{\rm det}\, g^*_{\mu\nu}|$. On the other
hand, the action describing the $n$ scalar fields $\varphi^a$ reads
\begin{equation}
S_{\rm spin 0} = {c^4\over 4\pi G_*}\int {d^4 x\over c} \sqrt{g_*}
\left(-{1\over 2} g_*^{\mu\nu}
\gamma_{ab}(\varphi^c)\partial_\mu\varphi^a
\partial_\nu\varphi^b\right)\ ,
\label{eq2.5}\end{equation}
where $g_*^{\mu\nu}$ is the inverse of $g^*_{\mu\nu}$, the indices
$a,b,c,\ldots$ vary from 1 to $n$, and $\gamma_{ab}(\varphi^c)$ is a
$n$-dimensional ($\sigma$-model) metric in the internal scalar space
spanned by the $\varphi^a$'s. [$\gamma_{ab}$ must be
positive-definite to get positive kinetic-energy terms.] A
tensor--multi-scalar theory contains in general $1+n(n-1)/2$
arbitrary functions of $n$ variables: the ``coupling function''
$A(\varphi^a)$ involved in Eq. (\ref{eq2.3}), and the $n(n+1)/2$
components of $\gamma_{ab}$ from which must be subtracted $n$
arbitrary functions parametrizing arbitrary changes of scalar-field
variables $\varphi'^a = f^a(\varphi^b)$. In the simplest case where
there is only one scalar field ($n=1$), the only arbitrary function
in the problem is $A(\varphi)$, the unique component of the
$\sigma$-model metric being always reducible to the trivial form
$\gamma_{11}(\varphi^1) = 1$. The reader should note that the
consideration of multiple scalar fields, far from complicating
uselessly our analysis, is in fact a technically powerful tool for
delineating the structure of the possible deviations from general
relativity. Once one is used to some notation, working with several
fields is anyway not more difficult than working with only one.

We could also have added a potential term $V(\varphi^a)$ in the action
(\ref{eq2.5}), but we will restrict our attention to infinite-range
fields in the present paper (see \cite{DEF1} for more details). Note
that $g^*_{\mu\nu}$ and the $n$ scalar fields $\varphi^a$ are
considered as forming the gravitational sector of the theory, by
contrast with the matter sector described by the fields $\psi_m$ of Eq.
(\ref{eq2.1}).

Although one should always keep in mind that the metric measured by
normal physical standards is $\widetilde g_{\mu\nu}$, it will be
convenient in the following to formulate the theory in terms of the
pure spin-2 and spin-0 fields $g^*_{\mu\nu}$ and $\varphi^a$. The
Einstein-frame infinitesimal lengths $\ell$, time-intervals $t$, and
masses $m$ will therefore be related to the physical (measured) ones by
\begin{equation}
\ell = A^{-1}(\varphi)\widetilde\ell\ ,\
t = A^{-1}(\varphi)\widetilde t\ ,\
m = A(\varphi) \widetilde m\ .
\label{eq2.6}\end{equation}
For instance, the action (\ref{eq2.2}) describing $N$
(non-self-gravitating) pointlike particles can be rewritten as
\begin{equation}
S_m = -\sum_{A=1}^{N}\int \widetilde m_A c\,
A\left(\varphi^a(x_A)\right)
\sqrt{-g^*_{\mu\nu}(x_A) dx_A^\mu dx_A^\nu}
=-\sum_{A=1}^{N}\int m_A\left(\varphi^a(x_A)\right) c\, ds^*_A\ ,
\label{eq2.7}\end{equation}
where the Einstein-frame masses $m_A(\varphi^a) \equiv A(\varphi^a)
\widetilde m_A$ are no longer constant, as opposed to the $\widetilde
m_A$'s.

The field equations deriving from the total action $S_{\rm spin
2}+S_{\rm spin 0}+S_m$ read
\begin{mathletters}
\label{eq2.8}
\begin{eqnarray}
&&R^*_{\mu\nu} =
2\gamma_{ab}(\varphi)\, \partial_\mu\varphi^a\partial_\nu\varphi^b +
{8\pi G^*\over c^4}\left(T^*_{\mu\nu}-{1\over 2}\, T^*g^*_{\mu\nu}
\right)\ ,
\label{eq2.8a}\\
&&\Box_{(g*,\gamma)}\varphi^a = -{4\pi G^*\over c^4}\,
\alpha^a(\varphi)
\, T^*\ ,
\label{eq2.8b}\\
&&\delta S_m[\psi_m,\widetilde g_{\mu\nu}]/\delta\psi_m = 0 \ .
\label{eq2.8c}
\end{eqnarray}
\end{mathletters}
Here $R^*_{\mu\nu}$ is the Ricci tensor of $g^*_{\mu\nu}$,
$T_*^{\mu\nu}\equiv(2c/\sqrt{g_*})\, \delta
S_m[\psi_m,A^2g^*_{\mu\nu}]/\delta g^*_{\mu\nu}$ is the Einstein-frame
energy tensor (related to the conserved ``physical'' energy tensor
$\widetilde T^{\mu\nu}$ by $T_*^{\mu\nu} = A^6 \widetilde T^{\mu\nu}$,
see \cite{DEF1}), and the d'Alembertian $\Box_{(g*,\gamma)}$ is
covariant with respect to both space-time and $\sigma$-model indices,
{\it i.e.}, involves the Levi-Civita connections of both $g^*_{\mu\nu}$
and $\gamma_{ab}(\varphi)$, denoted respectively as
$\Gamma^{*\lambda}_{\mu\nu}$ and $\gamma^a_{bc}(\varphi)$~:
\begin{equation}
\Box_{(g*,\gamma)}\varphi^a \equiv
g_*^{\mu\nu}\left[
\partial_\mu\partial_\nu\varphi^a
-\Gamma^{*\lambda}_{\mu\nu}\partial_\lambda\varphi^a
+\gamma^a_{bc}(\varphi)\partial_\mu\varphi^b\partial_\nu\varphi^c
\right]\ .
\label{eq2.9}\end{equation}
In Eqs. (\ref{eq2.8}) and everywhere else in this paper, the various
indices are moved by their corresponding metric, for instance
$T^*_{\mu\nu} = g^*_{\mu\alpha}g^*_{\nu\beta}T_*^{\alpha\beta}$, $T^* =
g^*_{\mu\nu} T_*^{\mu\nu}$, $\widetilde T_\mu^{\ \nu} = \widetilde
g_{\mu\alpha}\widetilde T^{\alpha\nu}$, {\it etc.}, but also
$\alpha^a(\varphi)=\gamma^{ab}(\varphi)\alpha_b(\varphi)$ where
$\gamma^{ab}$ is the inverse of $\gamma_{ab}$ and where we have
introduced the notation
\begin{equation}
\alpha_a(\varphi) \equiv {\partial\ln A(\varphi)\over
\partial\varphi^a}\ .
\label{eq2.10}\end{equation}
Note that, in view of the third equation (\ref{eq2.6}), the definition
of $\alpha_a(\varphi)$ can be rewritten as
\begin{equation}
\alpha_a(\varphi) \equiv {\partial\ln m(\varphi)\over
\partial\varphi^a}\ ,
\label{eq2.10bis}\end{equation}
where $m(\varphi)$ is the mass of any non self-gravitating particle in
the Einstein conformal frame. This second way of defining $\alpha_a$ is
quite general as it encompasses both self-gravity effects \cite{DEF1}
(see below) and possible composition-dependent effects \cite{DP}.
\subsection{1PN approximation}
As is clear from Eq.~(\ref{eq2.8b}), the quantity $\alpha^a(\varphi)$
(which is a vector field in the internal ``$\sigma$-model'' space
spanned by the $\varphi^a$'s) plays the crucial role of measuring
the {\it coupling strength\/} of the scalar fields to matter. As we
shall see below, all post-Newtonian deviations from general
relativity (of any post-Newtonian order) can be expressed in terms of
the asymptotic value of $\alpha^a(\varphi)$ at spatial infinity ({\it
i.e.}, far from all material sources) and of its successive
scalar-field derivatives. Denoting by $D_a$ the covariant derivative
with respect to the internal metric $\gamma_{ab}$, we define
\begin{mathletters}
\label{eq2.11}
\begin{eqnarray}
\beta_{ab} &\equiv & D_a\alpha_b =
{\partial\alpha_b\over\partial\varphi^a}
- \gamma^c_{ab}\alpha_c\ ,
\label{eq2.11a} \\
\beta'_{abc} &\equiv & D_aD_b\alpha_c\ .
\label{eq2.11b}
\end{eqnarray}
\end{mathletters}
Denoting by $\varphi_0^a$ the (cosmologically imposed) background
values of the scalar fields, we then set $\alpha_a^0\equiv
\alpha_a(\varphi_0)$, $\beta^0_{ab}\equiv\beta_{ab}(\varphi_0)$,
$\beta'^0_{abc}\equiv\beta'_{abc}(\varphi_0)$, {\it etc.}, the index
$\scriptstyle 0$ always meaning that these $\sigma$-model tensors are
calculated at $\varphi_0$. As shown in \cite{DEF1}, the effects of the
scalar fields on any observable effect at the first post-Newtonian
level depend only on two contractions of $\alpha^0_a$ and
$\beta^0_{ab}$, namely\footnote{Note that in the simplest one-scalar
case $\varphi = \varphi^1$, these contractions reduce to simple
products, $\alpha_0^2 = \alpha_0\times \alpha_0$,
$(\alpha\beta\alpha)_0 = \alpha_0\times \beta_0\times \alpha_0$, where
$\alpha_0 = \alpha(\varphi_0)= \partial\ln A(\varphi_0)/
\partial\varphi_0$, and $\beta_0 =
\partial\alpha(\varphi_0)/\partial\varphi_0$.}
\begin{mathletters}
\label{eq2.12}
\begin{eqnarray}
\alpha_0^2 &\equiv & \alpha_a^0\alpha_0^a = \alpha_a^0
\gamma^{ab}(\varphi_0) \alpha^0_b\ ,
\label{eq2.12a} \\
(\alpha\beta\alpha)_0 &\equiv & \alpha_0^a\beta^0_{ab}\alpha_0^b\ .
\label{eq2.12b}
\end{eqnarray}
\end{mathletters}
Indeed, the effective gravitational constant between two massive
particles is given by
\begin{equation}
\widetilde G \equiv G_* A_0^2 (1+\alpha_0^2)\ ,
\label{eq2.13}
\end{equation}
(with $A_0\equiv A(\varphi_0)$) instead of the bare constant $G_*$
involved in the action (\ref{eq2.4}), and the Eddington parameters
$\overline\beta$, $\overline\gamma$ read
\begin{mathletters}
\label{eq2.14}
\begin{eqnarray}
\overline\gamma &=& -2 \alpha_0^2/(1+\alpha_0^2)\ ,
\label{eq2.14a} \\
\overline\beta &=& {1\over 2}
(\alpha\beta\alpha)_0/(1+\alpha_0^2)^2\ ,
\label{eq2.14b}
\end{eqnarray}
\end{mathletters}
instead of their general relativistic values $\overline\beta=
\overline\gamma=0$. Let us recall again that $\overline\beta$ and
$\overline\gamma$ are usually denoted as $(\beta-1)$ and $(\gamma-1)$
in the literature. We will avoid the notation $\beta$, $\gamma$ in the
present paper to prevent a possible confusion with the $\sigma$-model
tensors $\beta_{ab}$ and $\gamma_{ab}$. Note, however, from
Eqs.~(\ref{eq2.12}) and (\ref{eq2.14}), that our notation has been
chosen so that $\overline\gamma\propto \gamma_{ab}\alpha^a\alpha^b$ and
$\overline\beta\propto \beta_{ab}\alpha^a\alpha^b$. As shown in
\cite{DEF1}, the results (\ref{eq2.13})-(\ref{eq2.14}) can be simply
interpreted (and remembered) in terms of the exchange of gravitons and
scalar particles between material sources. For instance,
Eq.~(\ref{eq2.13}) is the sum of the usual contribution of a graviton
exchange, $G_*$, together with the contributions of scalar exchanges,
$\sum_{a}G_*\alpha_a^0\alpha_0^a$, see Fig.~\ref{fig1}. The global
factor $A_0^2$ is due to the change of units between the Einstein
metric and the physical one $\widetilde g_{\mu\nu} =
A^2(\varphi)g^*_{\mu\nu}$ used to measure forces (see
Eqs.~(\ref{eq2.6}) above). Similarly, the parameter
$(\alpha\beta\alpha)_0$ involved in (\ref{eq2.14b}) corresponds to an
exchange of scalar particles between three massive bodies, as shown in
Fig.~\ref{fig2}. The method that we will use in section III below to
study the second post-Newtonian approximation is a straightforward
generalization of these diagrammatic observations.

\subsection{Self-gravitating bodies}
When studying the motion of massive bodies in the solar system, several
dimensionless ratios happen to be small. If we denote by $m$, $v$ and
$R$ the typical mass, orbital velocity and radius of a body, and by $r$
the typical distance between two bodies, we find $Gm/rc^2 \sim v^2/c^2
\lesssim 2\times 10^{-8}$ for the fastest planets, while $Gm/Rc^2 \sim
2\times 10^{-6}$ for the Sun and $\sim 7\times 10^{-10}$ for the Earth.
This is the reason why the formal ``post-Newtonian'' expansion in
powers of $1/c^2$ is so useful for analyzing the predictions of
relativistic theories of gravity in the solar system. However, in
situations involving compact bodies, like neutron stars (pulsars) in
binary systems, it is necessary to distinguish the self-gravity
parameter $s \sim Gm/Rc^2 \sim 0.2$ from the orbital parameters
$Gm/rc^2 \sim v^2/c^2 \ll 1$. In that case, one can still describe the
motion of the bodies by means of an expansion in powers of $Gm/rc^2
\sim v^2/c^2 \sim 1/c^2_{\rm orbital}$, but one must not expand in
powers of the compactnesses $s$. Such an expansion scheme has been
called ``post-Keplerian'' in \cite{DEF1}, since it is closely linked
with the phenomenological approach to the analysis of binary pulsar
data, introduced in \cite{D88,DT92} under the name of ``parametrized
post-Keplerian'' formalism.

In the present paper, our main goal is to analyze tensor--multi-scalar
theories of gravity at the second post-Newtonian level, {\it i.e.},
including all terms of formal order $1/c^4$ (be them of ``orbital'',
``self-gravity'' or mixed type). We found that the most efficient way of
doing so is to derive first the second post-Keplerian (2PK) limit of
these theories by means of a diagrammatic method. The 2PN approximation
is then obtained by expanding our general 2PK results in powers of the
compactnesses, up to the required order in $s$.

Let us recall how the motion of self-gravitating bodies is described in
tensor--scalar theories. Following a suggestion of Eardley \cite{E75},
one skeletonizes extended self-gravitating bodies as pointlike
particles whose inertial masses $\widetilde m_A$ depend on the scalar
fields $\varphi^a$, as opposed to the constant $\widetilde m_A$'s in
Eq.~(\ref{eq2.2}) describing non-self-gravitating bodies. This scalar
dependence of the inertial mass is due to the influence of the local
scalar field background on the equilibrium configuration of the body.
The action describing $N$ self-gravitating bodies is thus written as
\begin{equation}
S_m = -\sum_{A=1}^{N}\int \widetilde m_A(\varphi^a) c\, d\widetilde s_A
= -\sum_{A=1}^{N}\int m_A(\varphi^a) c\, ds^*_A\ ,
\label{eq2.15}\end{equation}
where $m_A(\varphi)\equiv A(\varphi)\widetilde m_A(\varphi)$. The
validity of the skeletonized action (\ref{eq2.15}) has been justified
in Appendix~A of \cite{DEF1} by a matching argument.

Note that the second expression (\ref{eq2.15}), in terms of the Einstein
line element $ds^*_A$, is formally identical to Eq.~(\ref{eq2.7})
describing non-self-gravitating bodies. However, the important
difference is that $m_A(\varphi)$ can now be a non-universal
(body-dependent) function of the scalar fields, instead of being
merely proportional to $A(\varphi)$. It is therefore convenient to
generalize to the case of self-gravitating bodies the $\sigma$-model
tensors defined in Eqs.~(\ref{eq2.10})--(\ref{eq2.11}) above:
\begin{mathletters}
\label{eq2.16}
\begin{eqnarray}
\alpha_a^A &\equiv & {\partial\ln m_A(\varphi)\over
\partial\varphi^a} =\alpha_a(\varphi) +{\partial\ln \widetilde
m_A(\varphi)\over \partial\varphi^a}
\ ,
\label{eq2.16a} \\
\beta^A_{ab} &\equiv & D_a\alpha^A_b\ ,
\label{eq2.16b} \\
\beta'^A_{abc} &\equiv & D_aD_b\alpha^A_c\ ,
\label{eq2.16c}
\end{eqnarray}
\end{mathletters}
and similarly for higher covariant derivatives. (As above, we will raise
and lower the $\sigma$-model indices $a,b,\dots$ with $\gamma^{ab}$ or
$\gamma_{ab}$.) Here again $\alpha_A^a$ plays the fundamental role of
measuring the coupling strength of the scalar fields to the
self-gravitating body $A$. Indeed, equation (\ref{eq2.8b}) now reads
\begin{equation}
\Box_{(g*,\gamma)}\varphi^a = -{4\pi
G_*\over c^4}\sum_A\alpha_A^aT^*_A\ ,
\label{eq2.17}\end{equation}
where $T_A^* = -m_Ac^2(ds^*_A/dx^0)\, \delta^{(3)}({\bf x}-{\bf
x}_A)/\sqrt{g^*(x_A)}$ is localized on the position of the $A$-th
``particle''. Note that the body-dependent quantities (\ref{eq2.16})
reduce to the definitions (\ref{eq2.10})--(\ref{eq2.11}) for
non-self-gravitating bodies, since the inertial masses $\widetilde m_A$
are constant in that case.

As shown in \cite{DEF1}, the 1PK approximation of tensor--multi-scalar
theories can then be expressed very simply in terms of contractions of
$\alpha_a^A$ and $\beta^A_{ab}$. More precisely, the Lagrangian
describing the motion of $N$ (spherical) self-gravitating bodies at the
1PK level reads (in Einstein-frame units)
\begin{equation}
L = \sum_A L_A^{(1)} + {1\over 2} \sum_{A\neq B} L_{AB}^{(2)} +
{1\over 2}\sum_{B\neq A\neq C} L^{(3) A}_{\ BC} + O\left({1\over
c^4}\right)\ ,
\label{eq2.18}\end{equation}
where the notation $B\neq A\neq C$ excludes $A=B$ and $A=C$ but
not $B=C$, and where
\begin{mathletters}
\label{eq2.19}
\begin{eqnarray}
L_A^{(1)} &=& -m_A^0 c^2 \sqrt{1-{\bf v}_A^2/c^2}
= -m_A^0 c^2 + {1\over 2} m_A^0 {\bf v}_A^2 + {1\over 8c^2} ({\bf
v}_A^2)^2 + O\left({1\over c^4}\right)\ ,
\label{eq2.19a} \\
L_{AB}^{(2)} &=& {G_{AB}m_A^0 m_B^0\over r_{AB}}\Bigl[
1+{3\over 2c^2} ({\bf v}_A^2+{\bf v}_B^2) -{7\over 2 c^2} ({\bf
v}_A\cdot {\bf v}_B)\nonumber\\
&& \hphantom{{G_{AB}m_A^0 m_B^0\over r_{AB}}}
-{1\over 2c^2}({\bf n}_{AB}\cdot{\bf v}_A)
({\bf n}_{AB}\cdot{\bf v}_B) +{\overline\gamma_{AB}\over c^2}
({\bf v}_A -{\bf v}_B)^2
\Bigr]\ ,
\label{eq2.19b} \\
L^{(3)A}_{\ BC} &=& -\, {G_{AB} G_{AC} m_A^0 m_B^0 m_C^0\over r_{AB}
r_{AC} c^2} (1+2\overline\beta^{\ A}_{BC})\ .
\label{eq2.19c}\end{eqnarray}
\end{mathletters}
Here we have set $m_A^0 \equiv m_A(\varphi_0)$, and we
have denoted by ${\bf n}_{AB}\equiv {\bf r}_{AB}/r_{AB}$ (with ${\bf
r}_{AB}\equiv {\bf x}_A-{\bf x}_B$) the unit vector directed from
body $B$ to body $A$. This Lagrangian involves three body-dependent
parameters generalizing the effective gravitational constant
(\ref{eq2.13}) and the 1PN Eddington parameters (\ref{eq2.14}), namely
\begin{mathletters}
\label{eq2.20}
\begin{eqnarray}
G_{AB} & \equiv & G_* [1+(\alpha_A\alpha_B)_0]\ ,
\label{eq2.20a} \\
\overline \gamma_{AB} & \equiv & -2 {(\alpha_A\alpha_B)_0\over
1+(\alpha_A\alpha_B)_0}\ ,
\label{eq2.20b} \\
\overline \beta^{\ A}_{BC} & \equiv & {1\over 2}
{(\alpha_B\beta_A\alpha_C)_0
\over [1+(\alpha_A\alpha_B)_0][1+(\alpha_A\alpha_C)_0]}\ ,
\label{eq2.20c}\end{eqnarray}
\end{mathletters}
where\footnote{Again, in the simplest case of one scalar field
$\varphi\equiv\varphi^1$, these quantities reduce simply to
$\alpha_A\times\alpha_B$ and $\alpha_B\times\beta_A\times\alpha_C$,
where $\alpha_A = \partial\ln m_A/\partial\varphi$,
$\beta_A=\partial\alpha_A/\partial\varphi$.} $(\alpha_A\alpha_B) \equiv
\alpha_A^a \gamma_{ab} \alpha_B^b$, $(\alpha_B\beta_A\alpha_C) \equiv
\alpha^a_B \beta^A_{ab} \alpha^b_C$, and the index $\scriptstyle 0$
means that these contractions are calculated at $\varphi^a =
\varphi^a_0$. Of course, when using physical units related to the
asymptotic metric $\widetilde g_{\mu\nu}$, the effective
gravitational constant reads $\widetilde G_{AB} = G_* A_0^2
[1+(\alpha_A\alpha_B)_0]$, and the parameters $\overline
\gamma_{AB}$, $\overline \beta^{\ A}_{BC}$ do not change since they
are dimensionless. Note that Figures \ref{fig1} and
\ref{fig2} give again a diagrammatic interpretation of $G_{AB}$ and
$(\alpha_B\beta_A\alpha_C)_0$, with the coupling coefficients
$\alpha^a$ and $\beta^{ab}$ of each body being replaced by their
strong-field counterparts $\alpha_A^a$ and $\beta^{ab}_A$.

The pointlike description (\ref{eq2.15}) neglects all finite size
effects. One might {\it a priori\/} worry that scalar interactions
might introduce relativistic couplings to ``scalar'' multipole moments,
starting with the spherical inertia moments $I\sim \int d^3{\bf x}\,
\sigma({\bf x}) {\bf x}^2$. If that were the case, that would introduce
in the equations of motion fractional corrections of order
$\overline\gamma (v/c)^2 (R/r)^2$ to the leading Newtonian force. Such
terms would be important notably in the Moon--Earth relative motion. In
fact, no such terms exist. This can be seen in two ways: either by
examining the exact external tensor--scalar field of a spherical
extended body (see below) which is found to depend on the structure of
the body only through $m_A$ and $\alpha_A$, or from considering the
general multipole expansion of the retarded field generated by an
arbitrary scalar source $S(x)$ which contains only one monopole moment
$M_0(u) = {1\over 2}\int d^3{\bf x}\int_{-1}^{+1}dz S({\bf x},u+zr/c)$
and higher ($\ell\geq 1$) multipole moments (see {\it e.g.}
\cite{DI91}). Because the bodies making up the solar system are nearly
spherical and in inner equilibrium\footnote{However, if one wanted to
deal with spherical bodies which are not in equilibrium, say a
collapsing star, one should carefully take into account the effects
associated with the intrinsic time variation of the scalar monopole
moment $m_A \alpha_A$.}, one easily checks that the corrections induced
by scalar-mediated couplings to the scalar multipole moments of order
$\ell\geq 1$ are negligible compared to the interactions described by
the action (\ref{eq2.15}). As for the corrections induced by
tensor-mediated couplings to the mass and spin multipole moments of
order $\ell \geq 1$, we assume that they are properly added, following
for instance the recent papers \cite{DSX} which worked out a general
relativistic description of multipole interactions.

\section{$N$-body Lagrangian}
Before entering the technical details of our derivation of the $N$-body
Lagrangian, let us clarify the approximation methods we shall employ.
As discussed in the previous section, we shall first study the second
post-Keplerian approximation of tensor--multi-scalar theories, before
particularizing our results to the second post-Newtonian case. In other
words, the compactnesses $s\sim Gm/Rc^2$ of the bodies will not be
considered {\it a priori\/} as small parameters. In order to construct
the Lagrangian describing the motion of $N$ self-gravitating bodies, we
will eliminate the field degrees of freedom, {\it i.e.}, we will
solve for the Einstein metric $g^*_{\mu\nu}$ and the scalar fields
$\varphi^a$ in terms of the material sources, using the field
equations (\ref{eq2.8a}), (\ref{eq2.8b}). To perform this
elimination, we shall consider that the interaction is propagated by
a time-symmetric (half-retarded--half-advanced) Green's function.
This leaves out radiation damping effects. However, the latter are
negligible when studying weakly self-gravitating bodies at 2PN order.
Indeed, in general relativity, the leading dissipative effects occur
at order
$1/c^5$, because of the well-known quadrupolar radiation of
gravitational waves. In tensor--scalar theories of gravity, the leading
radiation emitted by systems of {\it compact\/} bodies is dipolar and
occur at order $s^2/c^3$. In the solar system case, $s\sim 1/c^2$ and
the dipolar radiation is negligible compared to the $O(1/c^5)$ effects
due to the spin-2 quadrupolar waves and spin-0 monopolar and
quadrupolar waves \cite{DEF1}. In either case, we compute the
conservative part of the gravitational interaction and assume that
(tensor and scalar) radiation damping effects are added separately when
they are not negligibly small.

\subsection{Diagrammatic expansion}
We want to construct a ``Fokker'' Lagrangian \cite{Fokker,DS85}
describing the motion of $N$ massive bodies by eliminating the field
degrees of freedom from the total Lagrangian of the theory. In order to
prove formally that such a construction reproduces the correct dynamics
of the bodies, let us introduce a global notation for the fields
$\Phi\equiv (g^*_{\mu\nu}-f_{\mu\nu}, \varphi^a-\varphi_0^a)$, where
$f_{\mu\nu} = {\rm diag}(-1,1,1,1)$ is the flat metric and
$\varphi_0^a$ are the background values of the scalar fields.
Similarly, we denote globally by $\sigma$ the matter variables, {\it
i.e.}, the $N$ massive worldlines $x_A$ involved in the matter action
(\ref{eq2.15}). The total action of the theory can therefore be written
in the schematic form
\begin{equation}
S_{\rm tot}[\sigma,\Phi] = S_\Phi[\Phi]+S_m[\sigma,\Phi]\ ,
\label{eq3.1}\end{equation}
from which we want to eliminate the fields $\Phi$ by expressing them in
terms of the matter variables $\sigma$. This elimination cannot be done
when working directly with the Einstein-Hilbert action (\ref{eq2.4})
because of its invariance under diffeomorphisms (technically, the
kinetic term of the gravitons is non-invertible). We need to reduce the
field equations by means of a specific coordinate condition in order to
solve for $g^*_{\mu\nu}$. In the language of particle physics, we need
to fix the gauge in order to define the propagator of the gravitons. We
will choose the $g^*$-harmonic gauge for our explicit calculations. We
then replace $S_\Phi[\Phi]$ by its gauge-fixed version $S_\Phi^{\rm
g.f.}[\Phi]=S_\Phi[\Phi]+$ gauge-fixing terms (see Appendix~A below),
and the field equations deriving from the total action $S_{\rm
tot}^{\rm g.f.}[\sigma,\Phi]$ read:
\begin{mathletters}
\label{eq3.2}
\begin{eqnarray}
{\delta S_{\rm tot}^{\rm g.f.}[\sigma,\Phi] \over
\delta\Phi} & = & {\delta S_\Phi^{\rm g.f.}[\Phi] \over
\delta\Phi}+{\delta S_m[\sigma,\Phi] \over
\delta\Phi} = 0\ ,
\label{eq3.2a} \\
{\delta S_{\rm tot}^{\rm g.f.}[\sigma,\Phi] \over
\delta\sigma} & = & {\delta S_m[\sigma,\Phi] \over
\delta\sigma} = 0 \ .
\label{eq3.2b}\end{eqnarray}
\end{mathletters}
Eq. (\ref{eq3.2a}) can now be solved perturbatively, and we denote by
$\overline \Phi[\sigma]$ its solution. The Fokker action (which is a
functional of the matter variables only) is then defined as
\begin{equation}
S_F[\sigma] \equiv S_{\rm tot}^{\rm g.f.}\left[\sigma,\overline
\Phi[\sigma]\right] = S_\Phi^{\rm g.f.}\left[\overline
\Phi[\sigma]\right]+S_m\left[\sigma,\overline
\Phi[\sigma]\right]\ ,
\label{eq3.3}\end{equation}
and its variation with respect to $\sigma$ reads
\begin{equation}
{\delta S_F[\sigma]\over \delta\sigma}
= \left({\delta S_{\rm tot}^{\rm g.f.}[\sigma,\Phi] \over
\delta\sigma}\right)_{\Phi=\overline\Phi[\sigma]}
+\left({\delta S_{\rm tot}^{\rm g.f.}[\sigma,\Phi] \over
\delta\Phi}\right)_{\Phi=\overline\Phi[\sigma]}
\times {\delta\overline\Phi[\sigma]\over \delta\sigma}\ .
\label{eq3.4}\end{equation}
Since $\overline\Phi[\sigma]$ is precisely the solution which annuls the
functional derivative $\delta S_{\rm tot}^{\rm g.f.}/\delta\Phi$, the
second term on the right-hand side vanishes, and one finally gets
\begin{equation}
{\delta S_F[\sigma]\over \delta\sigma}
= \left({\delta S_{\rm tot}^{\rm g.f.}[\sigma,\Phi] \over
\delta\sigma}\right)_{\Phi=\overline\Phi[\sigma]}
= \left({\delta S_m[\sigma,\Phi] \over
\delta\sigma}\right)_{\Phi=\overline\Phi[\sigma]}\ .
\label{eq3.5}\end{equation}
Comparing (\ref{eq3.5}) with (\ref{eq3.2b}), we see that we have
formally proved that the Fokker action (\ref{eq3.3}) gives indeed the
correct equations of motion, {\it i.e.}, describes the actual dynamics
of the matter variables $\sigma$ in presence of the fields
$\overline\Phi[\sigma]$. It should be noted that $S_F[\sigma]$ is {\it
not\/} simply given by the matter action $S_m\left[\sigma,\overline
\Phi[\sigma]\right]$ embedded in the self-consistent background field
$\overline \Phi[\sigma]$, but that the field action $S_\Phi^{\rm
g.f.}\left[\overline \Phi[\sigma]\right]$ also contributes to the
dynamics of the material bodies.

To start with, we do not need to assume that the bodies are moving
slowly with respect to the velocity of light. In technical terms, we
shall expand the Lagrangian in powers of $G$, {\it i.e.} $Gm/rc^2$
(``nonlinearity expansion''), while keeping unrestricted the magnitudes
of $v/c$ and $s$. We need to retain the terms up to order $G^3$ in the
nonlinearity expansion of the Fokker action (\ref{eq3.3}). The zeroth
order term in $S_F$ is of course the action describing free bodies
$S_0[\sigma] \equiv S_m[\sigma, \Phi = 0]$, {\it i.e.}, the matter
action computed for $g^*_{\mu\nu} = f_{\mu\nu}$ and $\varphi^a =
\varphi^a_0$. It can be written explicitly as $S_0 = \sum_A\int
dt\,L_A^{(1)}$, where $L_A^{(1)}$ was given in Eq.~(\ref{eq2.19a}) and
has the structure $mc^2(1+v^2/c^2+v^4/c^4+v^6/c^6+\cdots)$. The next
approximation, first power of $G$, has the structure
$mc^2\times(Gm/rc^2)\times(1+v^2/c^2+v^4/c^4+ \cdots)$ and describes
the two-body interaction, starting with the Newtonian term $G_{AB} m_A
m_B/r_{AB}$ (with the self-gravity-modified effective gravitational
constant $G_{AB}$ of Eq.~(\ref{eq2.20a})). The second power of $G$,
$mc^2\times(Gm/rc^2)^2\times(1+v^2/c^2+ \cdots)$, describes three-body
interactions starting with the self-gravity-modified 1PK term $\propto
G_{AB}G_{AC}m_A m_B m_C/r_{AB} r_{AC}c^2$. Finally, the $G^3$ level
corresponds to four-body interactions: $G^3 m_A m_B m_C m_D /r^3c^4 +$
self-gravity and velocity modifications. Instead of counting powers of
$G$, we can count the number of matter source terms involved: we must
keep up to four powers of $\sigma$, {\it i.e.}, of the masses. Since
the matter action $S_m[\sigma,\Phi]$ is linear in $\sigma$ (see
Eq.~(\ref{eq2.15})), the solution $\overline\Phi[\sigma]$ of the field
equations (\ref{eq3.2a}) starts at order $\sigma$. We need therefore to
expand the total action $S_{\rm tot}^{\rm g.f.}[\sigma,\Phi]$ up to
orders $O(\Phi^4)$ and $O(\sigma\Phi^3)$ included, before replacing
$\Phi$ by its solution $\overline\Phi[\sigma]$.

Let us first expand $S_{\rm tot}^{\rm g.f.}$ in powers of the fields
$\Phi$, and define \begin{equation} S_{\rm tot}^{\rm g.f.}[\sigma,\Phi]
= S_0[\sigma] +S_1[\sigma,\Phi]
+S_2[\sigma,\Phi]+S_3[\sigma,\Phi]+\cdots \ ,
\label{eq3.6}\end{equation}
where the term $S_i[\sigma,\Phi]$ involves the $i$-th power of $\Phi$
(and zero or one power of the material sources $\sigma$). For instance,
$S_1$ is the linear interaction term between the fields and matter, and
has the formal structure $S_1 = \alpha \sigma \Phi$, where $\alpha$ is
a coupling constant. On the other hand, the term quadratic in $\Phi$
has the form $S_2 = -{1\over 2}\Phi {\cal P}^{-1} \Phi+{1\over 2}\beta
\sigma \Phi^2$, and involves both the kinetic operator $\propto \Box$
(or ``inverse propagator'' ${\cal P}^{-1}$) of the fields and a vertex
describing the interaction of matter with two fields (with a coupling
constant $\beta$). It will be convenient to introduce a diagrammatic
notation for this expansion. [Bertotti and Plebanski \cite{BP60} were
the first to introduce a similar diagrammatic notation for solving
Einstein's equations perturbatively.] Let us denote the propagator
$\cal P$ by a straight line, the material source $\sigma$ by a white
blob, and the term $({\cal P}^{-1}\Phi)$ by a black one; see
Fig.~\ref{fig3}. As this diagrammatic representation will be an
important tool in the present paper, let us explain it in detail with a
simpler example.

Let us consider the action (in Minkowski spacetime)
\begin{equation}
S[\varphi] = \int d^4x\left[
-{1\over 8\pi}(\partial \varphi)^2 +{g\over 3}\varphi(\partial\varphi)^2
+{\lambda\over 4}\varphi^4 +\sigma(x)\varphi(x)
\right]\ ,
\label{eq3.7}\end{equation}
where $(\partial\varphi)^2\equiv f^{\mu\nu}\partial_\mu\varphi
\partial_\nu\varphi$ and where $\sigma(x)$ is a given (spacetime
distributed) source for $\varphi(x)$. Integrating by parts, the kinetic
terms for $\varphi(x)$ read $+(1/8\pi)\int d^4x\, \varphi(x)\Box_f
\varphi(x)$, where $\Box_f$ is the flat d'Alembertian. Identifying this
with $-{1\over 2}\varphi{\cal P}^{-1}\varphi$ [which is a symbolic
notation\footnote{Note that in the operator notation used here, any
``contraction of spacetime indices'' means an integration over the
corresponding spacetime coordinates, {\it e.g.} $({\cal P}\varphi)(x)
\equiv \int d^4y\, {\cal P}_{xy}\varphi(y)$.} for $-{1\over
2}\int\!\!\int d^4x\, d^4y\, \varphi(x) {\cal P}^{-1}_{xy}\varphi(y)$,
where ${\cal P}^{-1}_{xy}$ is the kernel of an operator acting on
functions of $x^\mu$], we get ${\cal P}^{-1}_{xy} =
-(4\pi)^{-1}\Box_{x}\delta^{(4)}(x-y)$. The inverse of the operator
${\cal P}^{-1}_{xy}$ is the propagator ${\cal P}_{xy} = {\cal G}(x-y)$,
where ${\cal G}(x-y)$ is a (translation-invariant) Green function,
solution of $\Box_x{\cal G}(x-y) = -4\pi\delta^{(4)}(x-y)$. The cubic
vertex $V_3$ is defined as the distributional kernel entering the
term $3S_3 \equiv \int dx\, g \varphi(\partial\varphi)^2$, {\it i.e.},
$3 S_3 = \int dx_1 dx_2 dx_3
V_3(x_1,x_2,x_3)\varphi(x_1)\varphi(x_2)\varphi(x_3)$. Requiring this
kernel to be symmetric in its arguments leads to the explicit
expression
\begin{eqnarray}
V_3(x_1,x_2,x_3) & = & {g\over 3}\biggl[
{\partial\over\partial x_1^\mu} \delta(x_1-x_2)
{\partial\over\partial x_1^\mu} \delta(x_1-x_3)
+{\partial\over\partial x_2^\mu} \delta(x_2-x_3)
{\partial\over\partial x_2^\mu} \delta(x_2-x_1)
\nonumber\\
&& +{\partial\over\partial x_3^\mu} \delta(x_3-x_1)
{\partial\over\partial x_3^\mu} \delta(x_3-x_2)
\biggr]\ ,
\label{eq3.8}\end{eqnarray}
where the factor $1/3$ comes from the average over the three different
permutations needed to symmetrize $V_3(x_1,x_2,x_3)$. Similarly, the
quartic vertex, defined by $4S_4 \equiv \int dx_1 dx_2 dx_3 dx_4
V_4(x_1,x_2,x_3,x_4) \varphi(x_1) \varphi(x_2) \varphi(x_3)
\varphi(x_4)$, is
\begin{equation}
V_4(x_1,x_2,x_3,x_4)=
\lambda\delta(x_1-x_2)\delta(x_1-x_3)\delta(x_1-x_4)\ .
\label{eq3.9}\end{equation}
In the diagrammatic notation of Fig.~\ref{fig3}, the blobs denote
some spacetime functions ($\sigma(x)$ for the white blob and
$-(4\pi)^{-1}\Box\varphi(x)$ for the black one), and a line denotes a
propagator ${\cal P}_{xy} = {\cal G}(x-y)$. Connecting a line to a
blob or to a vertex (which is a cluster of several infinitesimally
close points) means that one ``contracts'' ({\it i.e.}, integrates)
over the points at the extremities of the line. For instance the
{\sf T}-shaped diagram on the left of the third line of
Fig.~\ref{fig7} below would represent, in the model (\ref{eq3.7}),
\begin{eqnarray}
&& \int dx_1 dx_2 dx_3 dy_1 dy_2 dy_3 V_3(x_1,x_2,x_3){\cal
G}(x_1-y_1) {\cal G}(x_2-y_2) {\cal G}(x_3-y_3) \sigma(y_1)
\sigma(y_2)
\sigma(y_3)
\nonumber\\
&& = g \int dx\, dy_1 dy_2 dy_3 {\cal G}(x-y_1)
{\partial\over
\partial x^\mu}{\cal G}(x-y_2){\partial\over \partial x^\mu}
{\cal G}(x-y_3) \sigma(y_1) \sigma(y_2) \sigma(y_3) \ .
\label{eq3.10}\end{eqnarray}
Similarly, the {\sf X} diagram on the left of the last line of
Fig.~\ref{fig7} would denote
\begin{eqnarray}
&& \int dx_1 dx_2 dx_3 dx_4 dy_1 dy_2 dy_3 dy_4 V_4(x_1,x_2,x_3,x_4)
{\cal G}(x_1-y_1){\cal G}(x_2-y_2){\cal G}(x_3-y_3){\cal G}(x_4-y_4)
\nonumber\\
&&\quad\times\,
\sigma(y_1)\sigma(y_2)\sigma(y_3)\sigma(y_4)
\nonumber \\
&& = \lambda \int dx\, dy_1 dy_2 dy_3 dy_4 {\cal G}(x-y_1)
{\cal G}(x-y_2) {\cal G}(x-y_3) {\cal G}(x-y_4) \sigma(y_1)
\sigma(y_2) \sigma(y_3) \sigma(y_4) \ .
\label{eq3.11}\end{eqnarray}
Having explained the precise meaning of our symbolic notation, let us
come back to the general action (\ref{eq3.6}).

The different terms of the expansion (\ref{eq3.6}) can be
represented as in Fig.~\ref{fig4}, which {\it defines\/} the different
diagrams. Note that, as in our example, we have conventionally
factorized a coefficient $1/i$ in defining the $i$-linear vertex $V_i$
from the $O(\Phi^i)$ action: $S_i = V_i/i$. This coefficient is chosen
in order to simplify the field equations (\ref{eq3.2a}), whose
diagrammatic expansion is displayed in Fig.~\ref{fig5}. [The reader is
invited to derive for himself the field equations of Fig.~\ref{fig5}
from the action of Fig.~\ref{fig4}, keeping in mind that the
multilinear forms in $\Phi$ appearing in Fig.~\ref{fig4} are supposed
to be symmetric in $\Phi(x_1),\ldots,\Phi(x_i)$.] Thanks to Euler's
theorem on homogeneous functions, the field equations imply the useful
result
\begin{equation}
\Phi\times {\delta S_{\rm tot}^{\rm
g.f.}[\sigma,\Phi]\over\delta\Phi} = S_1+2S_2+3S_3+\cdots+iS_i+\cdots
= 0 \ .
\label{eq3.12}\end{equation}
In diagrammatic terms, this corresponds to inserting a black blob at the
free ends of the propagators in Fig.~\ref{fig5}, {\it i.e.}, to express
the kinetic term of the fields, ${1\over2}\Phi {\cal
P}^{-1}\Phi\equiv{1\over2} ({\cal P}^{-1}\Phi){\cal P}({\cal
P}^{-1}\Phi)$, in terms of the other diagrams.

The Fokker action (\ref{eq3.3}) can now be written straightforwardly by
replacing in the total action (\ref{eq3.6}) the fields $\Phi$ by their
solution $\overline\Phi[\sigma]$, {\it i.e.}, by replacing the black
blobs in terms of the white ones through an iterative use of
Fig.~\ref{fig5}. The most delicate term to compute would be the
contribution due to the kinetic term of the fields in $S_2$, because
one must expand up to order $\sigma^3$ the two fields $\Phi$ it
involves. Fortunately, one can avoid estimating this term by using the
Euler identity (\ref{eq3.12}) to eliminate it from the Fokker action:
\begin{eqnarray}
S_F[\sigma] & = & \left[
(S_0+S_1+S_2+\cdots)-{1\over2}(S_1+2S_2+3S_3+\cdots)
\right]_{\Phi = \overline\Phi[\sigma]}\nonumber\\
& = & S_0 + \left[
{1\over 2}S_1-{1\over 2} S_3-S_4\right]_{\Phi = \overline\Phi[\sigma]}
+O(\sigma^5)\ .
\label{eq3.13}\end{eqnarray}
The result of inserting Fig.~\ref{fig5} into Eq.~(\ref{eq3.13}) is
displayed in Fig.~\ref{fig6}. [The different diagrams have been drawn
so that angles appear only at the vertices involving matter sources.]
In the following, we will designate these diagrams by the letter they
most naturally evoke, so that the final result for the Fokker action
reads
\begin{equation}
S_F[\sigma] = S_0[\sigma] + \left({1\over 2}\, {\rm I}\right)
+\left({1\over 2}\, {\sf V}+{1\over 3}\, {\sf T}\right)
+\left({1\over 3} \in +{1\over 2}\, {\sf Z}+ {\sf F}+{1\over 2}\,
{\sf H}+ {1\over 4}\, {\sf X}\right) + O(\sigma^5)\ .
\label{eq3.14}\end{equation}
The explicit form of this action is now only a question of
straightforward algebra: one must expand $S_{\rm tot}^{\rm
g.f.}[\sigma,\Phi]$ up to order $\Phi^4$ included to get the
expressions of the field propagator ${\cal P}$ and of the vertices
defined by Fig.~\ref{fig4}, and merely replace them in Fig.~\ref{fig6},
{\it i.e.}, Eq.~(\ref{eq3.14}). In doing so, one must take into account
the fact that $\Phi$ is a global notation for both the gravitons
$h_{\mu\nu}\equiv g^*_{\mu\nu} - f_{\mu\nu}$ and the scalar fields
$\varphi^a-\varphi^a_0$, and therefore that each propagator link in
Fig.~\ref{fig6} is to be replaced by a sum of terms corresponding to
the different fields. To simplify the expansion of the
scalar-field action (\ref{eq2.5}), it is convenient to choose Riemann
normal coordinates at $\varphi_0^a$, so that the metric
$\gamma_{ab}(\varphi)$ can be written as
\begin{equation}
\gamma_{ab}(\varphi) = \gamma_{ab}(\varphi_0) +
0\times(\varphi^a-\varphi_0^a)
-{1\over 3}R_{acbd}(\varphi_0)(\varphi^c-\varphi_0^c)
(\varphi^d-\varphi_0^d) +O(\varphi-\varphi_0)^3\ ,
\label{eq3.15}\end{equation}
where $R_{abcd}$ is the Riemann curvature of $\gamma_{ab}$. This choice
cancels the term of order $\varphi\partial\varphi\partial\varphi$ in
$S_{\rm spin0}$, {\it i.e.}, the ``{\sf T}'' vertex connecting three
scalar fields. The different diagrams of Eq.~(\ref{eq3.14}) can thus be
decomposed into the elementary diagrams displayed in Fig.~\ref{fig7},
where curly and straight lines represent the graviton and scalar
propagators respectively. The coefficients appearing in this figure are
simple binomial coefficients coming from the various ways of choosing
the lines (see Appendix~A). The diagrams involving only gravitons give
the $O(G^3)$ approximation of general relativity, which has been
studied in the literature (see notably \cite{DD,DS85}). The diagrams
involving at least one scalar propagator give therefore all the looked
for deviations from general relativity predicted by tensor--scalar
theories at this order. Their explicit calculation is performed in
Appendix~A below. Their global structure is easy to grasp. Denoting by
$\cal G$ a Green function, solution of $\Box_f{\cal G}(x) =
-4\pi\delta^{(4)}(x)$ where $\Box_f$ is the flat d'Alembertian, each
scalar propagator is ${\cal P}_{\varphi}^{ab} \propto {\cal
G}(x_A-x_B)\gamma^{ab}$, while each graviton propagator is ${\cal
P}^{h}_{\alpha\beta\gamma\delta} \propto {\cal G}(x_A-x_B)
(f_{\alpha\gamma}f_{\beta\delta}+ f_{\alpha\delta}f_{\beta\gamma} -
f_{\alpha\beta}f_{\gamma\delta})$. Each matter vertex containing a
single scalar coupling brings a factor $\alpha^A_a$, those containing
double scalar couplings bring a factor
$\beta^A_{ab}+\alpha^A_a\alpha^A_b$, while triple scalar couplings
bring a factor $\beta'^{abc}_A$ together with extra $\beta^A_{ab}$ and
$\alpha^A_a$ terms [see Eq.~(\ref{eqA11})--(\ref{eqA14}) in Appendix~A
below]. Each link implies a contraction over the internal
indices\footnote{Those internal contractions make it very easy to work
with $n$ scalar fields. Actually, once one is used to the notation, it
is easier to see which new 2PN parameters can occur in the multi-scalar
case rather than in the mono-scalar one.}. Finally, the global
structure of the action is
\begin{eqnarray}
\sum_A\sum_B\cdots
&&\int d\tau_A\int d\tau_B\cdots\
{\cal G}(x_A-x_B)\ {\cal G}\cdots
\nonumber \\
&&\times\
f\left(
u^\mu_{A,B,\ldots}; \alpha^a_{A,B,\ldots}(\varphi_0),
\beta^{ab}_{A,B,\ldots}(\varphi_0),
\beta'^{abc}_{A,B,\ldots}(\varphi_0), R_{abcd}(\varphi_0)
\right)\ ,
\label{eq3.16}\end{eqnarray}
where $\tau_A$ is the Minkowski proper time along the worldline
$x_A^\mu(\tau_A)$, and $u_A^\mu\equiv dx^\mu_A/cd\tau_A$. We use in our
calculations the symmetric Green function ${1\over 2}( {\cal G}_{\rm
retarded}+{\cal G}_{\rm advanced})$, given by \cite{DSX}
\begin{eqnarray}
{\cal G}_{\rm sym}(x_A-x_B) &=& \delta\left(f_{\mu\nu}(x_A^\mu-x_B^\mu)
(x_A^\nu-x_B^\nu)
\right)\nonumber \\
&=& \left[
\delta(x_A^0-x_B^0-r_{AB})
+\delta(x_A^0-x_B^0+r_{AB})
\right]/2r_{AB}\ .
\label{eq3.17}\end{eqnarray}

The power of our diagrammatic approach shows up in the fact that one can
identify the theory-dependent parameters appearing in each term of the
action without doing any calculation. For instance, the interaction
between two bodies $A$ and $B$ is described by the first two ``I''
diagrams of Fig.~\ref{fig7}, and involves therefore (because of the
first diagram) the contraction $(\alpha_A\alpha_B)_0$. This explains
why the effective gravitational constant $G_{AB}$ and the 1PK Eddington
parameter $\overline\gamma_{AB}$ appearing in $L^{(2)}_{AB}$,
Eq.~(\ref{eq2.19b}), depend only on this contraction. Similarly, the
interaction between three bodies is described by the ``{\sf V}'' and
``{\sf T}'' diagrams of Fig.~\ref{fig7}, and we see that it depends not
only on contractions like $(\alpha_A\alpha_B)_0$ [{\it cf.} the second
{\sf V} diagram and the first {\sf T} diagram], but also on
$(\alpha_A\beta_B\alpha_C)_0$ [{\it cf.} the first {\sf V} diagram,
involving only scalar propagators]. Here again, we understand very
simply why the 3-body interaction Lagrangian $L^{(3) A}_{\ BC}$ of
Eq.~(\ref{eq2.19c}) depends only on these two types of contractions,
involved in $G_{AB}$ and $\overline\beta^{\ A}_{BC}$.

The new parameters appearing at order $G^3$ can now be found by
examining the diagrams connecting four bodies in Fig.~\ref{fig7}. While
the {\sf H} and {\sf F} diagrams depend only on the previous
contractions of $\alpha^a_{A,B,\dots}$ and $\beta^{ab}_{A,B,\dots}$,
three new contractions occur in the first $\in$, {\sf Z} and {\sf X}
diagrams. Indeed, they involve respectively the contractions
$(\beta'^A_{abc}\alpha_B^a\alpha_C^b\alpha_D^c)_0$,
$(\alpha_A^a\beta_{ab}^B\beta_C^{bc}\alpha_c^D)_0$ and
$(R_{abcd}\alpha_A^a\alpha_B^b\alpha_C^c\alpha_D^d)_0$, which are
independent from $(\alpha_A\alpha_B)_0$ and
$(\alpha_A\beta_B\alpha_C)_0$ in generic tensor--multi-scalar theories.
It is convenient to introduce compact notations for these parameters;
we define
\begin{mathletters}
\label{eq3.18}
\begin{eqnarray}
\varepsilon^{\ \ A}_{BCD} & \equiv &
{(\beta'_A\alpha_B\alpha_C\alpha_D)_0 \over
[1+(\alpha_A\alpha_B)_0]\,
[1+(\alpha_A\alpha_C)_0]\,
[1+(\alpha_A\alpha_D)_0]
}\ ,
\label{eq3.18a} \\
\zeta_{ABCD} & \equiv & {(\alpha_A\beta_B\beta_C\alpha_D)_0 \over
[1+(\alpha_A\alpha_B)_0]\,
[1+(\alpha_B\alpha_C)_0]\,
[1+(\alpha_C\alpha_D)_0]
}\ , \label{eq3.18b} \\
\chi_{ABCD} & \equiv &
(R_{abcd}\alpha_A^a\alpha_B^b\alpha_C^c\alpha_D^d)_0\ ,
\label{eq3.18c}\end{eqnarray}
\end{mathletters}
where the choice of the carrier letters $\varepsilon$, $\zeta$, $\chi$
is made by pictorial analogy with the corresponding diagrams $\in$,
{\sf Z} and {\sf X}. The introduction of factors
$[1+(\alpha_A\alpha_B)_0]^{-1}$ is made to simplify some equations
below, where we shall factorize the effective gravitational constant
$G_{AB}$, Eq.~(\ref{eq2.20a}).

The main conclusion of our diagrammatic analysis of tensor--multi-scalar
theories is therefore that only {\it three} new self-gravity-dependent
parameters appear at the $G^3$ level. By contrast, previous studies in
the literature suggested the need for a much larger number of
parameters in the general framework of Lorentz-invariant theories of
gravity \cite{B92,N93}. Let us note that if we had restricted ourselves
to theories involving only one scalar field, only two
self-gravity-dependent parameters $\varepsilon^{\ \ A}_{BCD}$ and
$\zeta_{ABCD}$ would have showed up, the Riemann tensor of the scalar
manifold being then identically zero\footnote{In the one scalar case,
we could also express formally $\zeta_{ABCD}$ in terms of 1PK
parameters, using $(\alpha_A\beta_B\beta_C\alpha_D)=
(\alpha_A\beta_B\alpha_C)(\alpha_B\beta_C\alpha_D)
/(\alpha_B\alpha_C)$, but this expression is singular when $\alpha_B$
or $\alpha_C \rightarrow 0$.}. We are going to see that the parameter
$\chi_{ABCD}$ disappears anyway when considering the weak self-gravity
limit.

\subsection{2PK approximation}
The diagrammatic results of the previous subsection gave the structure
of the non-linearity expansion, in powers of $Gm/rc^2$, of the $N$-body
Lagrangian, keeping unexpanded all powers of $v/c$ and of the
self-gravity $s$. The second post-Keplerian limit of
tensor--multi-scalar theories can now be obtained by expanding the
results above in powers of $v^2/c^2$ (still keeping $s$ unexpanded).
First of all, the Minkowski proper time $\tau_A$ involved in
(\ref{eq3.16}) is obviously expanded as
\begin{equation}
d\tau_A = dt_A\sqrt{1-{\bf v}_A^2/c^2} = dt_A\left[
1-{1\over 2}{v_A^2\over c^2}-{1\over 8} \left({v_A^2\over
c^2}\right)^2-{1\over 16} \left({v_A^2\over
c^2}\right)^3+O\left({v_A^2\over c^2}\right)^4 \right]\ .
\label{eq3.19}\end{equation}
The $(v_A^2/c^2)^3$ term is necessary to write $S_0[\sigma]$ at the 2PK
order, but the expansion (\ref{eq3.19}) can be truncated to order
$(v_A^2/c^2)^2$ for the 2-body diagram ``I'', and to order
$(v_A^2/c^2)$ for the 3-body diagrams {\sf V} and {\sf T}. In the 2PK
diagrams $\in$, {\sf Z}, {\sf F}, {\sf H}, {\sf X}, it is enough to
replace the proper time $\tau_A$ by the coordinate time $t_A$.

Similarly, one must expand in powers of $v/c$ the unit 4-velocity
$u_A^\mu = dx_A^\mu/cd\tau_A = (1,v_A^i/c)/\sqrt{1-{\bf v}_A^2/c^2}$,
and in particular the contraction $2(f_{\mu\nu}u_A^\mu u_B^\nu)^2-1$
which appears for each graviton propagator connecting two bodies (see
Appendix~A). One gets easily
\begin{equation}
2(u_A u_B)^2-1 = 1+2\,{({\bf v}_A-{\bf v}_B)^2\over c^2}
+2\,{({\bf v}_A-{\bf v}_B)^2({\bf v}_A^2+{\bf v}_B^2)-
({\bf v}_A\times{\bf v}_B)^2 \over c^4}
+O\left({1\over c^6}\right)\ ,
\label{eq3.20}\end{equation}
where ${\bf v}_A\times{\bf v}_B$ denotes the usual vector skew product.
Finally, the Green function (\ref{eq3.17}) can be expanded as
\begin{eqnarray}
c\,{\cal G}_{\rm sym}(x_A-x_B) & = & [\delta(t_A-t_B-r_{AB}/c) +
\delta(t_A-t_B+r_{AB}/c)]/2r_{AB}
\nonumber \\
& = & \sum_{n=0}^\infty {|{\bf x}_A-{\bf
x}_B(t_B)|^{2n-1}\over(2n)!\, c^{2n}}\, {\partial^{2n}
\over (\partial t_B)^{2n}}\, \delta(t_A-t_B)
\nonumber \\
& = & {\delta(t_A-t_B)\over r_{AB}}
+{|{\bf x}_A-{\bf
x}_B(t_B)|\over 2c^2}\, {\partial^2\delta(t_A-t_B)\over \partial
t_B^2}
\nonumber \\
&& +\,{|{\bf x}_A-{\bf
x}_B(t_B)|^3\over 24c^4}\, {\partial^4\delta(t_A-t_B)\over \partial
t_B^4} +\left({1\over c^6}\right)\ .
\label{eq3.21}\end{eqnarray}
The three terms in (\ref{eq3.21}) are needed for the ``I'' diagram,
while only the first two are needed for the {\sf V} and {\sf T}
diagrams, and only the instantaneous Green function
$\delta(t_A-t_B)/r_{AB}$ for the 2PK diagrams $\in$, {\sf Z}, {\sf F},
{\sf H}, {\sf X}. The time derivatives of $\delta$ functions in
(\ref{eq3.21}) imply that the 2PK Lagrangian depends not only on the
positions ${\bf x}_A$ and the velocities ${\bf v}_A$ of the bodies, but
also on their accelerations ${\bf a}_A$. (Higher derivatives can be
eliminated, at 2PK order, by means of suitable integrations by
parts.) As discussed in \cite{DS85}, this is a consequence of our
choice of gauge for the Einstein-Hilbert action $S_{\rm spin2}$.
Indeed, all the accelerations can be eliminated by choosing for
instance the ``ADM'' (Arnowitt-Deser-Misner) gauge \cite{OOKH},
instead of the harmonic one we use in Appendix~A below to simplify
our calculations. (Actually, the simplest practical way of going from
the harmonic gauge to a higher-derivatives-free gauge is to
``wrongly'' eliminate the higher derivatives by using the equations
of motion in the Lagrangian
\cite{DS85}.)

The 2PK Lagrangian describing the motion of $N$ self-gravitating bodies
is hence obtained straightforwardly from the action (\ref{eq3.14}). We
refer the reader to Appendix~A below for its explicit derivation, and
quote here only its general structure. It can be written as a sum of
$i$-body interaction terms ($1\leq i\leq 4$)
\begin{equation}
L = L^{(1)}+L^{(2)}+L^{(3)}+L^{(4)}+O(1/c^6)\ ,
\label{eq3.22}\end{equation}
where $L^{(1)} = \sum_A L_A^{(1)}= -\sum_A m_A^0c^2\sqrt{1-{\bf
v}_A^2/c^2}$ is the Lagrangian describing free bodies. Its 2PK
expansion, generalizing (\ref{eq2.19a}) to order
$m_Ac^2\times(v_A^2/c^2)^3$, is given by (\ref{eq3.19}). The 2-body
interaction Lagrangian $L^{(2)}$, given by the ``I'' diagrams of
Fig.~\ref{fig7}, has the structure
\begin{eqnarray}
L^{(2)} = {1\over 2}\sum_{A\neq B}\Biggl[
&& {G_{AB}\over G_*}L_{AB}^{(2){\rm G.R.}}
\nonumber \\
&& + {G_{AB}m_A^0m_B^0\overline\gamma_{AB}\over c^2}\left(
{({\bf v}_A-{\bf v}_B)^2\over r_{AB}}
+{f_1({\bf r}_{AB},{\bf v}_A,{\bf v}_B,{\bf a}_A,{\bf a}_B)\over c^2}
\right)
\Biggr]\ ,
\label{eq3.23}\end{eqnarray}
where $L_{AB}^{(2){\rm G.R.}}$ is the expression obtained in general
relativity (pure spin-2 interaction):
\begin{equation}
L_{AB}^{(2){\rm G.R.}} =
G_* m_A^0 m_B^0\left[
{1\over r_{AB}} +
{f_2({\bf r}_{AB},{\bf v}_A,{\bf v}_B)\over c^2}
+{f_3({\bf r}_{AB},{\bf v}_A,{\bf v}_B,
{\bf a}_A,{\bf a}_B)\over c^4}
\right]\ ,
\label{eq3.24}\end{equation}
and where the expression of the function $f_1$ entering the
$\overline\gamma_{AB}/c^4$ term in (\ref{eq3.23}) is given in
Eqs.~(\ref{eqA17})--(\ref{eqA18}).

In other words, the 2-body interaction Lagrangian $L^{(2)}$ in
tensor--multi-scalar theories presents two main differences with
respect to the general relativistic result: (i)~the bare gravitational
constant $G_*$ is replaced by the effective one $G_{AB}$,
Eq.~(\ref{eq2.20a}); (ii)~a correcting term proportional to
$\overline\gamma_{AB}/c^2$ must be added.

The 3-body interaction Lagrangian, corresponding to the {\sf V} and {\sf
T} diagrams of Fig.~\ref{fig7}, has the structure
\begin{equation}
L^{(3)} = {1\over 2}\!\sum_{B\neq A\neq C}\!\!\!
{G_{AB}G_{AC}m_A^0m_B^0m_C^0\over c^2}
\left[
- {\left(1+2\overline\beta^{\ A}_{BC}\right)\over
r_{AB}r_{AC}} +{f_4\over
c^2}+{\overline\beta^{\ A}_{BC} f_5\over c^2}
\right]
+\!\!\sum_{A,B,C}\!\!
{O(\overline\gamma_{AB},\overline\gamma_{AC},
\overline\gamma_{BC})\over c^4}
\, ,
\label{eq3.25}\end{equation}
where the functions $f_4$ and $f_5$ depend only on the positions and the
velocities of the three bodies. The last term in Eq.~(\ref{eq3.25})
denotes a sum of terms which are at least linear in the indicated
$\overline\gamma_{AB}$, and which depend also on positions and
velocities.

Finally, the 4-body interaction Lagrangian, corresponding to the $\in$,
{\sf Z}, {\sf F}, {\sf H} and {\sf X} diagrams of Fig.~\ref{fig7}, has
the structure
\begin{eqnarray}
L^{(4)} & = & \sum_{A,B,C,D} L^{(4){\rm R.G.}}_{ABCD}
+{1\over 6}\sum_{A\neq(B,C,D)}
{G_{AB}G_{AC}G_{AD}m_A^0m_B^0m_C^0m_D^0\over
r_{AB}r_{AC}r_{AD}c^4}\, \varepsilon^{\ \ A}_{BCD}
\nonumber \\
& & +{1\over 2}\sum_{A\neq B\neq C\neq D}
{G_{AB}G_{BC}G_{CD}m_A^0m_B^0m_C^0m_D^0\over
r_{AB}r_{BC}r_{CD}c^4}\, \zeta_{ABCD}
\nonumber \\
& & +{1\over 24\pi}\sum_{A,B,C,D}
{G_*^3m_A^0m_B^0m_C^0m_D^0\over c^4}\, \chi_{ABCD}
\int d^3{\bf x}\,
{({\bf x}-{\bf x}_A)\cdot({\bf x}-{\bf x}_C)
\over
|{\bf x}-{\bf x}_A|^3|{\bf x}-{\bf x}_B|\,|{\bf x}-{\bf x}_C|^3
|{\bf x}-{\bf x}_D|}
\nonumber \\
& & +
\sum_{A,B,C,D}
{O\left(\overline\gamma_{AB},\overline\gamma_{AC},\cdots\right)\over
c^4} + \sum_{A,B,C,D} {O\left(\overline\beta^{\ A}_{BC},
\overline\beta^{\ B}_{AD},\cdots\right)\over c^4}\ ,
\nonumber \\
\label{eq3.26}\end{eqnarray}
where $L^{(4){\rm R.G.}}_{ABCD} \propto G_*^3m_A^0m_B^0m_C^0m_D^0$ is
the general relativistic result, and where the 2PK parameters
$\varepsilon^{\ \ A}_{BCD}$, $\zeta_{ABCD}$ and $\chi_{ABCD}$ have been
defined in Eqs.~(\ref{eq3.18}) above. Note that the notation $A\neq B
\neq C \neq D$ excludes only the equalities $A=B$, $B=C$ or $C=D$.

\subsection{2PN approximation}
The expression of the second post-Newtonian approximation can now be
obtained by expanding the results of the previous subsection in powers
of the compactnesses $s\sim Gm/Rc^2$ of the bodies. To do this, we make
use of results derived in \cite{DEF1}. Following section 8 of this
reference, we define
\begin{mathletters}
\label{eq3.27}
\begin{eqnarray}
c_A \equiv & \displaystyle -2\,{\partial\ln \widetilde m_A\over
\partial
\ln \widetilde G}\ & = O(s)\ ,
\label{eq3.27a} \\
c'_A \equiv & \displaystyle {\partial c_A\over \partial \ln
\widetilde G}\ & = O(s)\ ,
\label{eq3.27b} \\
a_A \equiv &\displaystyle -4\,{\partial\ln \widetilde m_A\over
\partial\overline\gamma}\ & =O(s^2)\ ,
\label{eq3.27c} \\
b_A \equiv &\displaystyle {\partial\ln \widetilde m_A\over
\partial\overline\beta}\ & =O(s^2)\ .
\label{eq3.27d}\end{eqnarray}
\end{mathletters}
Using the results (3.4) of \cite{DEF1} and neglecting the rotational
kinetic energy terms with respect to the pressure (as is appropriate
for solar system bodies), we can write these compactness parameters
explicitly at 2PN order:
\begin{mathletters}
\label{eq3.28}
\begin{eqnarray}
c_A & = & {1\over c^2}\langle \widetilde U\rangle_A
+{2\over c^4}\left\langle
3\overline \gamma {\widetilde p\over \widetilde\sigma}\, \widetilde U
-2\overline\beta \widetilde U^2\right\rangle_A
+O\left({1\over c^6}\right)\ ,
\label{eq3.28a} \\
c'_A & \sim & {1\over c^2}\langle \widetilde U\rangle_A
+O\left({1\over c^4}\right)\ ,
\label{eq3.28b} \\
a_A & = & {12\over c^4}\left\langle {\widetilde p\over \widetilde\sigma}
\, \widetilde U\right\rangle_A
+O\left({1\over c^6}\right)\ ,
\label{eq3.28c} \\
b_A & = & {1\over c^4}\langle \widetilde U^2\rangle_A
+O\left({1\over c^6}\right)\ .
\label{eq3.28d}\end{eqnarray}
\end{mathletters}
Here $\widetilde\sigma \equiv (\widetilde T^{00}+\widetilde T^{ii})/c^2$
denotes the mass density, $\widetilde p$ the pressure, $\widetilde U$
the potential satisfying $\widetilde\Delta \widetilde U = -4\pi
\widetilde G \widetilde\sigma$, and the angular brackets denote an
average weighted by $\widetilde\sigma$~: $\langle f\rangle \equiv \int
\widetilde\sigma f d^3\widetilde{\bf x}/ \int \widetilde\sigma
d^3\widetilde{\bf x}$. The tilde decorating the various quantities
means that they are measured in physical units {\it and\/} expressed in
terms of physically rescaled local coordinates $\widetilde x^\mu =
A(\varphi(x_A)) x^\mu$ adapted to describing the neighborhood of body
$A$. The mass density $\widetilde\sigma$ can be expressed in terms of
the coordinate-conserved baryonic-rest-mass density, say
$\widetilde\mu$, as $\widetilde\sigma = \widetilde\mu +
(\widetilde\mu\widetilde h +2\widetilde p - \widetilde\mu\widetilde
U)/c^2 +O(1/c^4)$, where $\widetilde h$ denotes the enthalpy. Note that
Eq.~(\ref{eq3.28a}) is more accurate than the usually employed 1PN
expression for the compactness\footnote{Note also that it simplifies
very much in the general relativistic case where the $1/c^4$ correction
vanishes.}: $c_A = \langle \widetilde U\rangle_A /c^2 + O(1/c^4) =
-2\widetilde E_A^{\rm grav}/\widetilde m_Ac^2 + O(1/c^4)$. A precise
value of $c'_A$ is not needed for the following as it will only appear
through the combination $(4\overline\beta-\overline\gamma)^2 c'_A$.

As shown in section 8 of \cite{DEF1}, the body-dependent parameters
(\ref{eq2.20}) can then be expanded as
\begin{mathletters}
\label{eq3.29}
\begin{eqnarray}
G_{AB}/G = \widetilde G_{AB}/\widetilde G & = & 1-{\eta\over
2}(c_A+c_B) +\left(\zeta+4\overline\beta-{\overline\gamma\over
2}\right)c_Ac_B +\overline\beta(2+\overline\gamma)(a_A+a_B)
\nonumber \\
&& +\left({\varepsilon\over
2}+\zeta-8\overline\beta^2\right)(b_A+b_B) +O(s^3)\ ,
\label{eq3.29a} \\
\overline\gamma_{AB} & = &
\overline\gamma+\eta\left(1+{\overline\gamma\over 2}\right)(c_A+c_B)
+O(s^2)\ ,
\label{eq3.29b} \\
\overline\beta^{\ A}_{BC} & = &
\overline\beta - \left({\varepsilon\over
2}+{\zeta\over 2}+\overline\beta
-8\overline\beta^2+\overline\beta\overline\gamma\right)c_A
-\left({\zeta\over
2}+\overline\beta-{\eta\overline\beta\over 2}\right)(c_B+c_C)
\nonumber \\
&&-{\eta^2\over 4}\, c'_A +O(s^2)\ .
\label{eq3.29c}\end{eqnarray}
\end{mathletters}
Here we have set $G\equiv G_* (1+\alpha_0^2)$ for
the Einstein-frame effective gravitational constant, and, as usual,
$\eta\equiv4\overline\beta-\overline\gamma$. We have also introduced
the notation
\begin{mathletters}
\label{eq3.30}
\begin{eqnarray}
\varepsilon & \equiv & {(\beta'_{abc}\alpha^a\alpha^b\alpha^c)_0
\over(1+\alpha_0^2)^3}\ ,
\label{eq3.30a}\\
\zeta & \equiv & {(\alpha_a\beta^a_b\beta^b_c\alpha^c)_0\over
(1+\alpha_0^2)^3}\ .
\label{eq3.30b}
\end{eqnarray}
\end{mathletters}
These parameters are the weak-self-gravity limits of the parameters
(\ref{eq3.18a})-(\ref{eq3.18b}) corresponding to the diagrams $\in$ and
{\sf Z}~:
\begin{equation}
\varepsilon^{\ \ A}_{BCD} = \varepsilon + O(s)\quad , \quad
\zeta_{ABCD} = \zeta+O(s) \ .
\label{eq3.31}\end{equation}
The parameter $\chi_{ABCD}$, Eq.~(\ref{eq3.18c}), vanishes in the
weak-self-gravity limit because of the antisymmetry of the Riemann
tensor $R_{abcd}(\varphi) = -R_{bacd}(\varphi)$~:
\begin{equation}
\chi_{ABCD} = (R_{abcd}\alpha^a\alpha^b\alpha^c\alpha^d)_0+O(s) = 0+
O(s^2) \ .
\label{eq3.32}\end{equation}
Indeed, for the same reason, the first correction of order $O(s)$ is
easily seen to vanish too.

Our final conclusion is therefore that the 2PN limit of
tensor--multi-scalar theories involves only the {\it two\/} new
parameters $\varepsilon$ and $\zeta$, Eqs.~(\ref{eq3.30}), besides the
usual 1PN Eddington parameters $\overline\beta$ and $\overline\gamma$.
These parameters consistently enter into several 2PN effects. First,
they parametrize two new independent contributions to the 4-body
interaction Lagrangian:
\begin{equation}
L^{(4)}_{\rm 2PN}(\varepsilon,\zeta) =
{\varepsilon\over 6}
\sum_{A\neq(B,C,D)}{G^3m_A^0m_B^0m_C^0m_D^0\over r_{AB}r_{AC}r_{AD}c^4}
+{\zeta\over 2}
\sum_{A\neq B\neq C\neq D}{G^3m_A^0m_B^0m_C^0m_D^0\over
r_{AB}r_{BC}r_{CD}c^4} \ .
\label{eq3.33}\end{equation}
Second, they parametrize the dependence upon self-gravity effects of the
effective gravitational constant and of the 1PK parameters $\overline
\gamma_{AB}$ and $\overline\beta^{\ A}_{BC}$. Discarding from
Eqs.~(\ref{eq3.29}) the 2PN corrections proportional to the already
experimentally constrained 1PN parameters $\overline \beta$ and
$\overline \gamma$, we can rewrite them in the simpler form
\begin{mathletters}
\label{eq3.34}
\begin{eqnarray}
G_{AB}/G & = & 1+\eta\left(
{E_A^{\rm grav}\over m_Ac^2}
+{E_B^{\rm grav}\over m_Bc^2}
\right)
+4\zeta\left({E_A^{\rm grav}\over m_Ac^2}\right)
\left({E_B^{\rm grav}\over m_Bc^2}\right)
\nonumber \\
& & +\left({\varepsilon\over 2}+\zeta\right){
\langle U^2\rangle_A+\langle U^2\rangle_B\over
c^4} +O(\overline\beta,\overline\gamma)\times O(s^2) +O(s^3)\ ,
\label{eq3.34a} \\
\overline \gamma_{AB} & = & \overline \gamma +
O(\overline\beta,\overline\gamma)\times O(s) +O(s^2)\ ,
\label{eq3.34b} \\
\overline\beta^{\ A}_{BC} & = & \overline\beta +(\varepsilon+\zeta)
{E_A^{\rm grav}\over m_Ac^2}
+\zeta\left(
{E_B^{\rm grav}\over m_Bc^2}
+{E_C^{\rm grav}\over m_Cc^2}
\right)
+O(\overline\beta,\overline\gamma)\times O(s) +O(s^2)\ .
\label{eq3.34c}\end{eqnarray}
\end{mathletters}
Eq.~(\ref{eq3.34a}) shows that $\varepsilon$ and $\zeta$ are two
independent 2PN generalizations of the well-known Nordtvedt 1PN
modification of the gravitational coupling [second term of the
right-hand side of (\ref{eq3.34a}), $\eta(E_A^{\rm grav}/m_Ac^2
+E_B^{\rm grav}/m_Bc^2)$]. In Eqs.~(\ref{eq3.34}), we have simplified
the writing by dropping all tildes in dimensionless ratios such as
$E_A^{\rm grav}/ m_Ac^2$ or $\langle U^2\rangle_A/c^4$. We shall
continue doing so in the following each time this does not lead to any
ambiguity.

In fact the two roles (\ref{eq3.33}), (\ref{eq3.34}) of $\varepsilon$
and $\zeta$ are deeply connected. Indeed, the Lagrangian
(\ref{eq3.33}) represents the interaction between an arbitrary number
of non-self-gravitating mass points. From it we can formally
reconstruct the Lagrangian representing the interaction between $N$
weakly-self-gravitating bodies, $\cal A$, $\cal B$, \dots\ by
considering each body as a collection of mass points $m_A$, $m_B$,
\dots\ (held together by a slow orbital motion within the volume $\cal
A$). Let us denote by $\sigma({\bf x})$ the average mass density within
$\cal A$, such that $\int_{\cal A}d^3{\bf x}\,\sigma({\bf x})=m_{\cal
A}^0= \sum_{A\in {\cal A}} m_A^0$ gives the total Einstein mass of body
$\cal A$. Then, if $A$ and $B$ label two point masses belonging to the
same body $\cal A$, one can rewrite the sum $\sum_{A\neq B}G m_A^0
m_B^0/r_{AB}$ as an integral
\begin{equation}
\int_{\cal A} {G \sigma({\bf x}_1)\sigma({\bf x}_2) d^3{\bf x}_1
d^3{\bf x}_2\over
|{\bf x}_1-{\bf x}_2|} = m^0_{\cal A} \langle U\rangle_{\cal A}+O\left(
{1\over c^2}\right)\ ,
\label{eq3.35}\end{equation}
where $U({\bf x}) \equiv \int d^3{\bf x}\,G\sigma({\bf x}')/|{\bf
x}-{\bf x}'|$ and where the angular brackets denote an average weighted
by $\sigma$. Similarly, if $A$, $B$ and $C$ belong to the same body
$\cal A$, the sum $\sum_{A\neq B\neq C}
G^2m_A^0m_B^0m_C^0/r_{AB}r_{BC}$ can be rewritten as an integral
\begin{equation}
\int_{\cal A} {G^2 \sigma({\bf x}_1)\sigma({\bf x}_2)\sigma({\bf x}_3)
d^3{\bf x}_1 d^3{\bf x}_2 d^3{\bf x}_3\over
|{\bf x}_1-{\bf x}_2|\, |{\bf x}_2-{\bf x}_3|} =
m^0_{\cal A} \langle U^2\rangle_{\cal A}+O\left(
{1\over c^2}\right)\ .
\label{eq3.36}\end{equation}
Using such results, the interaction Lagrangian (\ref{eq3.33}) leads to
\begin{eqnarray}
L(\varepsilon,\zeta) & = &
\sum_{{\cal A}\neq({\cal B},{\cal C},{\cal D})}
{G^3m_{\cal A}^0m_{\cal B}^0m_{\cal C}^0m_{\cal D}^0\over
r_{\cal AB}r_{\cal AC}r_{\cal AD}c^4}\,
{\varepsilon\over 6}
+\sum_{{\cal A}\neq {\cal B}\neq {\cal C}\neq {\cal D}}
{G^3m_{\cal A}^0m_{\cal B}^0m_{\cal C}^0m_{\cal D}^0\over
r_{\cal AB}r_{\cal BC}r_{\cal CD}c^4}\,
{\zeta\over 2}
\nonumber \\
& & +\sum_{{\cal B}\neq {\cal A}\neq {\cal C}}
{G^2m_{\cal A}^0m_{\cal B}^0m_{\cal C}^0\over
r_{\cal AB}r_{\cal AC} c^4}
\left[
\left(3\times{\varepsilon\over 6}+{\zeta\over 2}\right)
\langle U\rangle_{\cal A} +{\zeta\over 2}\left(
\langle U\rangle_{\cal B}+\langle U\rangle_{\cal C}
\right)\right]
\nonumber \\
& & +\sum_{{\cal A}\neq {\cal B}}{Gm_{\cal A}^0m_{\cal B}^0\over
r_{\cal AB} c^4}
\left[
{\zeta\over 2}\langle U\rangle_{\cal A} \langle U\rangle_{\cal B}
+\left({3\over 2}\times{\varepsilon\over 6}+{\zeta\over 2}\right)
\left(
\langle U^2\rangle_{\cal A}+\langle U^2\rangle_{\cal B}
\right)\right]\ ,
\label{eq3.37}\end{eqnarray}
where we have taken into account the different ways to choose
infinitesimal elements in the same body to compute the correct
multiplicities. [In pictorial language, if we denote the first two
terms by $({\in}^{\ \ {\cal A}}_{\cal BCD}/3)$ and $({\sf Z}_{\cal
ABCD}/2)$ respectively\footnote{The normalization of the $\in$ term is
chosen for consistency with Appendix~A.}, the following ones correspond
to $3\times({\in}^{\ \ {\cal A}}_{\cal ABC}/3)+({\sf Z}_{\cal
BAAC}/2)$, $({\sf Z}_{\cal BBAC}/2+{\sf Z}_{\cal BACC}/2)$, $({\sf
Z}_{\cal AABB}/2)$, ${3\over 2}\times({\in}^{\ \ {\cal A}}_{\cal
AAB}/3+ {\in}^{\ \ {\cal B}}_{\cal BBA}/3)+ ({\sf Z}_{\cal AAAB}/2+{\sf
Z}_{\cal ABBB}/2)$, where a repeated index means infinitesimal volume
elements inside the same body.]

The results (\ref{eq3.34}) can be directly read off the Lagrangian
(\ref{eq3.37}). Indeed, the last sum in (\ref{eq3.37}) leads to the
$\varepsilon$ and $\zeta$ renormalizations of the effective
gravitational constant (\ref{eq3.34a}), while the 3-body sum
$\sum_{{\cal B}\neq {\cal A}\neq {\cal C}}$ yields the renormalizations
(\ref{eq3.34c}) of the body-dependent Eddington parameter
$\overline\beta^{\ {\cal A}}_{\cal BC}$.

\section{Second post-Newtonian order metric}
\subsection{$N$-body physical metric}
As is well known, the $N$-body Lagrangian contains enough information to
derive the physical metric $\widetilde g_{\mu\nu}$ at any spacetime
point $x^\lambda$ outside the bodies generating $\widetilde
g_{\mu\nu}$. Deriving this metric is important because it allows one to
compute 2PN effects on, for instance, clock comparison, light
deflection or time-delay experiments. To compute $\widetilde
g_{\mu\nu}(x^\lambda)$, one introduces a test particle of negligible
mass\footnote{The index $\scriptstyle 0$ of this test mass should not
be confused with the time component $x^0$. It has been chosen so that
the body-dependent quantities $\alpha_A^a$, $\beta_A^{ab}$, \dots\ of
Eqs.~(\ref{eq2.16}) reduce to their background values $\alpha_0^a\equiv
\alpha^a(\varphi_0)$, $\beta_0^{ab}\equiv \beta^{ab}(\varphi_0)$,
\dots\ for the test mass $\widetilde m_0$.} $\widetilde m_0$ located at
$x^\lambda = (x^0, x^i)$, with an arbitrary 4-velocity $v^\mu = (c,
v^i)$. We can now write the $N+1$-body Lagrangian describing the $N$
massive bodies and this test particle, and identify its $\widetilde
m_0$-dependent part with the individual Lagrangian of the latter,
namely
\begin{eqnarray}
L^{\rm test{\scriptscriptstyle -}particle}(x^\lambda, v^i)
&=& - \widetilde m_0 c \sqrt{-\widetilde
g_{\mu\nu}(x^\lambda) v^\mu v^\nu}\nonumber\\
&=& -\widetilde m_0 c^2 \left[-\widetilde
g_{00}(x^\lambda) - 2 \widetilde
g_{0i}(x^\lambda) v^i/c - \widetilde g_{ij}(x^\lambda) v^i v^j / c^2
\right]^{1\over 2} \ .
\label{eq4.1}
\end{eqnarray}
This identification therefore allows us to compute $\widetilde
g_{\mu\nu}(x^\lambda)$ up to order $1/c^6$ included for $\widetilde
g_{00}$ [{\it i.e.} up to $O(Gm/rc^2)^3$ included], $1/c^5$ for
$\widetilde g_{0i}$, and $1/c^4$ for $\widetilde g_{ij}$.

If we are concerned only by the dependence upon the new parameters
$\varepsilon$ and $\zeta$ appearing at the 2PN order,
Eq.~(\ref{eq3.37}), we can consider a particle at rest, since the
$\varepsilon$ and $\zeta$ terms are not velocity dependent. We
therefore conclude immediately that $\widetilde g_{0i}$ and $\widetilde
g_{ij}$ do not involve the parameters $\varepsilon$ and $\zeta$ at the
2PN approximation ({\it i.e.}, at order $1/c^5$ for $\widetilde g_{0i}$
and $1/c^4$ for $\widetilde g_{ij}$), and that $\widetilde g_{00} = -1
+\delta \widetilde g_{00}$ can be deduced from the Lagrangian
\begin{equation}
L^{\rm test{\scriptscriptstyle -}particle}(x^\lambda) =
-\widetilde m_0c^2\sqrt{1-\delta\widetilde g_{00}(x^\lambda)}
= -\widetilde m_0c^2
+{1\over 2}\widetilde m_0c^2 \delta\widetilde g_{00}
+O(\delta\widetilde g_{00})^2\ .
\label{eq4.2}\end{equation}
Comparing (\ref{eq4.2}) with (\ref{eq3.37}) yields
\begin{eqnarray}
\delta_{\rm 2PN} \widetilde g_{00}(\varepsilon,\zeta) & = &
{\varepsilon\over 3} \sum_{A,B,C} { G^3 m_A m_B m_C\over
r_{0A} r_{0B} r_{0C} c^6}
+\varepsilon \sum_{A\neq (B,C)}
{G^3 m_A m_B m_C\over r_{0A} r_{AB} r_{AC}c^6}
+2\varepsilon\sum_{A\neq B}{G^2 m_A m_B\over r_{0A} r_{AB} c^6}
\langle U \rangle_A
\nonumber \\
&& + 2\zeta\sum_A\sum_{B\neq C}{G^3 m_A m_B m_C\over
r_{0A} r_{0B} r_{BC}c^6}
+2\zeta \sum_{A\neq B\neq C}
{G^3 m_A m_B m_C\over r_{0A} r_{AB} r_{BC}c^6}
\nonumber \\
&& +\zeta \sum_{A,B} {G^2 m_A m_B\over
r_{0A} r_{0B} c^6}(\langle U \rangle_A +\langle U \rangle_B)
+2\zeta\sum_{A\neq B}{G^2 m_A m_B\over r_{0A} r_{AB} c^6}
(\langle U \rangle_A+\langle U \rangle_B)
\nonumber \\
&& +(\varepsilon+2\zeta)\sum_A{G m_A\over r_{0A} c^6}
\langle U^2 \rangle_A\ ,
\label{eq4.3}
\end{eqnarray}
where $r_{0A}\equiv |{\bf x} - {\bf x}_A|$. [All the masses entering
this equation are evaluated at $\varphi=\varphi_0$, although we have
dropped their index $\scriptstyle 0$ for easier reading.] When a
continuous description of the source bodies is used, this takes the
simpler form
\begin{eqnarray}
\delta_{\rm 2PN} \widetilde g_{00}(\varepsilon,\zeta) & = &
{\varepsilon\over 3c^6} U^3({\bf x})
+{\varepsilon\over c^6}\!\int\! d^3{\bf x}'\,
{G \sigma({\bf x}')
U^2({\bf x}') \over
|{\bf x}-{\bf x}'|}
+{2\zeta\over c^6}\!\int\! d^3{\bf x}'\,
{G\sigma({\bf x}')\over
|{\bf x}-{\bf x}'|}
\!\int\! d^3{\bf x}''
{G\sigma({\bf x}'')
U({\bf x}'')\over
|{\bf x}'-{\bf x}''|}
\nonumber \\
& & + {2\zeta\over c^6} U({\bf x})
\!\int\! d^3 {\bf x}'{G\sigma
({\bf x}')
U({\bf x}')\over
|{\bf x}-{\bf x}'|}
\ .
\label{eq4.4}
\end{eqnarray}
Using a self-explanatory notation, the four terms of (\ref{eq4.4})
correspond to the diagrams ${\in}^{\ \ 0}_{AAA}$, ${\in}^{\ \
A}_{AA0}$, ${\sf Z}_{AAA0}$ and ${\sf Z}_{AA0A}$ respectively, the
index $\scriptstyle A$ meaning here any infinitesimal volume of matter,
and the index $\scriptstyle 0$ referring to the spacetime point
$x^\lambda$ where the metric is computed. It should be noted that
formulae (\ref{eq4.3}) and (\ref{eq4.4}) are valid in the
$g^*$-harmonic gauge used to derive the $N$-body Lagrangian in
subsection III--A. However, it is easy to see that the corrections
proportional to $\varepsilon$ and $\zeta$ are the same in all usual
coordinate systems (ADM-gauge, $\widetilde g$-harmonic, \dots), since
these parameters appear for the first time at order $1/c^6$ in the
time-time component of the metric\footnote{Of course, it is always
possible to introduce by hand a spurious dependence on these
parameters, by redefining for instance the spatial variables as $x'^i =
x^i + O(\varepsilon,\zeta)/c^4$, whereas any usual coordinate
transformation involves $\varepsilon$ and $\zeta$ at order $1/c^6$.}.
In particular, the fact that $\widetilde g_{0i}$ and $\widetilde g_{ij}$
do not involve $\varepsilon$ and $\zeta$ at the 2PN order is valid in
any of these coordinate systems.

Let us mention for completeness that the presence of $\varepsilon$ and
$\zeta$ modifies the total energy of a gravitating system by the amount
\begin{equation}
\delta_{\rm 2PN} \widetilde E(\varepsilon,\zeta) =
-{\varepsilon\over 6 c^4}\int d^3\widetilde{\bf x}\,
\widetilde\sigma(\widetilde{\bf x})
\widetilde U^3(\widetilde{\bf x})
-{\zeta\over 2 c^4}\int\!\!\int d^3\widetilde{\bf x}\,
d^3\widetilde{\bf x}'\,
{\widetilde\sigma(\widetilde{\bf x})
\widetilde U(\widetilde{\bf x})
\widetilde\sigma(\widetilde{\bf x}')
\widetilde U(\widetilde{\bf x}')
\over |\widetilde{\bf x}-\widetilde{\bf x}'|}\ ,
\label{eq4.5}\end{equation}
where the tilde decorating the various quantities means as before that
they are measured in the physical units (\ref{eq2.6}).

\subsection{Exact one-body metric}
In the present subsection, we verify and complement the above results by
using the exact static and spherically symmetric solution of the field
equations (\ref{eq2.8a})-(\ref{eq2.8b}) derived in section 2 of
\cite{DEF1}. In the coordinate system introduced by Just
\cite{J59,CEF1},
\begin{equation}
ds_*^2 = -e^\nu c^2dt^2 + e^{-\nu}\left[
dr^2+e^\lambda(d\theta^2+\sin^2\theta\, d\phi^2)
\right]\ ,
\label{eq4.6}\end{equation}
we found the solution
\begin{mathletters}
\label{eq4.7}
\begin{eqnarray}
e^\lambda & = & r^2-ar\ ,
\label{eq4.7a} \\
e^\nu & = & (1-a/r)^{b/a}\ ,
\label{eq4.7b}\end{eqnarray}
\end{mathletters}
where
\begin{equation}
b\equiv {2G_*m_A\over c^2} =
{2G_*\over c^4}\int\sqrt{g_*}d^3{\bf x}(-T_{*0}^0+T_{*i}^i)
\label{eq4.8}\end{equation}
is the Einstein-frame mass parameter, and $a$ is a constant of
integration. Introducing the parameter $p\equiv(1/a)\ln(1-a/r)$, the
equations controlling the radial variation of the scalar fields can be
expressed by saying that $\varphi^a(p)$ follows a geodesic of the
metric $\gamma_{ab}(\varphi^c)$, {\it i.e.}, $\delta\int
\gamma_{ab}(\varphi^c(p)) (d\varphi^a/dp)(d\varphi^b/dp)dp = 0$. In
particular, $p$ being an affine parameter, the norm of $d\varphi^a/dp$
is constant and given by
\begin{equation}
4\gamma_{ab}(\varphi){d\varphi^a\over dp}
{d\varphi^b\over dp} = a^2 - b^2
= {\rm const}\ .
\label{eq4.9}\end{equation}
The actual geodesic traced out by $\varphi^a(p)$ in the internal scalar
space is uniquely determined by the initial values at $p=0$ ({\it
i.e.}, at spatial infinity):
$\varphi^a(p=0)=\varphi^a(r=\infty)=\varphi^a_0$ and
$(d\varphi^a/dp)(p=0) = k^a$, say. Assuming $\varphi_0^a$ and $k^a$
given, we can choose field variables in the scalar space so that the
metric $\gamma_{ab}(\varphi)$ reduces to the constant $\delta_{ab}$ all
along the line $\varphi(r)_{r\in [0,\infty]}$ (Fermi coordinate system
in the $\varphi$-space). This allows us to write the solution of
(\ref{eq4.9}) in the simple form
\begin{equation}
\varphi^a = \varphi_0^a + p k^a = \varphi_0^a +{k^a\over
a}\, \ln\left(1-{a\over r}\right)\ .
\label{eq4.10}\end{equation}
Using then the results of \cite{DEF1}, we can relate
the $n$ integration constants $k^a$ to the matter distribution:
\begin{equation}
k^a = {G_*\over c^4} \int\sqrt{g_*}d^3{\bf x}\, \alpha^a(\varphi)
(-T_{*0}^0-T_{*i}^i)\ .
\label{eq4.11}\end{equation}
Note that (\ref{eq4.11}) is valid only if the $\varphi^a$-coordinates
are ``Fermi'' all along the solution $\varphi^a(r)$. When using generic
scalar variables, the quantity $\alpha^a(\varphi)$ appearing on the
right-hand side of (\ref{eq4.11}) must be replaced by the
parallel-transported value of the vector $\alpha^a(\varphi)$ up to the
point $\varphi_0^a$. To simplify, the reader can think that we work
from the beginning with a {\it flat\/} $\sigma$-model metric
$\gamma_{ab}(\varphi)=\delta_{ab}$. The only 2PN contribution that we
would forget in that case would derive from the first {\sf X} diagram
of Fig.~\ref{fig7}, and would involve the contraction
$R_{abcd}(\varphi)\alpha_A^a\alpha_A^b\alpha_A^c\alpha_0^d$ which
vanishes identically because of the antisymmetry of the Riemann
curvature tensor $R_{abcd}(\varphi)=-R_{bacd}(\varphi)$.

By comparing the behavior at infinity of the solution (\ref{eq4.10})
\begin{equation}
\varphi^a = \varphi_0^a - {k^a\over r} +O\left({1\over r^2}\right)
\label{eq4.12}\end{equation}
with the one obtained for an isolated pointlike body $A$ (see
Eq.~(\ref{eq2.17}))
\begin{equation}
\varphi^a = \varphi_0^a - {G_*\alpha_A^a m_A\over r c^2} + O\left(
{1\over r^2}\right)
= \varphi_0^a - \alpha_A^a{b\over 2r} + O\left({1\over r^2}\right)\ ,
\label{eq4.13}\end{equation}
we can deduce that $k^a = G_* m_A \alpha_A^a/c^2$ so that, using
(\ref{eq4.8}) and (\ref{eq4.11}), the actual expression of the
parameter $\alpha_A^a$ for extended bodies reads:
\begin{equation}
\alpha_A^a = {2k^a\over b}
={\int\sqrt{g_*}d^3{\bf x}\,\alpha^a(\varphi)(-T_{*0}^0-T_{*i}^i)
\over
\int\sqrt{g_*}d^3{\bf x}(-T_{*0}^0+T_{*i}^i)}\ .
\label{eq4.14}\end{equation}
Note the change of sign of the pressure term $T_{*i}^i$ between the
numerator and the denominator, which makes $\alpha_A^a$ differ from
$\alpha_0^a$ at order $1/c^2$ even if the function $\alpha^a(\varphi)$
is a constant, as in the Jordan--Fierz--Brans--Dicke theory. In the
latter case, one finds $\alpha_a^A =
\alpha_a^0\times(1-2\int\sqrt{g_*}d^3{\bf x}\, T_{*i}^i/m_A c^2)$. When
comparing this result with $\alpha_a^A = \alpha_a^0 + (\partial
\widetilde m_A/\partial\ln\widetilde G) (\partial\ln\widetilde
G/\partial\varphi^a) = \alpha_a^0 \times(1-c_A)$, we get the following
expression for the compactness in Jordan--Fierz--Brans--Dicke theory:
\begin{equation}
c_A = 2\int\sqrt{g_*}d^3{\bf x}\, T_{*i}^i/m_Ac^2\ .
\label{eq4.15}\end{equation}
This result is an (exact) generalization of the Newtonian virial theorem
$\int d^3{\bf x}\, \sigma U = 6\int d^3{\bf x}\, p +O(1/c^2)$ which is
valid in the Jordan--Fierz--Brans--Dicke theory and therefore also in
general relativity (in the limit $\alpha^a \rightarrow 0$).

The exact solution (\ref{eq4.7})-(\ref{eq4.10}) can now be expanded \`a
la Eddington in powers of $1/r$. Although the coordinate system
(\ref{eq4.6}) is as good as any other to verify the 2PN terms involving
the new parameters $\varepsilon$ and $\zeta$, it is convenient to
express our results in isotropic coordinates, to make better contact
with the general relativistic result. Let us then define a new radial
coordinate $\rho$ such that $\rho(1+a/4\rho)^2 = r$. The Einstein line
element (\ref{eq4.6}) now reads~\cite{CEF1}
\begin{eqnarray}
ds_*^2 & = & -\left(1-{a\over 4\rho}\right)^{2b/a}
\left(1+{a\over 4\rho}\right)^{-2b/a}c^2dt^2
\nonumber \\
& & +\left(1-{a\over 4\rho}\right)^{2-2b/a}
\left(1+{a\over 4\rho}\right)^{2+2b/a}
\left[d\rho^2+\rho^2(d\theta^2+\sin^2\theta d\phi^2)\right]
\nonumber \\
& = & -c^2dt^2\left[
1-{b\over \rho}+{b^2\over 2\rho^2}-{b^3\over 6\rho^3}\left(
1+{1+\alpha_A^2\over 8}\right)+O\left({1\over \rho^4}\right)
\right]
\nonumber \\
& & +\left[d\rho^2+\rho^2(d\theta^2+\sin^2\theta
d\phi^2)\right]\times
\left[
1+{b\over \rho}+{b^2\over 2\rho^2}\left(
1-{1+\alpha_A^2\over 4}\right)+O\left({1\over \rho^3}\right)
\right]\ ,
\label{eq4.16}\end{eqnarray}
where we have used Eqs.~(\ref{eq4.9}) and (\ref{eq4.14}) to replace the
constant $a^2$ in terms of $b^2$ and
$\alpha_A^2\equiv(\alpha_A^a\gamma_{ab}\alpha_A^b)_{\varphi=\varphi_0}$.
The metric which determines the dynamics of test particles in the
vicinity of $A$ is not the Einstein metric $g^*_{\mu\nu}$, but the
physical one $\widetilde g_{\mu\nu} = A^2(\varphi)g^*_{\mu\nu}$. We
must therefore expand also $A^2(\varphi)$ in powers of $1/\rho$, by
using the exact solution (\ref{eq4.10}) of the scalar fields. We get
\begin{eqnarray}
{A^2(\varphi)\over A^2(\varphi_0)} & = & 1 -(\alpha_A\alpha_0)
{b\over \rho} +\left[(\alpha_A\alpha_0)^2 +
{(\alpha_A\beta_0\alpha_A)\over 2}\right]{b^2\over 2\rho^2}
\nonumber \\
& & -\left[
(\beta'^{\ \ 0}_{AAA})+6(\alpha_A\alpha_0)(\alpha_A\beta_0\alpha_A)+
4(\alpha_A\alpha_0)^3 +{(\alpha_A\alpha_0)(1+\alpha_A^2)\over 2}\right]
{b^3\over 24\rho^3} +O\left({1\over \rho^4}\right)\ ,
\nonumber \\
&&\label{eq4.17}\end{eqnarray}
where we have set $(\alpha_A\alpha_0) \equiv
(\alpha_A^a\gamma_{ab}\alpha^b)_{\varphi=\varphi_0}$,
$(\alpha_A\beta_0\alpha_A) \equiv
(\alpha_A^a\beta_{ab}\alpha_A^b)_{\varphi=\varphi_0}$,
$(\beta'_{abc}\alpha_A^a\alpha_A^b\alpha_A^c)_{\varphi=\varphi_0}$, and
where the $\sigma$-model tensors $\alpha^a$, $\beta_{ab}$ and
$\beta'_{abc}$ have been defined in Eqs.~(\ref{eq2.11}) above. Finally,
when using the physical coordinates $\widetilde\rho\equiv
A(\varphi_0)\rho$ and $\widetilde t \equiv A(\varphi_0) t$ (see
Eq.~(\ref{eq2.6})), we find that the physical line element $d\widetilde
s^2 = A^2(\varphi)ds_*^2$ reads
\begin{eqnarray}
d\widetilde s^2 & = & -c^2d\widetilde t^2
\left[
1 -2{\widetilde\mu_A\over \widetilde\rho c^2}
+ 2 \left({\widetilde\mu_A\over \widetilde\rho
c^2}\right)^2 (1+\overline \beta^{\ 0}_{AA}) -{3\over 2}
\left({\widetilde\mu_A\over \widetilde\rho
c^2}\right)^3(1+\overline{\rm B})+O\left({1\over
\widetilde\rho^4}\right)\right]
\nonumber \\
& & +\left[d\widetilde\rho^2+\widetilde\rho^2
(d\theta^2+\sin^2\theta d\phi^2)\right]\times
\left[
1 +2{\widetilde\mu_A\over \widetilde\rho c^2}
(1+\overline\gamma_{A0})+{3\over 2}
\left({\widetilde\mu_A\over \widetilde\rho
c^2}\right)^2(1+\overline\Gamma)+O\left({1\over
\widetilde\rho^3}\right)\right]\ ,
\nonumber \\
&&\label{eq4.18}\end{eqnarray}
where
\begin{mathletters}
\label{eq4.19}
\begin{eqnarray}
\overline{\rm B} & \equiv & {2\over 9}\varepsilon^{\ \ 0}_{AAA}
+{8\over 3}\overline\beta^{\ 0}_{AA} +(4+\overline
\gamma_{A0}){\overline
\gamma_{A0}\over 36} - {(1+\overline \gamma_{A0}/2)^2\over
(1+\overline \gamma_{AA}/2)}\, {\overline \gamma_{AA}\over 18}\ ,
\label{eq4.19a} \\
\overline\Gamma & \equiv & {4\over 3} \overline \beta^{\ 0}_{AA} +
(28+15\overline \gamma_{A0}){\overline \gamma_{A0}\over 12}
+{(1+\overline \gamma_{A0}/2)^2\over
(1+\overline \gamma_{AA}/2)}\, {\overline \gamma_{AA}\over 6}\ ,
\label{eq4.19b}\end{eqnarray}
\end{mathletters}
and where we have set
\begin{mathletters}
\label{eq4.20}
\begin{eqnarray}
\widetilde\mu_A \equiv \mu_A/ A(\varphi_0)
& \equiv & \widetilde G_{A0} \widetilde m_A = G_* A^2(\varphi_0)
\left[1+(\alpha_A\alpha_0)\right] m_A\ ,
\label{eq4.20a} \\
\overline \gamma_{A0} & \equiv & -2{(\alpha_A\alpha_0)\over
1+(\alpha_A\alpha_0)}
\quad , \quad
\overline \gamma_{AA} \equiv -2{\alpha_A^2\over 1+\alpha_A^2}\ ,
\label{eq4.20b} \\
\overline \beta^{\ 0}_{AA} & \equiv & {1\over
2}(\alpha_A\beta_0\alpha_A)/\left[ 1+(\alpha_A\alpha_0)\right]^2\ ,
\label{eq4.20c} \\
\varepsilon^{\ \ 0}_{AAA} & \equiv &
(\beta'_{abc}\alpha_A^a\alpha_A^b\alpha_A^c)/\left[
1+(\alpha_A\alpha_0)\right]^3\ .
\label{eq4.20d}\end{eqnarray}
\end{mathletters}

As a check of the consistency of our method, it is useful to compare the
2PK Lagrangian of subsection III--B with the test-particle Lagrangian
(\ref{eq4.1}) written for the metric (\ref{eq4.18}). Let us first
rewrite this metric in $g^*$-harmonic coordinates. We find
\begin{mathletters}
\label{eq4.21}
\begin{eqnarray}
-\widetilde g_{00} & = &
1 - 2{\mu_A\over r c^2}
+ 2 \left({\mu_A\over r c^2}\right)^2
\left(1+\overline\beta^{\ 0}_{AA}\right)
\nonumber \\
&& - 2 \left({\mu_A\over r c^2}\right)^3
\left(
1+{3\overline{\rm B}\over 4}
+ {\overline\gamma_{A0}(4+\overline\gamma_{A0})
\over 8(2+\overline\gamma_{AA})}
- {\overline\gamma_{AA}\over
4(2+\overline\gamma_{AA})}
\right)
+ O\left({1\over c^8}\right)\ ,
\label{eq4.21a}\\
\widetilde g_{ij} & = &
\delta_{ij}\left[ 1+ 2{\mu_A\over r c^2}(1+\overline\gamma_{A0})
+ \left({\mu_A\over r c^2}\right)^2\left(
1+2\overline\beta^{\ 0}_{AA}
+3\overline\gamma_{A0}
+{7\over 4}(\overline\gamma_{A0})^2
+{\overline\gamma_{AA}(2+\overline\gamma_{A0})^2\over
4(2+\overline\gamma_{AA})}
\right)\right]
\nonumber \\
&& +{x^ix^j\over r^2}\left({\mu_A\over r c^2}\right)^2
{(2+\overline\gamma_{A0})^2\over
2(2+\overline\gamma_{AA})}
+O\left({1\over c^6}\right)\ ,
\label{eq4.21b}\end{eqnarray}
\end{mathletters}
where $r = \sqrt{\delta_{ij}x^ix^j}$ denotes the $g^*$-harmonic
radius, related to the isotropic radius $\rho$ by
\begin{equation}
r = \rho\left[1+\left({\mu_A\over 2\rho c^2}\right)^2
{(1+\overline\gamma_{A0}/2)^2\over
1+\overline\gamma_{AA}/2}
+O\left({1\over c^6}\right)
\right]\ .
\label{eq4.22}\end{equation}
[This $g^*$-harmonic radius $r$ should not be confused with the Just
radius used in Eqs.~(\ref{eq4.6})---(\ref{eq4.13}) above, also denoted
by $r$.] The test-particle Lagrangian deduced from the metric
(\ref{eq4.21}) agrees with the Lagrangian
(\ref{eq3.22})---(\ref{eq3.26}), written in the particular case of two
bodies $m_0$, $m_A$. It is notably easy to check the ${1\over
6}\varepsilon^{\ \ 0}_{AAA}$ contribution in (\ref{eq3.26}), which
comes from the ${3\over 4}\overline{\rm B}$ term of (\ref{eq4.21a}).
[The $\zeta_{A0A0}$ and $\chi_{A0A0}$ contributions in (\ref{eq3.26})
cannot be checked, because they involve the square of the test mass
$m_0$.] To ease the reading, we have used a slightly inconsistent
notation in expressing the metric coefficient $\widetilde g_{\mu\nu}$
(corresponding to the coordinate system $\widetilde x$) in terms of the
original (Einstein-frame) coordinates $x$. We use here, as we did
above, the simplifying fact that dimensionless ratios such as
$\mu/rc^2$ are numerically equal to their physical-units counterparts
$\widetilde\mu/ \widetilde rc^2$.

The 2PN (weak-self-gravity) limit of the one-body
metric (\ref{eq4.18}) [or (\ref{eq4.21})] can be obtained by
expanding the body-dependent parameters in powers of the compactness
of body $A$, as in Eqs.~(\ref{eq3.29}) above:
\begin{mathletters}
\label{eq4.23}
\begin{eqnarray}
\mu_A & = & G m_A\left[
1-{\eta\over 2}\, c_A +\overline\beta(2+\overline \gamma) a_A
+\left({\varepsilon\over 2}+\zeta-8\overline \beta^2 \right)b_A
\right]+O(s^3) \ ,
\label{eq4.23a} \\
\overline\beta^{\ 0}_{AA} & = & \overline\beta
-(\zeta+2\overline\beta-\eta\overline\beta)c_A +O(s^2)
\nonumber\\
&& \Leftrightarrow\quad
\mu_A^2(1+\overline\beta^{\ 0}_{AA}) =
(G m_A)^2\left(
1+\overline\beta-(\zeta+6\overline\beta-\overline\gamma)
c_A+O(s^2)\right)\ ,
\label{eq4.23b} \\
\overline{\rm B} & = & {2\over 9}\, \varepsilon +{8\over
3}\, \overline\beta +{\overline\gamma\over 18} +O(s)\ ,
\label{eq4.23c}\\
\overline \gamma_{A0} & = & \overline\gamma +\eta\left(
1+{\overline \gamma\over 2}\right)c_A +O(s^2)
\nonumber \\
&& \Leftrightarrow\quad
\mu_A(1+\overline\gamma_{A0}) =
G m_A\left(
1+\overline\gamma +{\eta\over 2}\, c_A+O(s^2)\right)\ ,
\label{eq4.23d} \\
\overline\Gamma & = & {4\over 3}\, \overline\beta +
{15+8\overline\gamma\over 6}\, \overline \gamma +O(s)\ .
\label{eq4.23e}
\end{eqnarray}
\end{mathletters}
We thus recover that the spatial metric $\widetilde g_{ij}$ does not
depend on the parameters $\varepsilon$ and $\zeta$ at the 2PN order, as
shown by Eqs.~(\ref{eq4.23d}) and (\ref{eq4.23e}). On the other hand,
the $\varepsilon$ and $\zeta$ contributions to $\widetilde g_{00}$ can
easily be deduced from the above results,
\begin{eqnarray}
\delta_{\rm 2PN}\widetilde g_{00}(\varepsilon,\zeta)
& = & {(\varepsilon+2\zeta)\over c^6}\,
{G m_A\over \rho}\langle U^2\rangle_A
+{2\zeta\over c^6}\left( {G m_A\over
\rho} \right)^2
\langle U\rangle_A
+{\varepsilon\over 3c^6}
\left({G m_A\over \rho} \right)^3
\nonumber \\
& & +O(\overline\beta,\overline\gamma)
+O\left({1\over c^8}\right)
\ ,
\label{eq4.24}\end{eqnarray}
consistently with Eqs.~(\ref{eq4.3})--(\ref{eq4.4}).

\section{2PN experiments}
\subsection{Constraints from 1PN experiments}
Solar-system experiments impose tight bounds on the Eddington parameters
$\overline\gamma$ and $\overline\beta$ \cite{R79,Di94}. Using
Eqs.~(\ref{eq2.14b}), they can be written as
\begin{mathletters}
\label{eq5.1}
\begin{eqnarray}
|\overline\gamma|<2\times10^{-3}&\quad\Leftrightarrow\quad&
-10^{-3}< \displaystyle -{\overline\gamma\over 2+\overline\gamma} =
(\alpha^a\gamma_{ab}\alpha^b)_0 <10^{-3}\ ,
\label{eq5.1a} \\
|\overline\beta|<6\times10^{-4}&\quad\Leftrightarrow\quad&
-1.2\times 10^{-3}< \displaystyle {8\overline\beta\over
(2+\overline\gamma)^2} = (\alpha^a\beta_{ab}\alpha^b)_0 <1.2\times
10^{-3}\ .
\label{eq5.1b}\end{eqnarray}
\end{mathletters}
Let us first discuss the lessons one can draw from the bounds
(\ref{eq5.1}) if one interprets them within the most natural
theoretical framework. Since we assume that the ($\sigma$-model) metric
$\gamma_{ab}$ is positive definite, so that the theory does not
incorporate any negative-energy excitation, Eq.~(\ref{eq5.1a})
constrains the magnitude of all the components of $\alpha_0^a$. 1PN
experiments thereby constrain the linear interaction of the scalar
fields to matter to be small. On the other hand, our diagrammatic
analysis of section III shows that any observable deviation from
general relativity involves at least two factors $\alpha_0^a$, to fill
the end blobs connected to scalar propagators, in diagrams such as
Fig.\ref{fig7} or any higher order ones. Moreover, theoretical
``naturalness'' leads us to expect the coupling function $A(\varphi)$
entering Eq.~(\ref{eq2.3}) not to involve any large dimensionless
number, so that the successive derivatives of $\ln A(\varphi)$ [such as
$\beta^0_{ab}$, $\beta'^0_{abc}$, \dots] are {\it a priori\/} expected
to be of order unity. The conclusion is that we anticipate the new
parameters entering any post-Newtonian order to be at most of the same
order of magnitude as $\overline\gamma$. In particular, the 2PN
parameters (\ref{eq3.30}), which involve respectively 3 and 2 factors
$\alpha^a_0$, are expected to be of order $|\varepsilon|\sim
|\alpha_0|^3\lesssim 3 \times 10^{-5}$ and $|\zeta|\sim
\alpha_0^2\lesssim 10^{-3}$. The corresponding 2PN deviations from
general relativity are thus generically constrained at the level
$\lesssim 10^{-3}\times (Gm/Rc^2)^2\sim 10^{-15}$, too small to be
detectable in the solar system, even in future high-precision
experiments. From this point of view, our conclusion is therefore that
solar-system tests cannot probe the 2PN structure of gravity. However,
they can give a clean access to the Eddington parameter
$\overline\gamma$, which is of greatest significance as it measures the
basic coupling strength between matter and the scalar fields, and
constrains all the other PN parameters. Within this viewpoint, the only
meaningful testing ground for 2PN and higher-order effects are
binary-pulsar experiments. Indeed, since the self-energy $Gm/Rc^2$ of a
neutron star is typically of order $0.2$, the 2PN effects $\lesssim
10^{-3}\times(Gm/Rc^2)^2$ can yield significant deviations on timing
data. Moreover, in such strong-field conditions, the sum of the series
of all higher post-Newtonian orders may be large enough to compensate
the global $\alpha_0^2$ factor of all the parameters. We have indeed
shown in \cite{DEF3} that non-perturbative strong-field effects can
compensate even a vanishingly small $\alpha_0^2\sim -{1\over
2}\overline\gamma$. The overall conclusion is that two different
regimes of gravity must be distinguished: (i)~the weak-field conditions
of the solar system can give a clean access to the fundamental
parameter $\overline\gamma$; (ii)~higher post-Newtonian effects can be
tested in the strong-field regime of binary pulsars.

These conclusions, derived from a theory-motivated point of view, happen
to be valid if we adopt a more phenomenological point of view, {\it
i.e.}, if the coupling function $A(\varphi)$ is now supposed to have
{\it a priori\/} any possible shape. It can thus involve in principle
large dimensionless numbers, and the 2PN parameters (\ref{eq3.30}) can
be of order unity in spite of the bounds (\ref{eq5.1a}) on
$\overline\gamma$. This is particularly true of the parameter $\zeta$,
which can be of order unity even if $A(\varphi)$ does not involve
numbers larger than\footnote{Let us note in this respect that in the
string-derived model of Ref.~\cite{DP}, the equivalent of $\beta_{11}$
is expected to be of order 40.} $\sim 30$. For instance, in a model
with two scalar fields and a flat $\sigma$-model metric
$\gamma_{ab}=\delta_{ab}$, the values
\begin{equation}
\alpha_1^2 \approx 10^{-3}\quad , \quad \alpha_2= 0 \quad ,
\quad \beta_{11} \approx 1 \quad , \quad \beta_{12}\approx 30\quad ,
\label{eq5.2}\end{equation}
give $(\alpha^2)_0 = \alpha_1^2\approx 10^{-3}$ and
$(\alpha\beta\alpha)_0= \alpha_1\beta_{11}\alpha_1 \approx 10^{-3}$,
consistently with (\ref{eq5.1}), but $\zeta\sim
(\alpha\beta\beta\alpha)_0 = \alpha_1\beta_{11}^2\alpha_1
+\alpha_1\beta_{12}^2\alpha_1\approx 1$. By contrast, getting
$\varepsilon\sim 1$ seems less natural, as some components of
$\beta'^0_{abc}$ would have to be as large as $3\times 10^4$ to
compensate the $\alpha_0^a\alpha_0^b\alpha_0^c$ factor.

More generally, it is easy to see that $\zeta$ is {\it a priori\/}
unconstrained by the solar-system bounds (\ref{eq5.1}). Indeed, from
the positive-definiteness of the metric $\gamma_{ab}$, we can use the
Cauchy--Schwarz inequality to obtain a lower bound\footnote{The fact
that we are able to derive relations between 1PN and 2PN parameters is
a further illustration of the power of our field-theory approach.} for
$\zeta$:
\begin{equation}
(\alpha\beta\beta\alpha)_0
\geq\left[(\alpha\beta\alpha)_0\right]^2/\alpha_0^2
\quad\Leftrightarrow\quad
\zeta \geq 8\overline\beta^2 / |\overline\gamma| \ .
\label{eq5.3}\end{equation}
On the other hand, $\zeta$ has {\it a priori\/} no upper bound, and we
can have $\zeta \gg 8\overline\beta^2 / |\overline\gamma|$ if the
($\varphi$-space) vector $(\beta^a_{\
b}\alpha^b)_0\equiv(\beta\alpha)^a_0$ is almost orthogonal to
$\alpha^a_0$, as in the example (\ref{eq5.2}) above.

In Ref.~\cite{DEF1}, we adopted an even more ``phenomenological'' point
of view, by assuming that the $\sigma$-model metric $\gamma_{ab}$ could
have {\it a priori\/} any signature ({\it i.e.}, that the theory could
involve negative-energy scalar fields). In that case, the bounds
(\ref{eq5.1a}) on $\overline\gamma$ do not constrain the magnitude of
$\alpha_0^a$, but only its direction: It should be almost a null vector
in the $\varphi$-space. Then, it is easy to see that $\varepsilon$ is
not constrained at all by the experimental limits (\ref{eq5.1}), and
that $\zeta$ is not constrained either if the theory involves at least
three scalar fields. However, from a theoretical point of view, it
seems more plausible that the gravitational interaction involves only a
small number of massless scalar fields. If it involves only two of them
(or less), the Cayley-Hamilton theorem can be written as
\begin{equation}
(\alpha \beta^2 \alpha) = ({\rm Tr} \beta)
(\alpha \beta \alpha) - ({\rm det} \beta) (\alpha^2)\ ,
\label{eq5.4}
\end{equation}
where the trace and determinant are taken for the $2\times 2$ matrix
$(\beta^a_{\ b})$. This relation shows that even in the case of a
hyperbolic metric $\gamma_{ab}$, the parameter $\zeta$ is
constrained to vanish with $\overline\beta$ and $\overline\gamma$.

The conclusion of the above discussion is therefore that the 1PN
experimental limits (\ref{eq5.1}) most plausibly constrain the 2PN
parameters $\varepsilon$ and $\zeta$ to be very small, but that there
is no theoretical impossibility that the latter be of order unity. In
the following, we will adopt a phenomenological point of view, and
consider that these parameters are {\it a priori\/} unconstrained to
discuss the possible experiments which may measure them. Let us
underline, however, that this ``phenomenological'' attitude is still in
the framework of our field-theory approach, and should not be confused
with previous studies in the literature \cite{BN88,B92,N93} which aimed
at describing any conceivable deviation from general relativity.

\subsection{Solar-system experiments}
\subsubsection{Light deflection and time delay}
As discussed in the previous subsection, the most important parameter to
determine experimentally is $\overline\gamma$, as it measures the basic
coupling strength and constrains {\it a priori\/} all the possible
deviations from general relativity. It has been shown in
\cite{DN93,DP} that the cosmological evolution of tensor--multi-scalar
theories generically drives $\overline\gamma$ towards 0, and that its
present value is expected to be $\lesssim 10^{-7}$. The 1PN deviations
proportional to $\overline\gamma$, entering light-deflection and
time-delay experiments, are thus expected to be comparable to 2PN
effects, if we adopt the phenomenological point of view that the 2PN
parameters $\varepsilon$ and $\zeta$ are {\it a priori\/} unconstrained
(and that $\overline\beta$ may be much larger than $\overline\gamma$).
It is therefore important to determine whether these effects can be
distinguished from each other. We shall see here that
light-deflection and time-delay experiments are ideally suited
to an accurate determination of $\overline\gamma$ in the solar system.

First of all, it is immediate to see that these experiments do not
depend on $\varepsilon$ and $\zeta$ at the 2PN level. Indeed, our
results of section III and IV show that these parameters appear for the
first time at order $O(1/c^6)$, in the time-time component of the
metric $\widetilde g_{00}$. Therefore, they do not enter the 2PN
geodesic equation satisfied by light, which involves only the
$O(1/c^2)$ and $O(1/c^4)$ orders of the metric. Consequently, any
experiment probing the propagation of light is independent from
$\varepsilon$ and $\zeta$ at the 2PN level.

We compute in Appendix~C the 2PN light deflection by a massive body
and find
\begin{equation}
\Delta\phi = {2\mu_A(2+\overline\gamma_{A0})\over\rho_0 c^2}
\left[1 - {\mu_A(2+\overline\gamma)\over \rho_0c^2}
\right]
+ {\pi\over 4}\left({\mu_A\over \rho_0c^2}
\right)^2
\left[15+{31\over 2}\overline\gamma +4 \overline\gamma^2\right]
+O\left({1\over c^6}\right)\ .
\label{eq5.5}\end{equation}
Here $\rho_0$ denotes the minimal distance between the light ray and the
center of body $A$ measured in isotropic coordinates. The body-dependent
parameter $\overline\gamma_{A0}$ entering the first term of
(\ref{eq5.5}) can be expanded in powers of the self-gravity of body
$A$, as in Eq.~(\ref{eq4.23d}) above:
\begin{equation}
\overline\gamma_{A0} = \overline\gamma-(4\overline\beta-\overline\gamma)
(2+\overline\gamma)E_A^{\rm grav}/m_A c^2 +O(1/c^4)\ .
\label{eq5.6}\end{equation}
Note that $\overline\beta$ appears in the self-energy corrections to the
1PN effects, whereas it is absent from the $O(Gm/rc^2)^2$ contributions.

Contrary to the hopes of Refs.~\cite{ES80,FF80,RM82,RM83}, a conclusion
of our approach is that improved light-deflection (or time-delay)
experiments do not give access to any theoretically significant 2PN
parameter. In fact, the formal 2PN generalization of the Eddington
$\gamma$ parameter introduced in Refs.~\cite{ES80,FF80} (under the
different notations $\epsilon$ and $\delta$; see also \cite{RM82} where
it is denoted $\Lambda$) is equal to $1+\overline\Gamma$, where
$\overline\Gamma$ is the function of 1PN parameters given in
Eq.~(\ref{eq4.23e}) above:
\begin{equation}
\epsilon_{\rm Epstein-Shapiro}
= \delta_{\rm Fischbach-Freeman}
= \Lambda_{\rm Richter-Matzner}
= 1+ {4\over 3}\overline\beta +{15+8\overline\gamma\over
6}\overline\gamma\ .
\label{eq5.7}\end{equation}

Our conclusion concerning the impossibility to probe significant 2PN
deviations from general relativity in light-deflection and time-delay
experiments should not be viewed only negatively. The positive aspect
is that these experiments can give a clean access to the fundamental
parameter $\overline\gamma$, {\it i.e.}, that no 2PN deviation from
general relativity can complicate their interpretation in the case
where the data are too scarce to allow a clean separation of $1/r$ and
$1/r^2$ effects. Indeed, at the level $\overline\gamma\lesssim 10^{-6}$
that these experiments aim for, the 2PN effects proportional to this
parameter in Eq.~(\ref{eq5.5}) are totally irrelevant\footnote{At the
qualitative level, it remains that checking the coefficient $15\pi/4$
appearing in the second term of (\ref{eq5.5}) will be a new
confirmation of the nonlinear structure of general relativity, even if
it does not constrain any plausible theoretical alternatives.}. On the
other hand, one remarks that the $\overline\gamma$ parameter which will
be measured by these experiments is the body-dependent quantity
$\overline\gamma_{A0}$ rather than its weak-field limit
$\overline\gamma$. However, the self-gravity renormalization of
$\overline\gamma$ displayed in Eq.~(\ref{eq5.6}) is already strongly
constrained by the LLR (Lunar Laser Ranging) bounds on
$\eta\equiv4\overline\beta-\overline\gamma$ ($|\eta|\lesssim 2\times
10^{-3}$ \cite{Di94}). Using $|E_\odot^{\rm grav}|/m_\odot c^2\sim
2\times 10^{-6}$ for the Sun, one gets
\begin{equation}
(4\overline\beta-\overline\gamma)(2+\overline\gamma)
|E_\odot^{\rm grav}|/m_\odot c^2 < 10^{-8}\ .
\label{eq5.8}\end{equation}
Therefore, an experimental determination of $\overline\gamma_{\odot 0}$
at the $10^{-6}$ or $10^{-7}$ level, which is expected to be reachable
in missions presently considered by the European Space Agency such as
GAIA (Global Astrometric Interferometer for Astrophysics) or SORT
(Solar Orbit Relativity Test), will indeed give a clean measurement of
$\overline\gamma$.

Finally, let us remark that the value of $\overline\gamma$ we are
talking about in this work is the one corresponding to the value of the
scalar-field background $\varphi_0^a$ around the solar system. As
remarked in \cite{N93}, it differs from, say, the cosmological average
of $\overline\gamma$ (which is the one discussed in
Refs.~\cite{DN93,DP}) by the effect of the spatial fluctuation $\Delta
U$ of the gravitational potential. However, we show in Appendix~B that
the corresponding change in $\overline\gamma$ is given by
$4\overline\beta(2+\overline\gamma)\Delta U/c^2$ and is therefore
expected to represent only a very small {\it fractional\/} change of
$\overline\gamma$ of order\footnote{Even if $|\beta^a_{\ b}|\sim 30$,
the fact that $\Delta U_{\rm cosmo}/c^2 \sim 10^{-5}$ indicates a
negligible fractional change.} $|\beta^a_{\ b}| \Delta U/c^2$, where
$|\beta^a_{\ b}|$ is the norm of the matrix $\beta^a_{\ b}$ (say the
modulus of its largest eigenvalue).

\subsubsection{Other possible 2PN experiments}
The previous subsection showed that experiments on the propagation of
light in the solar system cannot give access to any post-Newtonian
parameter but $\overline\gamma$. The question that we will address now
is whether the 2PN parameters $\varepsilon$ and $\zeta$ can be measured
in the solar system, or at least if it is possible to constrain their
magnitudes at an interesting level.

One of the most famous tests of post-Newtonian gravity is the perihelion
shift of Mercury, and more generally tests obtained through a global
fit to the orbital motion of the planets. Before any calculation, it is
clear that we cannot hope to measure the Mercury perihelion advance by
traditional means to an accuracy of 2PN significance. Indeed, the 2PN
effect is smaller than the 1PN prediction by a factor $\sim(Gm_\odot
/rc^2)$, and corresponds to an advance of $\sim 10^{-6}$ arcsecond per
century. However, it may be, one day, possible to reach this level by
using an artificial satellite orbiting around Mercury and tracked with
very high accuracy from Earth. To increase the parameter
$Gm_\odot/rc^2$, one could also observe the perihelion shift of a
drag-free satellite in close elliptical orbit around the Sun, but the
construction of such a satellite seems unrealistic with present
technology. Anyway, several reasons show that such experiments could
not constrain $\varepsilon$ and $\zeta$ at an interesting level.
Indeed, we compute in Appendix~C below the perihelion shift per orbit
for a test mass $m_0\ll m_\odot$, and we get in isotropic coordinates
\begin{equation}
\Delta\phi = {6\pi G m_\odot\over a(1-e^2)c^2}
\left[1+{2\overline\gamma-\overline\beta+\zeta c_\odot\over3}
+{Gm_\odot\over a c^2}\left({7\over 2}
+{\varepsilon\over 6}+O(e^2)\right)
\right]
+{O(\overline\beta,\overline\gamma)\over c^4}
+O\left({1\over c^6}\right) \ ,
\label{eq5.9}\end{equation}
where $a$ is the coordinate semi-major axis of the orbit, $e$ its
coordinate eccentricity (in isotropic coordinates), and $c_\odot\approx
4\times 10^{-6}$ is the compactness of the Sun. Although $\zeta$ enters
this expression, it cannot be distinguished from a small contribution
of $2\overline\gamma-\overline\beta$ at the 1PN level. In fact, it
comes from the expansion (\ref{eq4.23b}) of the body-dependent
parameter $\overline\beta^{\ 0}_{\odot\odot}=\overline\beta-\zeta
c_\odot +O(\overline\beta,\overline\gamma)/c^2+O(1/c^4)$, and is thus a
mere renormalization of $\overline\beta$. On the contrary, the
contribution proportional to $\varepsilon$ can in principle be
distinguished from the 1PN effects, since it involves a different power
of $(Gm/a)$. Numerically, we find that Mercury's perihelion is deviated
from the general relativistic prediction by $\approx 0.5\varepsilon$~mm
per year, too small to be of observational significance. In the totally
unrealistic situation of a drag-free satellite grazing the surface of
the Sun, this 2PN deviation would be of order $\approx 30\varepsilon\
{\rm m.yr}^{-1}$, which is {\it a priori\/} much easier to detect.
However, to distinguish the $O(Gm/ac^2)$ and $O(Gm/ac^2)^2$
contributions, it would be necessary to compare several satellites at
different distances from the Sun, or alternatively to look at periodic
effects on a given orbit, which may be much more difficult to observe
than secular effects. Moreover, one will have the difficulty of
separating the 2PN contribution from the Newtonian contribution due to
the quadrupole moment of the Sun, which as the same
dependence\footnote{The peculiar anisotropy of multipolar effects will,
however, help in this respect.} on $a$ (not to mention the huge thermal
and electromagnetic effects of the Sun on a close satellite). In
conclusion, perihelion-shift experiments in the solar system can in
principle give access to the 2PN parameter $\varepsilon$, but present
technology does not give the hope of measuring it to any significant
level.

We have seen in Eqs.~(\ref{eq3.29a}) and (\ref{eq3.34a}) above that the
effective gravitational constant $G_{AB}$ depends on the self-energies
of the bodies $A$ and $B$. In particular, the Earth ($\oplus$) and the
Moon ($\Bbb C$) do not fall with the same acceleration towards the Sun,
since $G_{\odot\oplus}\neq G_{\odot{\Bbb C}}$. This violation of the
strong equivalence principle implies a polarization of the Moon's orbit
(Nordtvedt effect), that can be tested in the Lunar Laser Ranging data.
The deviations from general relativity are proportional to the ratio
\begin{equation}
{G_{\odot\oplus}- G_{\odot{\Bbb C}}\over G}
=-{\eta\over 2}(c_\oplus- c_{\Bbb C}) +
\left(\zeta+O(\overline\beta,\overline\gamma)\right)
c_\odot(c_\oplus- c_{\Bbb C})
+ O(a_\oplus,b_\oplus,a_{\Bbb C},b_{\Bbb C})\ ,
\label{eq5.10}\end{equation}
where $c_{\Bbb C}\ll c_\oplus \ll c_\odot$ are the compactnesses of the
three bodies, and where we have neglected $a_\oplus\sim b_\oplus\sim
c_\oplus^2$ (as well as $a_{\Bbb C}$ and $b_{\Bbb C}$) with respect to
$c_\odot c_\oplus$. The first term, involving the parameter $\eta\equiv
4\overline\beta -\overline\gamma$, is the standard 1PN deviation which
has been constrained at the level $|\eta|<2\times 10^{-3}$ by LLR data.
The dominant 2PN contribution, $\zeta c_\odot(c_\oplus- c_{\Bbb C})$,
is equivalent to a renormalization of $\eta$ into $(\eta-2\zeta
c_\odot)\approx(\eta - 10^{-5} \zeta)$. From a theory-based viewpoint,
such a renormalization is of no consequence: as $\eta$ and $\zeta$ are
both proportional to the basic coupling strength $\overline\gamma$, we
can always neglect the {\it fractionally\/} small correction
$10^{-5}\zeta$ to $\eta$ (in the same way, as explained above, that one
does not have to worry about ``cosmic variance'' effects). However,
from a phenomenological viewpoint, $\zeta$ is an independent quantity
which would complicate the interpretation of a high-precision LLR
experiment reaching the $\eta\sim 10^{-5}$ level. [This level would
correspond to measuring the Earth--Moon distance with $0.1$ millimeter
accuracy.] In other words, the phenomenological point of view obliges
us to look for other independent experiments allowing one to separate
the effects of $\eta$ and $\zeta$.

The polarization of Mercury's orbit around the Sun
due to the presence of Jupiter ({\it i.e.}, the Nordtvedt effect for
Mercury) can give access to a different combination of $\eta$ and the
2PN parameters $\varepsilon$, $\zeta$. The corresponding deviation from
general relativity is proportional to
\begin{equation}
{G_{J\odot}-G_{JM}\over G}= -{\eta\over 2} c_\odot
+\left({\varepsilon\over 2}+\zeta\right)b_\odot
+\zeta\times O(c_\odot c_J) +O(\overline\beta,\overline\gamma)
O(c_M,c_\odot^2)\ ,
\label{eq5.11}\end{equation}
where $c_M\ll c_J \ll c_\odot$ are respectively the compactnesses of
Mercury, Jupiter and the Sun, and where we have neglected $c_\odot c_J$
with respect to $b_\odot \sim c_\odot^2$. The dominant 2PN contribution
plays again the role of a renormalization of $\eta$, but this time into
$[\eta-(\varepsilon+2\zeta)b_\odot/c_\odot]\approx[\eta-10^{-5}
(\varepsilon/2+\zeta)]$. Note that $\varepsilon$ enters now this
expression, whereas it was absent from the corresponding lunar result.
This complicates the problem of separating the contributions of $\eta$
and $\zeta$. It would be necessary to dispose of a third experiment
giving access to $\varepsilon$, for instance by using the perihelion
shift (\ref{eq5.9}). However, we have seen that such a measure is not
likely to be performed with current (or foreseeable) technology.

Another attempt to determine $\varepsilon$ could be to compare
ultra-stable clocks: one located on Earth and another one somewhere
close to the Sun. Indeed, the Einstein effect gives access to
$\sqrt{-\widetilde g_{00}}$, where $-\widetilde g_{00}$ is given by
Eq.~(\ref{eq4.21a}) in $g^*$-harmonic coordinates and by the first
bracket of Eq.~(\ref{eq4.18}) in isotropic coordinates. The general
relativistic prediction for the rates of clocks is thus multiplied
by a factor $[1+(Gm_\odot/rc^2)^2\overline\beta^{\ 0}_{\odot\odot} -
{1\over 6}(Gm_\odot/rc^2)^3 \varepsilon +O(\overline\beta,
\overline\gamma)/c^6]$. This gives (relative to a clock on Earth)
effects of order $\approx 1.6\times 10^{-18} \varepsilon$ for a clock
located at the surface of the Sun, and $\approx 3\times 10^{-24}
\varepsilon$ for a clock on Mercury. With foreseeable technology
($10^{-18}$ stability), one might barely be able to constrain
$\varepsilon$ at the $O(1)$ level. As for the perihelion shift
(\ref{eq5.9}), it should be noted that the parameter $\zeta$ contained
in $\overline\beta^{\ 0}_{\odot\odot} \approx \overline\beta - \zeta
c_\odot$ is a mere renormalization of $\overline\beta$ and cannot be
accessed independently.

A possible way to access this $\zeta$ contribution in
$\overline\beta^{\ 0}_{AA}$ would be to measure both
$\overline\beta^{\ 0}_{\odot\odot}\approx \overline\beta - \zeta
c_\odot$ and $\overline\beta^{\ 0}_{\oplus\oplus}\approx
\overline\beta - \zeta c_\oplus \approx \overline\beta$ at a level
$< c_\odot\approx 4\times 10^{-6}$. This would necessitate the
tracking of Mercury with few cm accuracy, and the observation of a
drag-free artificial Earth satellite at the $10^{-2}$~cm level
\cite{DEF6}. This second condition is two orders of magnitude smaller
than what can be done presently.

The conclusion of the above discussion is that the 2PN parameters
$\varepsilon$ and $\zeta$ are extremely difficult to measure in the
solar system. Within a phenomenological approach, the only role of
these parameters is negative: they complicate the interpretation of
high-precision 1PN experiments. Hence the solar system appears not to
be an appropriate testing ground for probing the 2PN structure of
gravity. On the contrary, it is perfectly suited for measuring the
fundamental parameter $\overline\gamma$, as underlined in subsection
V--B--1.

\subsection{Binary-pulsar experiments}
Let us now consider binary-pulsar experiments, which will turn out to
be much better testing grounds for the 2PN structure of gravity.
Since the self-energy of a neutron star is typically of order $0.2$ (as
compared to $\sim 10^{-6}$ for the Sun), the 2PN and higher-order
effects play an important role in the behavior of binary pulsars. In
the present subsection, we will show that the combined analysis of
several binary-pulsar data has the capability of constraining both
$\varepsilon$ and $\zeta$ at an interesting level\footnote{In the
language of \cite{DT92,TWDW}, we perform a combined theory-dependent
analysis of several independent pulsar data.}. Our aim is not to
perform a full statistical analysis, but rather to illustrate the
different types of constraints that can be obtained. We will consider
four binary pulsars, for which different observable quantities can be
measured.

The Hulse--Taylor binary pulsar PSR 1913+16 has been continuously
observed since its discovery in 1974. Besides the Keplerian orbital
parameters $P$ (orbital period) and $e$ (eccentricity), three
``post-Keplerian" \cite{D88,DT92} observables have been measured with
great accuracy: the periastron advance $\dot\omega$, the secular change
of the orbital period $\dot P$, and the time-dilation parameter
$\gamma_{\rm timing}$ which describes both the second-order Doppler
effect due to the velocity of the pulsar and the redshift due to the
presence of its companion. [This last parameter should not be confused
with the metric $\gamma_{ab}$, the Eddington parameter
$\overline\gamma$, nor its body-dependent generalization
$\overline\gamma_{AB}$.] Within a given theory of gravity, these three
timing observables\footnote{In order not to create any confusion with
our use of the word post-Keplerian in the present paper, we refer
to these quantities as ``timing observables'' instead of
``post-Keplerian'' parameters as used in \cite{D88,DT92}.} can be
predicted in terms of the masses $m_A$ and $m_B$ of the pulsar and its
companion, which are {\it a priori\/} unknown. The theory is consistent
with experiment if there exists a pair of masses ($m_A$,$m_B$) giving
the correct observed values for the three quantities $\dot P$,
$\dot\omega$ and $\gamma_{\rm timing}$. This is the so called ``$\dot
P$-$\dot\omega$-$\gamma$'' test, that general relativity passes with
flying colors, and which establishes the reality of gravitational
waves. We give below the experimental values quoted in \cite{T93}:
\begin{mathletters}
\label{eq5.12}
\begin{eqnarray}
P & = & 27906.9807804(6)\ {\rm s}\ ,
\label{eq5.12a} \\
e & = & 0.6171308(4)\ ,
\label{eq5.12b} \\
\dot P & = & -2.422(6)\times 10^{-12}\ ,
\label{eq5.12c} \\
\dot\omega & = & 4.226621(11)\ {}^\circ\ {\rm yr}^{-1}\ ,
\label{eq5.12d} \\
\gamma_{\rm timing} & = & 4.295(2)\times 10^{-3}\ {\rm s} \ ,
\label{eq5.12e}\end{eqnarray}
\end{mathletters}
where figures in parentheses represent 1$\sigma$ uncertainties in
the last quoted digits. In fact the determination of $\dot P$ is so
precise that it is necessary to take into account the small variable
Doppler effect due to the acceleration of the binary pulsar towards the
center of the Galaxy \cite{DT91}. This induces an
extra contribution to $\dot P$, which takes the value $\dot P_{\rm gal}
= -0.0124(64)\times 10^{-12}$ in general relativity. The intrinsic
variation of the orbital period (due to gravitational radiation damping)
is thus given by
\begin{equation}
\dot P_{\rm observed} - \dot P_{\rm gal} = -2.4101(85)\times 10^{-12}\ .
\label{eq5.13}\end{equation}
In tensor--scalar theories of gravity, $\dot P_{\rm gal}$ is modified by
a small contribution, $\delta\dot P_{\rm gal}$, that we will take into
account in our calculations below.

The binary pulsar PSR 1534+12 has been observed only since 1991, and the
present experimental accuracy on its parameter $\dot P$ is not
comparable with the one of Eq.~(\ref{eq5.12c}) above. However, this
pulsar is much closer to the Earth than PSR 1913+16, and it has been
possible to measure $\dot\omega$ and $\gamma_{\rm timing}$ (with very
good precision), as well as two new timing observables, $r$ and $s$,
measuring the amplitude and the shape of the Shapiro time-delay caused
by the companion. Out of these two parameters, only $s$ is measured
with precision. [Note that, geometrically, $s\equiv\sin i$ is the sine
of the angle between the orbit and the plane of the sky.] As above, the
three quantities $\dot\omega$, $\gamma_{\rm timing}$ and $s$ can be
predicted as functions of the masses $m_A$, $m_B$ within a given theory
of gravity. One can therefore test if this theory agrees with
experiment by looking for a pair of masses ($m_A$,$m_B$) consistent
with the three observed values of $\dot\omega$, $\gamma_{\rm timing}$
and $s$ (``$\dot\omega$-$\gamma$-$s$'' test). We shall make use of the
latest experimental data discussed in \cite{A95}:
\begin{mathletters}
\label{eq5.14}
\begin{eqnarray}
P & = & 36351.70267(2)\ {\rm s}\ ,
\label{eq5.14a} \\
e & = & 0.2736771(4)\ ,
\label{eq5.14b} \\
x & = & 3.729458(2)\ {\rm s} \ ,
\label{eq5.14c} \\
\dot\omega & = & 1.75573(4)\ {}^\circ\ {\rm yr}^{-1}
\label{eq5.14d} \\
\gamma_{\rm timing} & = & 2.081(16) \times 10^{-3} \ ,
\label{eq5.14e} \\
s & = & 0.981(8) \ .
\label{eq5.14f}\end{eqnarray}
\end{mathletters}
Here $x=a_1 s/c$ is the projection of the semi-major axis ($a_1$) of the
pulsar orbit on the line of sight (in light-seconds). A precision
should be given concerning the quoted experimental uncertainties.
Ref.~\cite{A95} gives (for the more reliable $1.4$~GHz data) the
$\Delta\chi^2 = 2.30$ and $\Delta\chi^2= 6.17$ contours in the $(r,c)$
plane with $c\equiv\sqrt{1-s^2}$. We deduced from these the
$\Delta\chi^2 = 4$ contour which defines, when projected onto the $c$
axis, a $2\sigma_{\rm stat}$ interval for $c$ considered alone. We use
the corresponding $2\sigma_{\rm stat}$ interval for $s=\sqrt{1-c^2}$ as
a realistic 1$\sigma$ interval (in other words, we double the
statistical 1$\sigma$ interval for $s$ to take into account possible
systematic effects). We similarly doubled the statistical 1$\sigma$
uncertainties obtained in \cite{A95} for $\dot\omega$ and $\gamma_{\rm
timing}$.

The binary pulsar PSR 0655+64 is composed of a neutron star of mass
$\approx 1.4 m_\odot$, and a light companion of mass $\approx 0.8
m_\odot$, which is probably a white dwarf. The gravitational waves
emitted by such a dissymmetrical system involve a large dipolar
contribution of order $1/c^3$ in tensor--scalar theories, whereas the
dominant radiation in general relativity is quadrupolar and of order
$1/c^5$. The fact that the observed value of $\dot P$ is very small
(and consistent with zero) constrains therefore the existence of scalar
fields, or more precisely the magnitude of their interaction with
matter. We will see below that this system imposes a tight bound on the
2PN parameter $\zeta$. The experimental data that we will need for our
analysis are taken from \cite{A95}:
\begin{mathletters}
\label{eq5.15}
\begin{eqnarray}
P & = & 88877.06194(4)\ {\rm s}\ ,
\label{eq5.15a} \\
e & < & 3\times 10^{-5}\ ,
\label{eq5.15b} \\
\dot P & = & (1\pm 4)\times 10^{-13}\ .
\label{eq5.15c}\end{eqnarray}
\end{mathletters}
The masses of the pulsar and its companion are not known independently;
several pairs ($m_A$,$m_B$) are thus {\it a priori\/} possible, such as
$(1.30m_\odot,0.7m_\odot)$, $(1.35m_\odot,0.8m_\odot)$ or
$(1.40m_\odot,0.9m_\odot)$. In our calculations below, we will choose
the mass pair which gives the most conservative bounds on the 2PN
parameters, namely $m_A=1.30 m_\odot$, $m_B= 0.7 m_\odot$. [Smaller
masses could be consistent with experimental data, but current lore
favors neutron star masses close to $1.4 m_\odot$.]

The fourth and last binary pulsar that we will consider, PSR 1800$-$27,
is also a dissymmetrical system, involving a neutron star of mass
$m_A\approx 1.4 m_\odot$ and a light companion of negligible
self-energy. The acceleration of the pulsar towards the center of the
Galaxy is therefore proportional to the self-gravity-modified effective
gravitational constant $\widetilde G_{A0}$ [{\it cf.}
Eqs.~(\ref{eq4.20a}) and (\ref{eq4.23a}) above], whereas the companion
is accelerated by a force proportional to the weak-field gravitational
constant $\widetilde G_{00}=\widetilde G$, Eq.~(\ref{eq2.13}). As shown
in \cite{DS91}, this violation of the strong equivalence principle
causes a ``gravitational Stark effect'' on the orbit of the system,
polarizing its eccentricity in a particular direction. A highly
circular orbit, like the one of PSR 1800$-$27, is therefore very
improbable. A statistical study can thus be performed to constrain the
magnitude of the matter--scalar field interaction. Following the method
of \cite{DS91}, Ref.~\cite{A95} has obtained the bound
\begin{equation}
|\delta_A| < 1.4 \times 10^{-3}
\label{eq5.16}\end{equation}
on the parameter $\delta_A$ characterizing this gravitational Stark
effect. This bound corresponds to the 90\% confidence level, which
plays the role of an ``effective 1$\sigma$ level'' for the non-Gaussian
statistics of this test. [Twice this value gives the 95\% C.L., {\it
i.e.}, the standard 2$\sigma$ level.] Note that it is more secure than
the standard 1$\sigma$ (68\%) level.

The analytic expressions of all the observable quantities discussed
above have been derived in \cite{DS91,DT92,DEF1}. The theoretical
prediction for the observed time derivative
of the orbital period has the form
\begin{equation}
\dot P = \dot P_{\rm spin0}^{\rm monopole}
+\dot P_{\rm spin0}^{\rm dipole}
+\dot P_{\rm spin0}^{\rm quadrupole}
+\dot P_{\rm spin2}^{\rm quadrupole}
+\dot P_{\rm gal}
+\delta\dot P_{\rm gal}
+O\left({1\over c^7}\right)\ ,
\label{eq5.17}\end{equation}
where the different contributions are given in Eqs.~(6.52) and (9.22) of
\cite{DEF1}. The same reference gives also the expressions of the
periastron shift $\dot\omega$ and of the time-dilation parameter
$\gamma_{\rm timing}$ in Eqs.~(9.20), as well as the ``Stark'' parameter
$\delta_A$ in Eq.~(9.16). Finally, the theoretical prediction for the
timing observable $s$ is given in Eq.~(3.15) of \cite{DT92}.
Introducing the notation
\begin{equation}
M\equiv m_A+m_B\quad,\quad X_A\equiv m_A/M\quad,\quad X_B \equiv m_B /M
\quad,\quad n\equiv 2\pi/P\ ,
\label{eq5.18}\end{equation}
it can be written as
\begin{equation}
{x\over s} = {X_B \over n}(G_{AB} M n /c^3)^{1/3}\ .
\label{eq5.19}\end{equation}
These theoretical predictions give (when written out in detail) the
various timing observables as theory-dependent functions of the masses
of the bodies. In general, because of possible nonperturbative
strong-field effects \cite{DEF3}, the latter functions should be
considered as {\it functionals\/} of the $1+n(n-1)/2$ arbitrary
functions entering the definition of a tensor--scalar theory (notably
$A(\varphi)$). However, if we assume the absence of genuine
nonperturbative effects, we can expand the functions giving the
observables in powers of the compactnesses $c_A$, $c_B$ of the bodies.
This has the effect of reducing the functional dependence of the
observables to a dependence upon a finite number of theory parameters.
At the lowest orders, there appear the 1PN Eddington parameters
$\overline\beta$ and $\overline\gamma$. As these are already tightly
constrained by solar-system data, Eqs.~(\ref{eq5.1}), we will neglect
them in the following, and investigate the limits that can be set on
further theory parameters. The 2PN parameters $\varepsilon$ and $\zeta$
appear precisely at the next significant order of the expansion in
compactnesses \cite{DEF1}. The deeper layers of theory-parameters
introduced in \cite{DEF1} always appear with higher powers of the
compactnesses. [As shown in Table~3 of \cite{DEF1}, this is true even
for the different contributions of the gravitational radiation, in
spite of their rather complicated structure.]

Here we shall estimate what constraints are imposed by binary-pulsar
data on $\varepsilon$ and $\zeta$, when restricting our attention to
the lowest-order terms in compactnesses involving them. From a
numerical point of view, our approximation is rather rough, as the next
orders that we neglect are only $O(c_A)\sim 0.3$ smaller than the terms
we will consider. A way of justifying our approach is to say that we
assume that there is no fine-tuned compensation between the
$\varepsilon$ and $\zeta$-dependent terms we retain and the
higher-order terms which involve new (3PN, 4PN, \dots) parameters. We
believe that our simplified analysis will give at least the right order
of magnitude for the constraints on $\varepsilon$ and $\zeta$ that
would follow from a more complete theory-dependent analysis.

To write in closed form the truncated expressions of the different
observables, let us introduce the notation\footnote{The notation $\cal
V$ is a reminder of the fact that this quantity measures some mean
orbital velocity.}
\begin{equation}
{\cal V} \equiv (G_* M n)^{1/3}\ ,
\label{eq5.20}\end{equation}
where $M$ and $n$ have been defined in Eq.~(\ref{eq5.18}) above, and
where $G_*$ is the bare gravitational constant. Note that since we
neglect here $\alpha_0^2\propto \overline\gamma$, the bare constant
$G_*$ can be identified with the gravitational constant
$G=G_*(1+\alpha_0^2)$ measured in weak-field conditions. The timing
observables can now be written as
\begin{mathletters}
\label{eq5.21}
\begin{eqnarray}
\dot P_{\rm spin0}^{\rm monopole} & = & -12\pi{{\cal V}^5\over c^5}
X_A X_B \zeta {e^2(1+e^2/4)\over (1-e^2)^{7/2}}
+O(s)\ ,
\label{eq5.21a} \\
\dot P_{\rm spin0}^{\rm dipole} & = & -2 \pi{{\cal V}^3\over c^3}
X_AX_B\left[(c_A-c_B)^2\zeta{1+e^2/2\over (1-e^2)^{5/2}}+O(s^3)\right]
\nonumber \\
& & + 4\pi {{\cal V}^5\over c^5} X_A X_B\left[
(X_A-X_B)(c_A-c_B)\zeta{1+3e^2+3e^4/8\over(1-e^2)^{7/2}}+O(s^2)
\right] ,
\label{eq5.21b} \\
\dot P_{\rm spin0}^{\rm quadrupole}\! & = & -{32\pi\over5}{{\cal
V}^5\over c^5}
X_AX_B\left[(c_BX_A+c_AX_B)^2\zeta+(b_BX_A+b_AX_B)(\varepsilon+2\zeta)
+O(s^3)\right]
\nonumber\\
&&\times{1+73e^2/24+37e^4/96\over (1-e^2)^{7/2}}\ ,
\label{eq5.21c} \\
\dot P_{\rm spin2}^{\rm quadrupole}\! & = & -{192\pi\over5}
{{\cal V}^5\over c^5}
X_AX_B\left[1+{2\over 3}c_Ac_B\zeta+{1\over 3}(b_A+b_B)
(\varepsilon+2\zeta)
+O(s^3)\right]
\nonumber\\
&&\times{1+73e^2/24+37e^4/96\over (1-e^2)^{7/2}}\ ,
\label{eq5.21d} \\
\delta\dot P_{\rm gal}^{\rm 1913+16} & = &
-{1.22\times10^{-18}\over n}(b_Ax_A+b_BX_B) (\varepsilon+2\zeta)
+O(s^3)\ ,
\label{eq5.21e} \\
\dot\omega & = & {{\cal V}^2\over c^2}\,{n\over 1-e^2}\left[
3+(c_AX_A+c_BX_B)\zeta
+{1\over 2}(c_BX_A+c_AX_B)(\varepsilon+\zeta)+O(s^2)
\right] ,
\label{eq5.21f} \\
\gamma_{\rm timing} & = & {{\cal V}^2\over c^2}\, {eX_B\over n}\left[
1+X_B-2\kappa_A c_B \zeta +O(s^2)\right] ,
\label{eq5.21g} \\
\delta_A & = & b_A\left({\varepsilon\over 2}+\zeta\right)+O(s^3)\ ,
\label{eq5.21h} \\
{x\over s} & = & {{\cal V}\over c}\, {X_B\over n}
\left[
1+{1\over 3}c_Ac_B\zeta + {1\over 3}(b_A+b_B)
\left({\varepsilon\over2}+\zeta\right) +O(s^3)
\right]\ .
\label{eq5.21i}\end{eqnarray}
\end{mathletters}
As in Eqs.~(\ref{eq3.27}) above, $s$ in the error terms on the
right-hand side is a global notation for the compactnesses of the
bodies, which should not be confused with the timing observable $s$ on
the left-hand side of (\ref{eq5.21i}). The $O({\cal V}^5/c^5)$
contribution in $\dot P_{\rm spin0}^{\rm dipole}$ can of course be
neglected with respect to its $O({\cal V}^3/c^3)$ term. On the other
hand, the monopolar and quadrupolar contributions should not be
neglected in spite of their being also of order $O({\cal V}^5/c^5)$.
Indeed, the dipolar term (\ref{eq5.21b}) can become very small if the
pulsar $A$ and its companion $B$ are almost identical, and the
monopolar and quadrupolar contribution then dominate. The galactic
contribution to $\dot P$ given in Eq.~(\ref{eq5.21e}) corresponds to
the case of PSR 1913+16, and should not be used for other pulsars. The
parameter $\kappa_A$ entering (\ref{eq5.21g}) is the logarithmic
derivative of the inertia moment $I_A$ of the pulsar with respect to
the gravitational constant: $\kappa_A\equiv\partial\ln I_A /
\partial\ln G$. It has been estimated in Ref.~\cite{WZ} to range
between $0.5$ and $1.7$, depending on the nuclear equation of state
used to describe the neutron star matter. The ``$\dot
P$-$\dot\omega$-$\gamma$'' test in PSR 1913+16 is almost
insensitive\footnote{This follows from the near parallelism of the
$\dot P$ and $\dot\omega$ curves in the mass plane.} to the value
chosen for $\kappa_A$. On the contrary, the
``$\dot\omega$-$\gamma$-$s$'' test in PSR 1534+12 gives slightly
tighter bounds on $\varepsilon$ and $\zeta$ for $\kappa_A=1.7$ than for
$\kappa_A = 0.5$. In order to derive conservative bounds for these
parameters, we will use the value $\kappa_A = 0.5$ in the following.

In all the equations (\ref{eq5.21}), the compactnesses of the bodies can
be estimated by using the results of Appendix~B of \cite{DEF1}. For a
realistic equation of state of matter inside a neutron star, we found
in this reference
\begin{mathletters}
\label{eq5.22}
\begin{eqnarray}
c_A & \approx & 0.21\ m_A/m_\odot\ ,
\label{eq5.22a} \\
a_A & \approx & 2.16\ c_A^2\ ,
\label{eq5.22b} \\
b_A & \approx & 1.03\ c_A^2 \ .
\label{eq5.22c}\end{eqnarray}
\end{mathletters}
The coefficient $0.21$ of Eq.~(\ref{eq5.22a}) can be lowered to $\sim
0.15$ for a stiff equation of state, and increased to $\sim 0.30$ for a
soft equation of state. We will choose the central value $0.21$ in our
calculations. When the companion of the pulsar is not itself a neutron
star, but a weakly self-gravitating body like a white dwarf, its
compactness $c_B\ll c_A$ can be neglected. This is the case for PSR
0655+64 and PSR 1800$-$27.

The predictions (\ref{eq5.21}) of tensor--scalar theories can now be
confronted to the experimental data (\ref{eq5.12})---(\ref{eq5.16}).
Since we are working at the first order in $\varepsilon$ and $\zeta$,
we can replace the masses $m_A$, $m_B$ of the bodies (as well as the
corresponding compactnesses $c_A$, $c_B$, $a_A$, $a_B$, $b_A$, $b_B$)
by their general relativistic predictions in all the terms involving
one of these parameters. By contrast, the dominant contributions in
$\dot P_{\rm spin2}^{\rm quadrupole}$, $\dot\omega$, $\gamma_{\rm
timing}$ and $x/s$ should be considered as functions of two {\it a
priori\/} unknown masses $m_A$, $m_B$. For each value of $\varepsilon$
and $\zeta$, we can thus determine if the above four tests can be
passed. More precisely, in the cases of PSR 0655+64 or PSR 1800$-$27,
the measurement of the observables $\dot P$ or $\delta_A$ directly
defines (when assuming the above indicated values of the masses) a
one-sigma constraint on some linear combination of $\zeta$ and
$\varepsilon$. In the cases of PSR 1913+16 or PSR 1534+12, the
simultaneous measurement of three observables ($\dot
P$-$\dot\omega$-$\gamma$ or $\dot\omega$-$\gamma$-$s$), which are
predicted to be some functions of the four quantities $m_A$, $m_B$,
$\zeta$ and $\varepsilon$, defines (by eliminating the two unknown
masses between the three equations and by adding in quadrature the
1$\sigma$ errors on the three observables) a one-sigma
constraint\footnote{An alternative method is to start from the full
$\chi^2(m_A,m_B,\zeta,\varepsilon) \equiv\sum_\alpha [q_\alpha^{\rm
obs} - q_\alpha^{\rm th} (m_A, m_B,\allowbreak \zeta, \varepsilon)]^2
/ \sigma_\alpha^2$ associated with the three measurements
$q_\alpha=q_\alpha^{\rm obs}\pm \sigma_\alpha$, and to reduce it to a
function of $\zeta$ and $\varepsilon$ by minimizing over $m_A$ and
$m_B$. Note that we neglect the correlations between the three
observables $q_\alpha$.} on some linear combination of $\zeta$ and
$\varepsilon$. Summarizing: each set of pulsar data leads to a
reduced $\chi^2$ of the form $\chi^2_P(\zeta,\varepsilon)=
(\zeta+\lambda_P\varepsilon-\mu_P)^2/\sigma_P^2$, equivalent to the
one-sigma constraint
$-\sigma_P<\zeta+\lambda_P\varepsilon-\mu_P<\sigma_P$. We find the
following bounds (at the 1$\sigma$ level for the first three tests,
and at the 90\% C.L. for PSR 1800$-$27)
\begin{mathletters}
\label{eq5.23}
\begin{eqnarray}
\hbox{PSR 1913+16:} & \quad\quad &
-4\times 10^{-4}< \zeta-5\times10^{-2}\varepsilon< 7\times
10^{-3}\ ,
\label{eq5.23a} \\
\hbox{PSR 1534+12:} & \quad\quad &
-8\times 10^{-2}< \zeta + 0.15\, \varepsilon < -10^{-4}\ ,
\label{eq5.23b} \\
\hbox{PSR 0655+64:} & \quad\quad &
-7\times 10^{-3}< \zeta< 4\times 10^{-3}\ ,
\label{eq5.23c} \\
\hbox{PSR 1800$-$27:} & \quad\quad &
|\zeta+\varepsilon/2|< 1.5\times 10^{-2}\ .
\label{eq5.23d}\end{eqnarray}
\end{mathletters}
These four allowed regions of the $\varepsilon$--$\zeta$ plane are
displayed in Fig.~\ref{fig8}. Clearly pulsar data favor only a small
neighborhood of the origin $\varepsilon=\zeta=0$, {\it i.e.}, of general
relativity. To combine the constraints on $\varepsilon$ and $\zeta$
coming from different pulsar experiments, we have added their individual
$\chi^2$ (as defined above) as if they were part of a total
experiment with uncorrelated Gaussian errors: $\chi^2_{\rm
tot}(\varepsilon,\zeta) = \chi^2_{1913+16}(\varepsilon,\zeta)
+ \chi^2_{1534+12}(\varepsilon,\zeta)
+ \chi^2_{0655+64}(\varepsilon,\zeta)
+ \chi^2_{1800-27}(\varepsilon,\zeta)$. [In spite of the
non-Gaussian statistics of the gravitational Stark effect in PSR
1800$-$27, we used the bound (\ref{eq5.16}) as an ``effective 1$\sigma$
level".] In the approximation explained above, each $\chi^2$ is
quadratic in $\varepsilon$ and $\zeta$. Therefore, the sum $\chi^2_{\rm
tot}(\varepsilon,\zeta)$ is a quadratic form in $\varepsilon$ and
$\zeta$. The contour level $\Delta\chi^2_{\rm tot}(\varepsilon,\zeta)=
2.3$, where\footnote{The nice global consistency of the independent
pulsar tests, proven by the overlapping of the strips in
Fig.~\ref{fig8}, means that the overall goodness-of-fit criterion
associated with $(\chi^2_{\rm tot})_{\rm min}$ is satisfied. This
entitles us to use the variation of $\chi^2_{\rm tot}$ above
$(\chi^2_{\rm tot})_{\rm min}$ to define meaningful error levels on
$\varepsilon$ and $\zeta$.} $\Delta\chi^2_{\rm tot}\equiv \chi^2_{\rm
tot} - (\chi^2_{\rm tot})_{\rm min}$, defines for two degrees of
freedom the 68\% C.L. (1$\sigma$ level) ellipse represented in
Fig.~\ref{fig9}.

In conclusion, our analysis of these four binary-pulsar tests yields
the bounds (at the combined 68\% C.L.)
\begin{equation}
|\varepsilon|< 7\times 10^{-2}\quad,\quad |\zeta|< 6\times 10^{-3}\ .
\label{eq5.24}\end{equation}
Because of our truncation of the observables (\ref{eq5.21}) to their
lowest order term in $\varepsilon$ and $\zeta$, these values should be
considered only as estimates of the constraints that binary-pulsar data
can provide. They show nevertheless that possible 2PN deviations from
general relativity can be tested with great accuracy in binary-pulsar
experiments, whereas we saw in subsection V--B that they are almost
impossible to detect in the solar system.

\section{Conclusions}
We proposed in this paper a new, theory-based framework for conceiving
and interpreting experimental tests of relativistic gravity. Previous
frameworks were characterized by a phenomenological attitude. Eddington
\cite{E23} initiated such an approach by assuming that the (static)
spherically symmetric one-body solution in a generalized relativistic
theory of gravity could differ from the Schwarzschild solution in having
arbitrary coefficients in front of the different powers of the small
parameter $Gm/\rho c^2$. Namely, he wrote
\begin{mathletters}
\label{eq6.1}
\begin{eqnarray}
-g_{00} & = & 1-2\alpha\,{Gm\over \rho c^2}
+2\beta\left({Gm\over \rho c^2}\right)^2+\cdots\ ,
\label{eq6.1a} \\
g_{ij} & = & \delta_{ij}\left[
1+2\gamma\,{Gm\over \rho c^2}+\cdots
\right]\ ,
\label{eq6.1b}\end{eqnarray}
\end{mathletters}
thereby introducing at the first post-Newtonian (1PN) level two
independent phenomenological parameters $\beta$ and $\gamma$ with
values one in general relativity [after having remarked that the
Newtonian level parameter $\alpha$ can be conventionally fixed to
unity]. The Eddington approach has been extended in several directions.
Following some pioneering work of Schiff \cite{S60} and Baierlein
\cite{B67}, Nordtvedt \cite{N68} and Will \cite{W71} introduced ten
independent phenomenological parameters, $\beta$, $\gamma$, $\xi$,
$\alpha_1$, $\alpha_2$, $\alpha_3$, $\zeta_1$, $\zeta_2$, $\zeta_3$,
$\zeta_4$, to describe the most general {\it $N$-body\/} metric at the
1PN level. [However, their subsequent work made it clear that $\beta$
and $\gamma$ played a key role among the ten parameters of their
extended ``PPN'' formalism.] Epstein and Shapiro \cite{ES80}, and
Fischbach and Freeman \cite{FF80}, extended the original Eddington
expansion (\ref{eq6.1}) of the {\it one-body\/} metric by introducing a
parameter $\epsilon_{ES} \equiv \delta_{FF}$ describing the {\it second
post-Newtonian\/} (2PN) order contribution to the (isotropic) spatial
metric:
\begin{equation}
g_{ij} = \delta_{ij}\left[
1+2\gamma\,{Gm\over \rho c^2}
+{3\over2}\,\epsilon_{ES}\left({Gm\over \rho c^2}\right)^2
+\cdots
\right]\ .
\label{eq6.2}\end{equation}
Benacquista and Nordtvedt \cite{BN88,B92,N93} tried to extend directly
the {\it $N$-body\/} PPN formalism to the {\it second post-Newtonian\/}
order by introducing a large number of {\it a priori\/} independent
parameters. Finally, in a somewhat different vein, Damour and Taylor
\cite{D88,DT92} introduced a phenomenological approach specifically
adapted to extracting the maximum possible number of relativistic
gravity tests from binary pulsar data (``parametrized post-Keplerian''
formalism).

By contrast with such phenomenological approaches, we have
systematically adopted in the present paper (which is an extension of
our previous work \cite{DEF1}) a {\it theory-based\/} approach. Instead
of trying to parametrize any conceivable phenomenological deviation
from general relativity, we work within the simplest and best motivated
class of non-Einsteinian theories: the tensor--multi-scalar theories in
which gravity is mediated by a tensor field ($g^*_{\mu\nu}$) together
with one or several scalar fields ($\varphi^a$; $a=1,2,\ldots,n$).
These theories are, in our opinion, preferred for three types of
reasons: (i)~massless scalars naturally appear as partners of the
graviton in most unified theories (from Kaluza--Klein to string
theory); (ii)~this is the only known class of theories respecting the
basic tenets of field theory (notably the absence of negative-energy
excitations) in which very high precision tests of the equivalence
principle can naturally be compatible with post-Newtonian deviations at
a measurable level\footnote{For instance, in all (positive-energy)
vector theories coupled (at the linear level) to some current, the
existing equivalence principle tests at the $10^{-12}$ level
necessarily constrain all post-Newtonian deviations at the $\sim
10^{-9}$ level. Though current unified models suggest the existence of
massless scalar fields with composition-dependent couplings, they leave
open the possibility of a very small parameter (e.g. a factor $\sim
10^{-5}$ was found in string-inspired models \cite{DP}) relating
post-Newtonian deviations to equivalence-principle tests.}; (iii)~they
naturally ``explain'' the key role played by the original Eddington
parameters $\beta$ and $\gamma$ at the 1PN level, and have a simple
enough structure (in spite of their great generality\footnote{Indeed,
we consider the most general $n$-scalar models described by
$1+n(n-1)/2$ arbitrary functions of $n$ variables.}) to allow one to
work out in detail their observational consequences at the 2PN level.

Our main results are the following. Two, and only two, new parameters
(beyond the non-Einsteinian 1PN parameters $\overline\beta\equiv\beta
-1$, $\overline\gamma\equiv \gamma -1$) quantify possible {\it
non-Einsteinian\/} effects at the 2PN level: $\varepsilon$ and $\zeta$.
The role of these 2PN parameters is threefold:

(i)~They parametrize 2PN deviations from the general relativistic
($N$-body) physical metric tensor:
\begin{mathletters}
\label{eq6.3}
\begin{eqnarray}
\delta g_{00}(x) & = & {\varepsilon\over 3c^6} U^3({\bf x})
+{\varepsilon\over c^6}\int d^3{\bf x}'\,
{G \sigma({\bf x}')U^2({\bf x}')
\over |{\bf x}-{\bf x}'|}
+{2\zeta\over c^6}\int d^3{\bf x}'\, {G\sigma({\bf x}')\over
|{\bf x}-{\bf x}'|}\int d^3{\bf x}''{G\sigma({\bf x}'')U({\bf x}'')\over
|{\bf x}'-{\bf x}''|}
\nonumber \\
& & + {2\zeta\over c^6} U({\bf x})\int d^3{\bf x}'{G\sigma({\bf x}')
U({\bf x}')\over |{\bf x}-{\bf x}'|}
+O\left({\overline\beta\over c^6},{\overline\gamma\over c^6}\right)
+O\left({1\over c^8}\right)
\ ,
\label{eq6.3a} \\
\delta g_{0i}(x) & = &
O\left({\overline\beta\over c^5},{\overline\gamma\over c^5}\right)
+O\left({1\over c^7}\right) \ ,
\label{eq6.3b} \\
\delta g_{ij}(x) & = &
O\left({\overline\beta\over c^4},{\overline\gamma\over c^4}\right)
+O\left({1\over c^6}\right)\ .
\label{eq6.3c}\end{eqnarray}
\end{mathletters}

(ii)~They determine the renormalizations, due to {\it self-gravity\/}
effects, of the various effective gravitational coupling parameters
between massive bodies:
\begin{mathletters}
\label{eq6.4}
\begin{eqnarray}
G_{AB}/G & = & 1+\eta\left(
{E_A^{\rm grav}\over m_Ac^2}
+{E_B^{\rm grav}\over m_Bc^2}
\right)
+4\zeta\left({E_A^{\rm grav}\over m_Ac^2}\right)
\left({E_B^{\rm grav}\over m_Bc^2}\right)
\nonumber \\
& & +\left({\varepsilon\over 2}+\zeta\right){
\langle U^2\rangle_A+\langle U^2\rangle_B\over
c^4} +O(\overline\beta,\overline\gamma)\times O(s^2) +O(s^3)\ ,
\label{eq6.4a} \\
\overline \gamma_{AB} & = & \overline \gamma +
O(\overline\beta,\overline\gamma)\times O(s) +O(s^2)\ ,
\label{eq6.4b} \\
\overline\beta^{\ A}_{BC} & = & \overline\beta +(\varepsilon+\zeta)
{E_A^{\rm grav}\over m_Ac^2}
+\zeta\left(
{E_B^{\rm grav}\over m_Bc^2}
+{E_C^{\rm grav}\over m_Cc^2}
\right)
+O(\overline\beta,\overline\gamma)\times O(s) +O(s^2)\ .
\label{eq6.4c}\end{eqnarray}
\end{mathletters}
Here, $G_{AB}$ denotes the effective Newtonian constant measuring the
strength of the $O(m_Am_B/r_{AB})$ leading gravitational coupling
between $A$ and $B$, while $\overline\gamma_{AB}$ and
$\overline\beta^{\ A}_{BC}$ denote effective Eddington parameters
measuring the strength of the $O(m_A m_B ({\bf v}_A-{\bf
v}_B)^2/r_{AB}c^2)$ and $O(m_A m_B m_C/ (r_{AB} r_{AC} c^2))$
post-Newtonian couplings. Moreover, $\eta$ stands for the combination
$4\overline\beta-\overline\gamma \equiv 4\beta-\gamma-3$, while $s\sim
E^{\rm grav}/ mc^2$ denotes the strength of self-gravity effects. The
more complete version of our results on self-gravity renormalizations
is given in Eqs.~(\ref{eq3.29}).

(iii)~They determine the renormalizations of the locally measured
coupling parameters (e.g. the local gravitational constant in physical
units $\widetilde G_{\rm loc}$) due to the presence of distant
``spectator'' matter (say, a mass $m_S$ at a distance $D$) around the
considered gravitating system:
\begin{mathletters}
\label{eq6.5}
\begin{eqnarray}
\widetilde G_{\rm loc} & = & \widetilde G_{\infty}\left[
1-{G m_S\over D c^2}\, \eta\right]
+ O\left({1\over D^2}\right)\ ,
\label{eq6.5a} \\
\overline\gamma_{\rm loc} & = & \overline\gamma_{\infty}
+4\, {G m_S\over D c^2}\, \overline\beta\,
(2+\overline\gamma) + O\left({1\over D^2}\right)\ ,
\label{eq6.5b} \\
\overline\beta_{\rm loc} & = & \overline\beta_{\infty}
- {G m_S\over D c^2}
\left({\varepsilon\over 2}+\zeta-8\overline\beta^2\right)
+ O\left({1\over D^2}\right)\ .
\label{eq6.5c}\end{eqnarray}
\end{mathletters}

On the other hand, we find that $\varepsilon$ and $\zeta$ {\it do not\/}
enter light-deflection nor time-delay experiments at the 2PN level
(see Appendix~C). In particular, we find that the ``post-Eddington''
formal 2PN parameter introduced in Refs.~\cite{ES80,FF80}, as
recalled in Eq.~(\ref{eq6.2}) above, is the following function of the
1PN parameters
$\overline\beta\equiv\beta-1$, $\overline\gamma\equiv\gamma-1$:
\begin{equation}
\epsilon_{ES} = 1 + {4\over 3}\, \overline\beta +
{15+8\overline\gamma\over 6}\, \overline \gamma\ .
\label{eq6.6}\end{equation}
This shows that second-order light-deflection (or time-delay)
experiments do not probe any theoretically-motivated 2PN deviations
from general relativity. More generally, after discussing observable
effects linked to $\varepsilon\neq 0$ or $\zeta\neq 0$ (in planetary
perihelia, Lunar Laser Ranging or other strong-equivalence-principle
tests, and clock experiments), we conclude that the solar system is not
an appropriate testing ground for probing possible 2PN deviations from
general relativity. More precisely, we find that in the best cases,
foreseeable technology might barely be able to constrain $\varepsilon$
and $\zeta$ at the order unity level, while, in the worst cases, these
parameters might complicate the interpretation of 1PN experiments by
contaminating some observables.

This seemingly negative conclusion is, however, to be tempered by the
following two positive conclusions of ours:

(a)~We identified binary pulsar experiments as an excellent testing
ground for the 2PN structure of relativistic gravity. By a simplified
(linearized) analysis of existing data on the binary systems PSR
1913+16, PSR 1534+12, PSR 0655+64 and PSR 1800$-$27, we were in fact
able to constrain already $\varepsilon$ and $\zeta$ at the level
\begin{equation}
|\varepsilon|< 7\times 10^{-2}\quad,\quad |\zeta|< 6\times 10^{-3}\ .
\label{eq6.7}\end{equation}

(b)~We stressed that solar-system experiments are well suited to
measuring with high precision the 1PN parameter $\overline\gamma$ which
is of greatest significance among all post-Einstein parameters. Indeed,
our theory-based analysis shows very clearly that $\overline\gamma$ is
a direct measure of the coupling strength of matter to the scalar
fields. All the other post-Einstein parameters ($\overline\beta$,
$\varepsilon$, $\zeta$, \dots) necessarily tend to zero with
$\overline\gamma$ in all theories having only positive-energy
excitations (more about this below). {}From this (theory-based) point
of view, the most important solar-system experiments would be
high-precision light-deflection or time-delay experiments reaching the
level, say, $\overline\gamma\lesssim 3\times 10^{-7}$, which comes out
naturally from mechanisms of cosmological attraction of tensor--scalar
models toward general relativity \cite{DN93,DP}\footnote{We note that
the space missions GAIA (Global Astrometric Interferometer for
Astrophysics) and SORT (Solar Orbit Relativity Test), proposed by the
European Space Agency, aim at such a level for $\overline\gamma$.}.

Let us end by stressing again some of the conceptual and technical
differences between the theory-based framework introduced here, and the
usual Eddington--Nordtvedt--Will PPN framework (and its extensions).
The natural range of values of the PPN parameters is uniformly supposed
to be of order unity, independently of the post-Newtonian order at
which they appear. In our approach, there is one basic set of
(experimentally small) coupling parameters,
\begin{equation}
\alpha_a = {\partial\ln m(\varphi)\over \partial\varphi^a}\ ,
\label{eq6.8}\end{equation}
where $m(\varphi)$ is the mass of a particle in the Einstein conformal
frame. [This way of writing the definition of $\alpha_a$ is of
great generality as it encompasses both self-gravity effects \cite{DEF1}
and possible composition-dependent effects \cite{DP}.]

All the phenomenological parameters $\overline\gamma$, $\overline\beta$,
$\zeta$, $\varepsilon$, \dots measuring the 1PN, 2PN, {\it etc.}
deviations from general relativity (see \cite{DEF1} for a list of the
3PN and higher parameters) are explicitly constructed by contracting at
least two of the basic $\alpha_a$'s with objects built from successive
covariant field derivatives of the $\alpha_a$'s: $\beta_{ab}\equiv
D_a\alpha_b$, $\beta'_{abc}\equiv D_aD_b\alpha_c$, {\it etc.} (see
section II--B for the definition of $D_a$). In particular, internal
indices in the scalar-field space being raised and lowered by means of
the positive-definite metric $\gamma_{ab}(\varphi^c)$ defining the
kinetic energy of the scalar fields,
\begin{mathletters}
\label{eq6.9}
\begin{eqnarray}
\overline\gamma & = &
-2\,{\alpha_a \alpha^a\over 1+\alpha_a \alpha^a}\ ,
\label{eq6.9a} \\
\overline\beta & = & {1\over2}\, {\alpha^a\beta_{ab}\alpha^b\over
(1+\alpha_a \alpha^a)^2}\ ,
\label{eq6.9b} \\
\zeta & = & {\alpha_a\beta^a_{\ b}\beta^b_{\ c}\alpha^c\over
(1+\alpha_a \alpha^a)^3}\ ,
\label{eq6.9c} \\
\varepsilon & = & {\beta'_{abc}\alpha^a\alpha^b\alpha^c\over
(1+\alpha_a \alpha^a)^3}\ .
\label{eq6.9d}\end{eqnarray}
\end{mathletters}
Therefore, contrary to the PPN philosophy where, say, the 1PN parameters
$\overline\gamma$ and $\overline\beta$ could be accidentally small and
the 2PN ones $\zeta$ and $\varepsilon$ of order unity, our approach
suggests that all of them tend to zero with $\overline\gamma$, or more
precisely with $\alpha^2 = -\overline\gamma/(2+\overline\gamma)$ which
is a positive measure of the total coupling strength of matter to the
scalar field. [The ones, like $\varepsilon$, which are cubic in the
$\alpha$'s are even expected to be much smaller than the quadratically
small ones\footnote{Note that the expected extra smallness of
$\varepsilon$ gives a mnemonic rule for remembering its definition.}
$\overline\gamma$, $\overline\beta$, $\zeta$.]

Finally, let us note that, as far as we are aware, the present work is
the first one to use in an effective way a direct classical
diagrammatic approach to the relativistic $N$-body
problem\footnote{Bertotti and Plebanski \cite{BP60} mainly discussed
general features of a classical diagrammatic expansion, while previous
work by the Japanese school \cite{OOKH,OOKH73b} had used quantum
diagrams and then converted $S$-matrix elements into an effective
$N$-body potential.}. We hope to have convinced the reader of the
technical power of this method. Indeed, once one is used to the
notation, our results (\ref{eq6.9}) on the complete set of parameters
entering the first two post-Newtonian levels can be obtained in drawing
just half a page of simple diagrams.

\acknowledgments
We thank Zaven Arzoumanian for communicating us, in advance of
publication, the results of his thesis work.

\appendix
\section{Explicit diagrammatic calculations}
In order to compute explicitly the diagrams of Fig.~\ref{fig7}, we must
first derive the expressions of the propagators and of the different
vertices of Fig.~\ref{fig4}. As discussed in subsection III--A, we need
to expand the gauged-fixed action of the theory up to the fourth order
in $h_{\mu\nu}\equiv g^*_{\mu\nu}-f_{\mu\nu}$ and
$\varphi^a-\varphi_0^a$. We need only the terms quadratic in
$h_{\mu\nu}$ in the Einstein--Hilbert action (\ref{eq2.4}) to define
the graviton propagator (see e.g. \cite{OOKH73b} for higher-order
terms). We get easily
\begin{equation}
S_{\rm spin2} = {c^3\over 16\pi G_*}\int
d^4x\left[
-{1\over 2}(\partial_\mu h_{\alpha\beta})Q^{\alpha\beta\gamma\delta}
(\partial^\mu h_{\gamma\delta})+{1\over 2}\left(\partial_\nu
h_\mu^\nu-{1\over 2}\partial_\mu h_\nu^\nu\right)^2
+O(h^3)\right]\ ,
\label{eqA1}\end{equation}
where the indices are raised with the flat metric $f^{\mu\nu}={\rm
diag}(-1,1,1,1)$, and where $Q^{\alpha\beta\gamma\delta}\equiv {1\over
4} (f^{\alpha\gamma}f^{\beta\delta}+f^{\alpha\delta}f^{\beta\gamma}
-f^{\alpha\beta}f^{\gamma\delta})$. To invert this kinetic term, it is
necessary to fix the gauge. The most convenient choice for our
calculations will be the harmonic gauge, corresponding to a
gauge-fixing term $-(c^3/32\pi G_*)\int d^4x(\partial_\nu
h_\mu^\nu-{1\over 2}\partial_\mu h_\nu^\nu)^2 +O(h^3)$ (see e.g.
\cite{DS85} for the exact harmonic-gauge-fixing term).
This cancels exactly
the second term of (\ref{eqA1}), and we get after an integration by
parts
\begin{equation}
S_{\rm spin2}^{\rm g.f.} = {c^3\over 16 \pi G_*}\int d^4 x\, {1\over 2}
h_{\alpha\beta} Q^{\alpha\beta\gamma\delta} \Box_f h_{\gamma\delta} +
{\rm tot.div.} +O(h^3)\ ,
\label{eqA2}\end{equation}
where $\Box_f\equiv f^{\mu\nu}\partial_\mu\partial_\nu$ is the flat
d'Alembertian. This kinetic term has therefore the form $\int
d^4x(-{1\over2}h{\cal P}^{-1}_h h)$, where ${\cal P}_h$ defines the
graviton propagator, represented as a curly line in Fig.~\ref{fig7}. In
terms of the Green function (\ref{eq3.17}), satisfying $\Box_f{\cal
G}(x) = -4\pi \delta^{(4)}(x)$, we thus get
\begin{equation}
{\cal P}^h _{\alpha\beta\gamma\delta}(x,y)
= {4 G_*\over c^3} P_{\alpha\beta\gamma\delta}\,
{\cal G}(x-y)\ ,
\label{eqA3}\end{equation}
where $P_{\alpha\beta\gamma\delta} \equiv
f_{\alpha\gamma}f_{\beta\delta}+f_{\alpha\delta}f_{\beta\gamma}
-f_{\alpha\beta}f_{\gamma\delta}$ is the inverse of
$Q^{\alpha\beta\gamma\delta}$.

Let us now expand the action (\ref{eq2.5}) of the scalar fields up to
the fourth order in $\varphi^a-\varphi_0^a$ or $h_{\mu\nu}$. To
simplify the expressions, it is convenient to choose Riemann normal
coordinates at $\varphi_0$ in the internal scalar space, so that the
metric $\gamma_{ab}$ can be expanded as in Eq.~(\ref{eq3.15}). It is
also useful to define the origin of this scalar space at $\varphi_0$,
{\it i.e.}, to choose the coordinates $\varphi^a$ of this space so that
$\varphi_0^a=0$. Then, the scalar-field action (\ref{eq2.5}) can be
expanded as
\begin{eqnarray}
S_{\rm spin0} & = & -{c^3\over 4 \pi G_*}\int d^4 x\,
{1\over 2}\partial_\mu\varphi^a\partial_{\nu}\varphi^b
\left(f^{\mu\nu}-h^{\mu\nu}+h^\mu_\rho h^{\rho\nu}+O(h^3)\right)
\nonumber \\
&& \times \left(1+{1\over2} h^\alpha_\alpha - {1\over 2}h_{\alpha\beta}
Q^{\alpha\beta\gamma\delta}h_{\gamma\delta}+O(h^3) \right)
\nonumber \\
&& \times \left(\gamma_{ab}(\varphi_0)-{1\over 3}R_{acbd}(\varphi_0)
\varphi^c \varphi^d + O(\varphi^3)\right)\ ,
\label{eqA4}\end{eqnarray}
where the terms inside parentheses are respectively the expansions of
$g_*^{\mu\nu}$, $\sqrt{g_*}$, and $\gamma_{ab}(\varphi)$. The kinetic
term of the scalar fields reads therefore $(c^3/4 \pi G_*)\int d^4 x\,
{1\over 2} \varphi^a \gamma_{ab}(\varphi_0) \Box_f \varphi^b + {\rm
tot.div.} = \int d^4x (-{1\over 2}\varphi{\cal
P}^{-1}_{\varphi}\varphi)$, where
\begin{equation}
{\cal P}_{\varphi}^{ab}(x,y)
= {G_*\over c^3} \gamma^{ab}(\varphi_0) {\cal G}(x-y)
\label{eqA5}\end{equation}
is the scalar propagator, represented as a straight line in
Fig.~\ref{fig7}. [As before, $\gamma^{ab}$ denotes the inverse of
$\gamma_{ab}$.]

We can also derive from (\ref{eqA4}) the expressions of the vertices
connecting scalar fields and gravitons. Our conventions for defining
vertices are the following: (i)~We first define some formal ``global''
vertices $V_i\equiv i S_i$ when considering (as in Figs.~\ref{fig3} and
\ref{fig4}) the gravitational sector as a whole, $\Phi = (\varphi,h)$.
(ii)~We then define the individual vertices $V_i(\varphi,h)$ by
formally expanding the global multilinear forms $V_i(\Phi)$ as if
$\Phi$ were equal to the sum $\varphi+h$, e.g.
\begin{equation}
V_3(\varphi+h,\varphi+h,\varphi+h) = V_3(\varphi,\varphi,\varphi)
+3 V_3(\varphi,\varphi,h)+ 3V_3(\varphi,h,h)+V_3(h,h,h)\ .
\label{eqA6}\end{equation}
This convention allows us to
use directly the simple global diagrams of Fig.~\ref{fig6} with the
intuitive replacements of Fig.~\ref{fig7}. For instance, the vertex of
order $O(\varphi\varphi h)$, entering the first {\sf T} diagram of
Fig.~\ref{fig7} as well as the first four {\sf F} diagrams and the
first three {\sf H} diagrams, reads
\begin{equation}
{\sf T}_{\varphi\varphi h} =
{c^3\over 4 \pi G_*} \int d^4x\, \gamma_{ab}(\varphi_0)
Q^{\mu\nu\alpha\beta}
\partial_\mu\varphi^a \partial_\nu\varphi^b h_{\alpha\beta}\ ,
\label{eqA7}\end{equation}
where $Q^{\mu\nu\alpha\beta}$ is the same tensor as in (\ref{eqA1})
above. In other words, the kernel defining the ${\sf T}_{\varphi\varphi
h}$ vertex is
\begin{equation}
({\sf T}_{\varphi\varphi h})^{\alpha\beta}_{ab}(x_1,x_2,x_3) =
{c^3\over 4 \pi G_*} \gamma_{ab}(\varphi_0) Q^{\mu\nu\alpha\beta}
{\partial\over\partial x_3^\mu}\delta(x_3-x_1)
{\partial\over\partial x_3^\nu}\delta(x_3-x_2)
\ .
\label{eqA8}\end{equation}
Note that it is symmetric in $x_1$, $x_2$ (scalar lines), but not in all
variables since they represent physically inequivalent lines.
The vertex connecting two scalar lines and two graviton lines,
entering the second {\sf X} diagram of Fig.~\ref{fig7}, reads
\begin{equation}
{\sf X}_{\varphi\varphi h h} =
-{c^3\over 12\pi G_*}\int d^4 x\,
\gamma_{ab}(\varphi_0)\left(2f^{\mu\alpha}Q^{\nu\beta\gamma\delta}
-{1\over 2} f^{\mu\nu}Q^{\alpha\beta\gamma\delta}\right)
\partial_\mu\varphi^a\partial_\nu\varphi^b h_{\alpha\beta}
h_{\gamma\delta}\ .
\label{eqA9}\end{equation}
Finally, the vertex connecting four scalar lines, entering the first
{\sf X} diagram of Fig.~\ref{fig7}, reads
\begin{equation}
{\sf X}_{\varphi^4} =
{c^3\over 6 \pi G_*} \int d^4 x R_{acbd}(\varphi_0) \varphi^c \varphi^d
\partial_\mu\varphi^a \partial^\mu \varphi^b\ ,
\label{eqA10}\end{equation}
where $R_{abcd}$ is the Riemann curvature tensor of $\gamma_{ab}$. It
should be noted that the four scalar lines of this vertex are not
equivalent, since two of them involve the derivatives
$\partial_\mu\varphi^a$ of the fields. The same remark holds also for
the multi-graviton vertices of order $O(h^3)$ and $O(h^4)$ in
(\ref{eqA1}). One should therefore symmetrize these vertices
[{\it i.e.}, write the distributional kernel ${\sf
X}_{\varphi^4}(x_1,x_2,x_3,x_4)$ read off (\ref{eqA10}) as a symmetric
function of $x_1$, $x_2$, $x_3$, $x_4$] before
computing the Fokker action (\ref{eq3.14}). Alternatively, one can also
use their non-symmetric form, but take into account the different ways
to choose the lines involving derivatives in the diagrams of
Fig.~\ref{fig7}. This does not change anything for the {\sf X}
diagrams, since the Lagrangian is anyway symmetrized over the 4 bodies
$A,B,C,D$, but this leads to the numerical weights displayed in
Fig.~\ref{fig10} for the {\sf F} and {\sf H} diagrams.

Let us now consider the matter action (\ref{eq2.15}), describing $N$
self-gravitating bodies. As shown in subsection III--A, we need to
expand it only up to the third order in $h_{\mu\nu}$ and $\varphi^a$.
Still using Riemann normal coordinates at $\varphi_0=0$ in the
$\varphi$-space, we get easily
\begin{eqnarray}
S_m & = & -\sum_{A=1}^N\int m_A^0 c^2 d\tau_A\,
\Biggl[1+ (\alpha_a^A)_0\, \varphi^a + {1\over 2}
(\beta_{ab}^A+\alpha_a^A\alpha_b^A)_0\, \varphi^a \varphi^b
\nonumber \\
&& +{1\over 6} (\beta'^A_{abc}+\beta_{ab}^A\alpha_c^A
+\beta_{bc}^A\alpha_a^A+\beta_{ca}^A\alpha_b^A
+\alpha_a^A\alpha_b^A\alpha_c^A)_0\,
\varphi^a \varphi^b \varphi^c
+O(\varphi^4)\Biggr]
\nonumber \\
&& \times\Biggl[
1-{1\over 2}h_{\mu\nu}u_A^\mu u_A^\nu
-{1\over 8}(h_{\mu\nu}u_A^\mu u_A^\nu)^2
-{1\over 16}(h_{\mu\nu}u_A^\mu u_A^\nu)^3
+O(h^4)
\Biggr]\ ,
\label{eqA11}\end{eqnarray}
where the first bracket comes from the expansion of $m_A(\varphi)$
around $\varphi_0$, and the second one from the expansion of $ds_A^*$
around the flat metric. All the fields appearing in (\ref{eqA11}) must
be evaluated on the worldline $x^\mu = x^\mu(\tau_A)$. As above,
$d\tau_A \equiv (1-{\bf v}_A^2/c^2)^{1/2}dt$ denotes the Minkowski
proper time of body $A$, $u_A^\mu\equiv dx_A^\mu/cd\tau_A$ denotes its
(Minkowski) unit 4-velocity, and the index $\scriptstyle 0$ means that
the corresponding quantities are evaluated at $\varphi^a = \varphi_0^a$
($=0$). The different vertices connecting material bodies to gravitons
or scalar fields can thus be read directly from (\ref{eqA11}), taking
into account both the factors 1, 2 or 3 entering their definition in
Fig.~\ref{fig4} and the binomial factors coming from our
convention~(ii) above, {\it cf.} Eq.~(\ref{eqA6}). The linear
interaction terms read
\begin{mathletters}
\label{eqA12}
\begin{eqnarray}
{\rm I}_\varphi & = & -\sum_A\int d\tau_A\, m_A^0 c^2
(\alpha_a^A)_0\, \varphi^a(x_A)\ ,
\label{eqA12a} \\
{\rm I}_h & = & \sum_A \int d\tau_A\, m_A^0 c^2
{1\over 2}\, u_A^\alpha u_A^\beta h_{\alpha\beta}(x_A)\ .
\label{eqA12b}\end{eqnarray}
\end{mathletters}
The corresponding spacetime sources (white blobs) read thus explicitly
\begin{mathletters}
\label{eqA13}
\begin{eqnarray}
\sigma_a(x) & = & -\sum_A\int d\tau_A\, m_A^0 c^2
(\alpha_a^A)_0\, \delta^{(4)}(x-x_A(\tau))\ ,
\label{eqA13a} \\
\sigma^{\alpha\beta}(x) & = & {1\over 2}\sum_A \int d\tau_A\, m_A^0 c^2
u_A^\alpha u_A^\beta \delta^{(4)}(x - x_A(\tau))\ .
\label{eqA13b}\end{eqnarray}
\end{mathletters}
As for the nonlinear interaction with the source (vertices connecting
matter to several field lines), they read, when omitting a common
$\sum_A\int d\tau_A\ m_A^0 c^2 \delta^{(4)}(x-x_A(\tau))$ in front:
\begin{mathletters}
\label{eqA14}
\begin{eqnarray}
{\sf V}_{\varphi\varphi} &\ :\qquad &
-(\beta^A_{ab}+\alpha_a^A\alpha_b^A)\ ,
\label{eqA14a} \\
{\sf V}_{\varphi h} &\ :\qquad &
{1\over 2} \alpha^A_a u_A^\alpha u_A^\beta\ ,
\label{eqA14b} \\
{\sf V}_{h h} &\ :\qquad &
{1\over 4}u_A^\alpha u_A^\beta u_A^\gamma u_A^\delta\ ,
\label{eqA14c} \\
\in_{\varphi\varphi\varphi} &\ :\qquad &
-{1\over 2}\left(
\beta'^A_{abc} + 3 \beta^A_{(ab}\alpha^A_{c)}
+\alpha^A_a \alpha^A_b \alpha^A_c
\right)\ ,
\label{eqA14d} \\
\in_{\varphi\varphi h} &\ :\qquad &
{1\over 4}(\beta^A_{ab}+\alpha^A_a\alpha^A_b)u_A^\alpha u_A^\beta\ ,
\label{eqA14e} \\
\in_{\varphi h h} &\ :\qquad &
{1\over 8} \alpha^A_a u_A^\alpha u_A^\beta u_A^\gamma u_A^\delta\ ,
\label{eqA14f} \\
\in_{h h h} &\ :\qquad &
{3\over 16}u_A^\alpha u_A^\beta u_A^\gamma u_A^\delta u_A^\epsilon
u_A^\zeta\ .
\label{eqA14g}\end{eqnarray}
\end{mathletters}

The calculation of the different diagrams of Fig.~\ref{fig7} can now be
performed straightforwardly. Let us summarize our diagrammatic rules:
The Fokker action is a sum of contributions, each of which is
represented by a diagram endowed with a numerical coefficient obtained
by combining the factors indicated in Fig.~\ref{fig6} and in
Fig.~\ref{fig7}. Each (bare) diagram is computed by the following
rules: (i)~replace the white blobs by Eqs.~(\ref{eqA13}) if they have
only one incident line, or by Eqs.~(\ref{eqA14}) [with an extra
$\sum_A\int d\tau_A\, m_A^0 c^2 \delta^{(4)}(x-x_A(\tau))$] if they
have several incident lines; (ii)~replace each internal line by the
appropriate propagator ${\cal P}(x,y)$; (iii)~replace each field vertex
by a suitably symmetrized distributional kernel; (iv)~integrate over
all the spacetime points. An additional rule is that infinite
self-interactions are discarded. (As they have the same structure as in
general relativity, this rule can probably be justified by methods
similar to the ones used in deriving the 2PN Lagrangian in Einstein's
theory \cite{DD}.) In order to save space, we will not compute each of
the diagrams in the present appendix, but rather show the technique on
the most important of them.

Let us start with the 2-body interaction term ${1\over 2}\,{\rm I}$
of Eq.~(\ref{eq3.14}). We get
\begin{eqnarray}
{1\over 2}\,{\rm I} & = &
{1\over 2}\int\!\!\int dx\, dy \left[
\sigma_a(x){\cal P}^{ab}_\varphi(x,y)\sigma_b(y)
+\sigma^{\alpha\beta}(x){\cal P}^h_{\alpha\beta\gamma\delta}(x,y)
\sigma^{\gamma\delta}(y)
\right]
\nonumber \\
& = & {1\over 2}\sum_{A\neq B} \int d\tau_A
\int d\tau_B\, (m_A^0 c^2) (m_B^0 c^2)
{G_*\over c^3} {\cal G}(x_A-x_B) \left[(\alpha_A\alpha_B)_0
+ 2 (u_A u_B)^2 - 1\right]\ ,
\label{eqA15}\end{eqnarray}
where $(\alpha_A\alpha_B) = \alpha_a^A \gamma^{ab} \alpha_b^B$ comes
from the scalar propagator (first {\rm I} diagram of Fig.~\ref{fig7}),
and $2(u_Au_B)^2 - 1 = u_A^\alpha u_A^\beta \,
P_{\alpha\beta\gamma\delta}\, u_B^\gamma u_B^\delta$ comes from the
graviton propagator (second {\rm I} diagram). Introducing
the notation $G_{AB}\equiv G_* [1+(\alpha_A\alpha_B)_0]$ and
$2+\overline\gamma_{AB}\equiv 2/[1+(\alpha_A\alpha_B)_0]$ as in
Eqs.~(\ref{eq2.20}) above, we thus get
\begin{equation}
{1\over 2}\,{\rm I} = {1\over 2}\sum_{A\neq B}\int d\tau_A \int
d\tau_B\ G_{AB} m_A^0 m_B^0 \left[1+(2+\overline\gamma_{AB})
\left((u_A u_B)^2-1\right)\right] c{\cal G}(x_A-x_B)\ .
\label{eqA16}\end{equation}
This action describes the 2-body interaction of self-gravitating bodies
having arbitrary velocities. The second post-Keplerian
approximation can now be obtained by expanding this result in powers of
$v/c$, using Eqs.~(\ref{eq3.19})--(\ref{eq3.21}) above. We get an
expression of the form (\ref{eq3.23}), where the corrections
proportional to $\overline \gamma_{AB}/c^4$ involve the function
\begin{eqnarray}
f_1({\bf r}_{AB},{\bf v}_A,{\bf v}_B,{\bf a}_A,{\bf a}_B,\dot{\bf a}_B)
& = &
\left[\left({\bf v}_A-{\bf v}_B\right)^2
\left({\bf v}_A^2+{\bf v}_B^2\right)/2
-\left({\bf v}_A\times{\bf v}_B\right)^2\right]/r_{AB}
\nonumber \\
&& +{1\over 2}\,{\partial^2\over \partial t_B^2}\left[\,
|{\bf x}_A-{\bf x}_B(t_B)|\, \left({\bf v}_A-{\bf
v}_B(t_B)\right)^2\right]_{t_B=t}\ .
\label{eqA17}\end{eqnarray}
Note that only ${\bf x}_B$ and ${\bf v}_B$ are time-differentiated in
the second term, before setting $t_B=t$. The derivative of ${\bf a}_B$
can be eliminated by an integration by parts, and up to a total
derivative, we finally get
\begin{eqnarray}
f_1({\bf r}_{AB},{\bf v}_A,{\bf v}_B,{\bf a}_A,{\bf a}_B) & = &
{{\bf v}_{AB}^2\over 2 r_{AB}}\left[
{\bf v}_A^2+{\bf v}_B^2+{\bf v}_A\cdot{\bf v}_B
-({\bf v}_A\cdot{\bf n}_{AB}) ({\bf v}_B\cdot{\bf n}_{AB})\right]
-{({\bf v}_A\times{\bf v}_B)^2\over r_{AB}}
\nonumber \\
&& +({\bf v}_A\cdot{\bf n}_{AB})({\bf a}_B\cdot{\bf v}_{AB})
+({\bf v}_B\cdot{\bf n}_{AB})({\bf a}_A\cdot{\bf v}_{AB})
+({\bf a}_A\cdot{\bf a}_B)r_{AB}\ ,
\nonumber \\
&&\label{eqA18}\end{eqnarray}
where ${\bf v}_{AB}\equiv {\bf v}_A-{\bf v}_B$ and
${\bf n}_{AB} \equiv ({\bf x}_A-{\bf x}_B)/r_{AB}$.

Let us now consider the 3-body interaction, written as ${1\over 2}{\sf
V}+{1\over 3}{\sf T}$ in (\ref{eq3.14}) and Fig.~\ref{fig6}. The three
{\sf V} diagrams of Fig.~\ref{fig7} give the contribution
\begin{eqnarray}
{1\over 2}{\sf V} & = & {1\over 2}\sum_{B\neq A\neq C}
\int\!\! d\tau_A \!\!\int\!\! d\tau_B \!\! \int\!\! d\tau_C
(m_A^0 c^2)(m_B^0 c^2)(m_C^0 c^2)
{G_*^2\over c^6} {\cal G}(x_A-x_B){\cal G}(x_A-x_C)
\nonumber \\
&& \times
\biggl[\left(2(u_A u_B)^2-1-(\alpha_A\alpha_B)_0\right)
\left(2(u_A u_C)^2-1-(\alpha_A\alpha_C)_0\right)
\nonumber\\
&& -(\alpha_B\beta_A\alpha_C)_0
-2(\alpha_A\alpha_B)_0(\alpha_A\alpha_C)_0
\biggr]\ .
\label{eqA19}\end{eqnarray}
Introducing as in Eqs.~(\ref{eq2.20}) the notation $G_{AB}$,
$\overline\gamma_{AB}$ and $\overline\beta^{\ A}_{BC}$, this
contribution can therefore be written as
\begin{eqnarray}
{1\over 2}{\sf V} & = & \sum_{B\neq A\neq C}\int d\tau_A
\int d \tau_B \int d\tau_C\ {G_{AB}G_{AC}m_A^0m_B^0m_C^0\over c^2}\,
c{\cal G}(x_A-x_B) \, c{\cal G}(x_A-x_C)
\nonumber \\
&& \times\biggl\{
\left[1+\overline \gamma_{AB}
+(2+\overline \gamma_{AB})\left((u_A u_B)^2-1\right)\right]
\left[1+\overline \gamma_{AC}
+(2+\overline \gamma_{AC})\left((u_A u_C)^2-1\right)\right]
\nonumber \\
&& -2\overline\beta^{\ A}_{BC}- {\overline \gamma_{AB}\overline
\gamma_{AC}\over 2}
\biggr\}\ .
\label{eqA20}\end{eqnarray}
The 2PK approximation can now be obtained easily by expanding this
expression in powers of $v/c$. Note that the Newtonian approximation of
the Green function (\ref{eq3.21}), $c{\cal
G}(x_A-x_B)=\delta(t_A-t_B)/r_{AB}+O(1/c^2)$, is sufficient for the
terms proportional to $(u_Au_B)^2-1 = ({\bf v}_A -{\bf
v}_B)^2/c^2+O(1/c^4)$, or $(u_Au_C)^2-1$. The accelerations introduced
by these Green functions in the other terms can be eliminated by
suitable integrations by parts. For instance, one can write
\begin{eqnarray}
{1\over r_{AB}}\times{1\over 2 c^2}\,{\partial^2\over
\partial t_C^2}
\left|{\bf x}_A-{\bf x}_C(t_C)\right| & = & {\rm tot.div.}
-{({\bf v}_{AB}\cdot{\bf n}_{AB})({\bf v}_C\cdot{\bf n}_{AC})\over
2 r_{AB}^2c^2}
\nonumber \\
&& +{{\bf v}_A\cdot{\bf v}_C
-({\bf v}_A\cdot{\bf n}_{AC})({\bf v}_C\cdot{\bf n}_{AC})
\over 2 r_{AB}r_{AC} c^2}\ .
\label{eqA21}\end{eqnarray}
The {\sf T} diagrams of Fig.~\ref{fig7} are a little more subtle to
compute, because the central vertex is not located on a material body,
and also because it involves derivatives of the fields. For instance,
the first {\sf T} diagram, involving one graviton and two scalar lines,
gives the contribution
\begin{eqnarray}
{\sf T}_1 & = & \sum_{A,B,C}\int\! d\tau_A\!\int\! d\tau_B\!\int
\!d\tau_C\!\int\! d^4 x\
{G_*^2 m_A^0 m_B^0 m_C^0\over 2\pi}\, (\alpha_A\alpha_B)_0
\nonumber \\
&& \times\, u_C^\mu u_C^\nu\, {\partial{\cal G}(x-x_A)\over \partial
x^\mu}
\, {\partial{\cal G}(x-x_B)\over \partial x^\nu}
\, {\cal G}(x-x_C)\ ,
\label{eqA22}\end{eqnarray}
where $x^\mu$ is the arbitrary spacetime location of the vertex. The
lowest order term of the post-Keplerian approximation reads thus
\begin{eqnarray}
{\sf T}_1 & = & \sum_{A,B,C}\int\! dt_A\!\int\! dt_B\!\int\!
dt_C\!\int\! dt
\!\int\! d^3{\bf x}\
{G_*^2 m_A^0 m_B^0 m_C^0\over 2\pi c^2}\, (\alpha_A\alpha_B)_0
\nonumber \\
&& \times\, u_C^\mu u_C^\nu\,
{\partial\over\partial x^\mu}\!\left[{\delta(t-t_A)\over
r_{xA}}\right]
{\partial\over\partial x^\nu}\!\left[{\delta(t-t_B)\over
r_{xB}}\right]
\, {\delta(t-t_C)\over r_{xC}}\times
\left(1+O\left({1\over c^2}\right)\right)
\nonumber \\
& = & -\sum_{A,B,C}\int\! dt\!\int\! d^3{\bf x}\
{G_* G_{AB} \overline \gamma_{AB} m_A^0 m_B^0 m_C^0\over 4\pi c^4}\,
{({\bf v}_{AC}\cdot{\bf n}_{xA})({\bf v}_{BC}\cdot{\bf n}_{xB})
\over r_{xA}^2 r_{xB}^2 r_{xC}}
+O\left({1\over c^6}\right)\ ,
\label{eqA23}\end{eqnarray}
where we have used $u_C^\mu\partial_\mu[\delta(t-t_A)/r_{xA}]
=\delta(t-t_A)\, {\bf v}_{AC}\cdot{\bf n}_{xA}/r_{xA}^2 c +O(1/c^3)$.
Therefore, although this {\sf T} diagram is of the same formal order
$G^2m^3/r^2c^2$ as the {\sf V} diagrams in the non-linearity expansion
(arbitrarily large velocities), it reduces to order $O(1/c^4)$ in the
post-Keplerian approximation ($|v/c|\ll 1$). This is due to the
particular form of the ${\sf T}_{\varphi\varphi h}$ vertex,
Eq.~(\ref{eqA7}), which involves a specific combination of derivatives
of the fields with the inverse of the tensor
$P_{\alpha\beta\gamma\delta}$ entering the graviton propagator
(\ref{eqA3}). It should be noted that the presence of derivatives is
not sufficient to conclude that the diagram is reduced by a factor
$(v/c)^2$. For instance, the second {\sf T} diagram of Fig.~\ref{fig7},
involving three graviton lines, does contribute at the first
post-Keplerian order,
\begin{equation}
{1\over 3} {\sf T}_2 = -\sum_{B\neq A\neq C} \int dt\ {G_*^2 m_A^0 m_B^0
m_C^0\over r_{AB} r_{AC} c^2} + O\left({1\over c^4}\right)\ ,
\label{eqA24}\end{equation}
although the 3-graviton vertex is also of the form $h\partial h\partial
h$. Indeed, the dominant contribution to this diagram is proportional
to the contraction $\partial_\mu{\cal G}(x-x_A)\partial^\mu{\cal
G}(x-x_B)= \delta(t-t_A) \delta(t-t_B) \,{\bf n}_{xA}\cdot{\bf
n}_{xB}/r_{xA}^2 r_{xB}^2 c^2 + O(1/c^4)$, which starts at order
$1/c^2$. The sum of (\ref{eqA24}) with the lowest order term of
(\ref{eqA20}) gives the 1PK contribution to the 3-body interaction
Lagrangian, displayed in Eqs.~(\ref{eq2.19c}) or (\ref{eq3.25}).

Let us turn now to the most important 4-body interaction terms, namely
the $\in$, {\sf Z}, and {\sf X} diagrams of Fig.~\ref{fig7}, which
involve the new parameters (\ref{eq3.18}). At the second post-Keplerian
approximation, it is sufficient to use the instantaneous Green function
$c{\cal G}(x_A-x_B) = \delta(t_A-t_B)/r_{AB}$ in these diagrams, and
one easily gets
\begin{eqnarray}
{1\over 3}\!\in\! & = & {1\over 2}\sum_{A\neq(B,C,D)}\int dt\
{G_{AB}G_{AC}G_{AD} m_A^0m_B^0m_C^0m_D^0\over r_{AB}r_{AC}r_{AD} c^4}
\biggl[1+{1\over 3}\varepsilon^{\ \ A}_{BCD} + {2\over 3}
\left(\overline\beta^{\ A}_{BC}+\overline\beta^{\ A}_{BD}
+\overline\beta^{\ A}_{CD}\right)
\nonumber \\
&& +{2\over 3}\left(\overline\gamma_{AB}+\overline\gamma_{AC}
+\overline\gamma_{AD}\right)
+{1\over 2}\left(\overline\gamma_{AB}\overline\gamma_{AC}
+\overline\gamma_{AB}\overline\gamma_{AD}
+\overline\gamma_{AC}\overline\gamma_{AD}\right)
+{1\over 3}\overline\gamma_{AB}\overline\gamma_{AC}\overline\gamma_{AD}
\biggr]
\nonumber \\
&& +O\left({1\over c^6}\right)\ ,
\label{eqA25} \\
{1\over 2}\, {\sf Z} & = & {1\over 2}\sum_{A\neq B\neq C\neq D}\int
dt\ {G_{AB}G_{BC}G_{CD} m_A^0m_B^0m_C^0m_D^0\over r_{AB}r_{BC}r_{CD}
c^4}
\biggl[1 + \zeta_{ABCD} + 2\left(\overline\beta^{\ B}_{AC}
+\overline\beta^{\ C}_{BD}\right)
\nonumber \\
&& + {1\over 2}\left(2+\overline\gamma_{BC}\right)
\left(\overline\gamma_{AB}+\overline\gamma_{CD}
+\overline\gamma_{AB}\overline\gamma_{CD}\right)\biggr]
+O\left({1\over c^6}\right)\ .
\label{eqA26}\end{eqnarray}
[Note that the $\varepsilon$ contribution to the $\in$ diagram has a
non-trivial normalization: $\in \sim {1\over 2}\varepsilon
G^3m^4/r^3c^4$, while ${\sf Z}\sim \zeta G^3m^4/r^3c^4$.] The
contribution of the first {\sf X} diagram of Fig.~\ref{fig7} reads
\begin{eqnarray}
{1\over 4}\,{\sf X}_1 & = & {1\over 4}\sum_{A,B,C,D}
\int\! dt_A\!\int\! dt_B\!\int
\! dt_C\!\int\! dt_D\!\int\! dt\!
\int\! d^3 {\bf x}\
{G_*^3 m_A^0m_B^0m_C^0m_D^0\over 6\pi c^4}
\left(R_{acbd}\alpha_A^a\alpha_B^b\alpha_C^c\alpha_D^d\right)_0
\nonumber \\
&& \times\ {\delta(t-t_C)\over r_{xC}}\,
{\delta(t-t_D)\over r_{xD}}\,
\partial_\mu\!\left[{\delta(t-t_A)\over r_{xA}}\right]
\partial^\mu\!\left[{\delta(t-t_B)\over r_{xB}}\right]
+O\left({1\over c^6}\right)
\nonumber \\
& = & {1\over 24\pi}\sum_{A,B,C,D}
\int\! d t\!\int\! d^3{\bf x}\
{G_*^3 m_A^0m_B^0m_C^0m_D^0 ({\bf n}_{xA}\cdot{\bf n}_{xB})
\over r_{xA}^2 r_{xB}^2 r_{xC} r_{xD}}
\, \chi_{ACBD}
+O\left({1\over c^6}\right)\ .
\label{eqA27}\end{eqnarray}
Note that although $\chi_{ACBD}$ is antisymmetric in $A,C$ (and $B,D$),
and although the sum is taken over all possible choices of $A,B,C,D$,
this contribution does not vanish identically since the integrand
$({\bf n}_{xA}\cdot{\bf n}_{xB})/r_{xA}^2 r_{xB}^2 r_{xC} r_{xD}$ is
not symmetric in $A,C$. This contribution vanishes nevertheless in the
2PN approximation, {\it cf.} (\ref{eq3.32}), or if the theory involves
only one scalar field ($R_{abcd}(\varphi)=0$). The contributions of the
two other {\sf X} diagrams of Fig.~\ref{fig7} are computed in the same
way. The last one gives the usual contribution obtained in general
relativity, and we find for the second {\sf X} diagram
\begin{equation}
{6\over 4}\,{\sf X}_2 = -{1\over 8\pi}
\sum_{A,B,C,D}
\int \! dt\! \int \! d^3{\bf x}\
{G_*^2 G_{AB} \overline \gamma_{AB} m_A^0 m_B^0 m_C^0 m_D^0\over c^4}
\ {({\bf n}_{xA}\cdot{\bf n}_{xB})\over
r_{xA}^2 r_{xB}^2 r_{xC} r_{xD}}
+O\left({1\over c^6}\right)\ .
\label{eqA28}\end{equation}

Let us end this appendix by a brief discussion of the {\sf F} and {\sf
H} diagrams of Fig.~\ref{fig7}, which are less important since the
deviations from general relativity they give at 2PN order are
proportional to $\overline \gamma$, $\overline \gamma^2$ or $\overline
\beta$, and therefore already tightly constrained by present
experimental data, Eqs.~(\ref{eq5.1}). Is is easy to see that the first
four {\sf F} diagrams do not contribute at the second post-Keplerian
level, because of the presence of the ${\sf T}_{\varphi\varphi h}$
vertex whose graviton is directly coupled to a material body. Indeed,
the calculation is similar to that of the first {\sf T} diagram,
Eqs.~(\ref{eqA22})--(\ref{eqA23}), and the factor
$(u^\mu\partial_\mu{\cal G})^2\propto v^2$ reduces these {\sf F}
diagrams to the order $G^3 m^4 v^2/r^3 c^6$. Similarly, the second {\sf
H} diagram is even of order $G^3 m^4 v^4/r^3c^8$, and contributes only
at the fourth post-Keplerian level. On the contrary, the first and
third {\sf H} diagrams do contribute at the 2PK level, because the
graviton of their ${\sf T}_{\varphi\varphi h}$ vertices is connected to
a second vertex, and not directly to a material body. These diagrams
involve therefore contractions of the form $\partial_\mu{\cal
G}\partial^\mu{\cal G}$, like in (\ref{eqA24}), and are indeed of order
$G^3m^4/r^3c^4$. Note that the four material bodies of these {\sf H}
diagrams are not supposed to be necessarily different from each other,
since they are not directly connected by a propagator (as opposed to
the {\rm I}, {\sf V}, $\in$, or {\sf Z} diagrams). Together with the
first {\sf T} diagram (\ref{eqA23}) and the second {\sf X} diagram
(\ref{eqA28}), they thus yield contributions proportional to
\begin{equation}
{\overline\gamma_{AA}\over 1+\overline\gamma_{AA}/2} =
{G_{AA}\over G_*}\, \overline \gamma_{AA}
= - 2 (\alpha_A\alpha_A)_0 \ .
\label{eqA29}\end{equation}
This explains the presence of this factor in the static and spherically
symmetric solution discussed in section IV, Eq.~(\ref{eq4.21}). [The
isotropic form (\ref{eq4.18}) contains extra $\alpha_A^2$ terms due to
the change of coordinates (\ref{eq4.22}).] In fact, the contribution
of the first {\sf H} diagram to this one-body metric is proportional to
$(\alpha_A\alpha_0)(\alpha_A^2)$: One of the white blobs, involving the
background value $\alpha(\varphi_0)\equiv \alpha_0$, corresponds to the
point where the metric is computed, and the three other blobs
correspond to body $A$. Note also that the first {\sf T} diagram,
Eq.~(\ref{eqA23}), does not contribute to the $\alpha_A^2$ terms of
$\widetilde g_{00}$, since it vanishes in the static case. Finally, let
us mention that the last {\sf F} and {\sf H} diagrams are the usual
contributions obtained in general relativity, and that the fifth {\sf
F} diagram is obviously proportional to the second {\sf T} diagram:
\begin{equation}
{\sf F}_5 = \sum_{D\neq C} {1\over 3}{\sf T}_2^{(A,B,C)}\times
{3G_{CD}\overline \gamma_{CD} m_D^0\over 2 r_{CD} c^2}
+ O\left({1\over c^6}\right)\ ,
\label{eqA30}\end{equation}
where ${1\over 3}{\sf T}_2^{(A,B,C)}$ is given by (\ref{eqA24}) above,
symmetrized over the three bodies $A,B,C$.

\section{2PN renormalizations of coupling parameters,
the strong equivalence principle, and all that}
As emphasized by Nordtvedt \cite{N93}, the coupling parameters
($\widetilde G$, $\widetilde m$, $\overline \beta$, $\overline \gamma$,
\dots) of Lorentz-invariant gravitational theories are influenced not
only by the self-gravity of the interacting bodies, but also by the
presence of distant ``spectator'' matter around the system. We have
already studied the self-gravity renormalizations in the main text. In
this Appendix, we relate our self-gravity results to the ones of
Ref.~\cite{N93} and show how, within the context to
tensor--multi-scalar theories, one can derive very easily the influence
of external matter.

Let us first mention that, from the 1PK Lagrangian (\ref{eq2.18}), one
can easily give explicit expressions for all the mass parameters
$M(G),M(\gamma),M(\beta),\ldots$ introduced in \cite{N93}. For the
convenience of the reader, we give in Table~\ref{tab1} a translation of
Nordtvedt's notation in terms of our body-dependent parameters
$\widetilde G_{AB}$, $\overline \gamma_{AB}$ and $\overline \beta^{\
A}_{BC}$ defined in Eqs.~(\ref{eq2.20}) above. As before an index
$\scriptstyle 0$ in one of these parameters corresponds to a
non-self-gravitating body, so that the $\sigma$-model tensors
$\alpha_A^a,\beta_A^{ab}$ of Eqs.~(\ref{eq2.16}) should be replaced by
their weak-field counterparts (\ref{eq2.10})--(\ref{eq2.11}). For
instance, $\overline\beta^{\ A}_{00}= {1\over 2}
(\alpha\beta_A\alpha)_0/ \left[1+(\alpha_A\alpha)_0\right]^2$ instead
of (\ref{eq2.20c}). The 2PN renormalizations of $\widetilde G_{AB}$,
$\overline \gamma_{AB}$ and $\overline \beta^{\ A}_{BC}$ due to the
self-energy of the bodies have been obtained in Eqs.~(\ref{eq3.29})
above. Using them in the translation of notation of Table~\ref{tab1},
we recover the corresponding (less complete) results of~\cite{N93}.

Let us now turn to the study of the influence of external matter. The
effect of a distant spectator body on local gravitational physics can
be analyzed straightforwardly in the context of tensor--multi-scalar
theories. Indeed, if this spectator $S$ is located at a distance $D$
from a local system, the local background values of the scalar fields
are changed from their values $\varphi_0^a$ far from the spectator to
\begin{equation}
\varphi_{\rm loc}^a = \varphi_0^a - {G_*\alpha_S^a m_S\over
Dc^2} +O\left({1\over D^2}\right)\ .
\label{eqB1}\end{equation}
All the physical quantities $f$ which depend on these background values
are therefore renormalized (compared to the value they would be
measured to have at infinity) by the presence of the spectator as
\begin{equation}
f \quad \rightarrow \quad f_{\rm loc} =
f- {G_*m_S\over Dc^2}\, \alpha_S^a\,
{\partial f\over \partial\varphi_0^a} +O\left({1\over D^2}\right) \ .
\label{eqB2}\end{equation}
In particular, if the spectator body is supposed to have a negligible
self-energy ($\alpha_S^a = \alpha_0^a$), we get for the effective
gravitational constant (\ref{eq2.13}) and the Eddington parameters
(\ref{eq2.14})
\begin{mathletters}
\label{eqB3}
\begin{eqnarray}
\widetilde G & \quad\rightarrow\quad & \widetilde G\left[
1-{G m_S\over D c^2}\, \eta\right]
+ O\left({1\over D^2}\right)\ ,
\label{eqB3a} \\
\overline\gamma & \quad\rightarrow\quad & \overline\gamma
+4\, {G m_S\over D c^2}\, \overline\beta\,
(2+\overline\gamma) + O\left({1\over D^2}\right)\ ,
\label{eqB3b} \\
\overline\beta & \quad\rightarrow\quad & \overline\beta
- {G m_S\over D c^2}
\left({\varepsilon\over 2}+\zeta-8\overline\beta^2\right)
+ O\left({1\over D^2}\right)\ ,
\label{eqB3c}\end{eqnarray}
\end{mathletters}
where $\eta\equiv4\overline\beta - \overline \gamma$ as before, and
where the dimensionless ratio $G m_S / D c^2$ may be replaced by its
expression in physical units $\widetilde G \widetilde m_S/\widetilde D
c^2$. Equation (\ref{eqB3a}) is the well-known renormalization of the
gravitational constant derived in \cite{W81}. The renormalizations of
$\overline\gamma$ and $\overline\beta$ have also been studied in
Eqs.~(5.15) and (3.7) of \cite{N93}, but they were expressed in terms
of several 2PN parameters, instead of the simple form (\ref{eqB3b}) and
(\ref{eqB3c}). In particular, the analysis of Ref.~\cite{N93} did not
suffice to prove that the 2PN renormalization of $\overline \gamma$ is
in fact proportional to $\overline \beta$, and therefore that it is
already constrained by the 1PN experimental bounds on $|\overline\beta|
< 6\times 10^{-4}$.

Of course, the renormalizations (\ref{eqB3}) can be generalized
straightforwardly to the strong-field regime, by considering the
body-dependent parameters $\widetilde G_{AB}(\varphi_0)$,
$\overline\gamma_{AB}(\varphi_0)$, $\overline \beta^{\
A}_{BC}(\varphi_0)$, and a compact spectator body ($\alpha_S^a\neq
\alpha_0^a$). This has been done in section 7.2 of Ref.~\cite{DEF1},
where we analyzed the consequences of the strong equivalence principle
(SEP) in tensor--multi-scalar theories. This principle states that
local gravitational physics is totally independent of the presence of
spectator bodies. In other words, the renormalizations (\ref{eqB3}) and
their strong-field analogues are supposed to vanish. Let us prove that
the only tensor--multi-scalar theories containing only positive-energy
excitations ($\gamma_{ab}>0$) and satisfying the strong equivalence
principle are perturbatively equivalent to general relativity (to all
orders). Indeed, we showed in \cite{DEF1}, using Eqs.~(\ref{eqB3}),
that any such theory must satisfy $\overline\gamma = 0$, among other
relations. From Eq.~(\ref{eq2.14a}) and the positivity of
$\gamma_{ab}$, this implies $\alpha_0^a= 0$. Using now the
diagrammatically evident\footnote{Note that this would not be true if
we were considering quantum (loop) diagrams.} fact that any observable
deviation from general relativity must involve at least two factors
$\alpha_0^a$ (to fill the end blobs connected to scalar propagators, in
diagrams such as Fig.~\ref{fig7} or any higher-order ones), we find
that all non-Einsteinian terms necessarily vanish. It should be noted
that this result cannot be extended to the pure scalar theories we also
considered in section 7.2 of \cite{DEF1}, in which the gravitational
interaction is mediated by one or several scalar fields without any
tensorial contribution. We showed that the SEP implies in that case
$\widetilde G_{AB} = \widetilde G$, $\overline \gamma_{AB} =
\overline\gamma = -2$, and $\overline\beta^{\ A}_{BC}= \overline\beta =
-{1\over2}$, {\it i.e.}, that the theory is equivalent to Nordstr\o m's
theory at the 1PK level. The assumption that the $\sigma$-model metric
is positive does not allow us to complement this result to higher
orders.

We can also prove that some assumptions made in Ref.~\cite{B92} are
inconsistent within the tensor--scalar framework. Indeed,
Ref.~\cite{B92} assumed one could work within a class of theories
containing no dipolar radiation and still differing from general
relativity at 2PN order. However, we can use the fact that the dominant
dipolar radiation emitted by a binary system is proportional to
$\gamma_{ab}(\alpha_A^a-\alpha_B^a)(\alpha_A^b-\alpha_B^b)$. Let us
first assume (as we generally do) that $\gamma_{ab}$ is positive
definite. Then the assumption that the dipolar radiation vanishes
implies that $\alpha_A^a= \alpha_B^a$ for any body $A$ and $B$, and in
particular that $\alpha_A^a = \alpha^a$ if we choose a
non-self-gravitating body $B$ ($\widetilde m_B = {\rm const}$). Using
now the expansion of $\alpha_A^a$ in powers of the compactness of body
$A$, that we derived in Eqs.~(8.3) and (8.4) of \cite{DEF1}, and the
fact that $\alpha_A^a=\alpha^a$ must be verified for any body $A$, we
can conclude that $\alpha^a=0$. In other words, the scalar fields are
totally decoupled from matter, and the theory is strictly equivalent to
general relativity. Even if we drop the assumption of a positive
definite $\gamma_{ab}$, {\it i.e.}, if we phenomenologically allow the
presence of scalar fields with negative energy (ghost modes), then the
same Eqs.~(8.3) and (8.4) of \cite{DEF1} can be used to show that the
theory is equivalent to general relativity up to the 2PN order
included, but not beyond (which means that it can differ from it at the
2PK order, as illustrated in section~9 of \cite{DEF1}).

The ease with which we derived the results of this appendix illustrates
again the power of our field-theoretical approach.

\section{Explicit expressions of the 2PN light deflection and
perihelion shift}
Before computing the explicit expression of the light-deflection angle
for the one-body metric (\ref{eq4.18}), let us show how our diagrammatic
approach allows one to get, without any calculations, the structure of
the final result. More precisely, let us show that the light-deflection
angle for a self-gravitating body has the structure
\begin{equation}
O(\overline\gamma_{A0})(G_{A0}m_A/rc^2)
+O(\overline\gamma_{A0},\overline\gamma_{AA})(G_{A0}m_A/rc^2)^2
+O(1/c^6)\ ,
\label{eqC1}\end{equation}
in which neither $\overline\beta^A_{BC}$ nor the new 2PK parameters we
introduced above enter. Indeed, the action describing the
electromagnetic field minimally coupled to the physical metric
$\widetilde g_{\mu\nu}$ is a conformal invariant, and can thus be
written as
\begin{equation}
S_{\rm EM} = -{1\over 4}\int {d^4x\over c}\sqrt{\widetilde g}\,
\widetilde g^{\lambda\nu} \widetilde g^{\mu\rho} F_{\lambda\mu}
F_{\nu\rho} =-{1\over 4}\int {d^4x\over c}\sqrt{g_*}\,
g_*^{\lambda\nu}g_*^{\mu\rho} F_{\lambda\mu} F_{\nu\rho}\ .
\label{eqC2}\end{equation}
The second expression, involving the Einstein metric $g^*_{\mu\nu}$,
shows that photons are coupled to the graviton $h_{\mu\nu}\equiv
g^*_{\mu\nu} -f_{\mu\nu}$, but not to the scalar fields
$\varphi^a-\varphi_0^a$. Therefore, the only diagrams describing the
2PK interaction of light with the gravitational field generated by
material bodies are those of Fig.~\ref{fig11}. [The $\in$, {\sf Z},
{\sf F}, {\sf H} and {\sf X} diagrams enter the metric at order
$O(1/c^6)$, and influence the propagation of light at the third
post-Keplerian level.] The dominant contribution does not involve any
scalar field, and is thus proportional to $G_* m_A/rc^2$. However, we
do not have a direct experimental access to $G_*$, and this
contribution should be rewritten in terms of the Newtonian potential
$G_{A0}m_A/r$ felt by a test mass in the vicinity of body $A$. This can
be done thanks to the identity $G_*\equiv
G_{A0}(1+\overline\gamma_{A0}/2)$, which derives from the definitions
(\ref{eq2.20a}) and (\ref{eq2.20b}). We have thus recovered without any
explicit calculation that the deflection of light and its time-delay
are both proportional to $2+\overline\gamma_{A0}$ at the 1PK level, and
to $2+\overline\gamma$ at the 1PN level. [Remember that
$\overline\gamma$ is usually denoted as $\gamma-1$ in the literature,
so that $2+\overline\gamma$ is the usual factor $1+\gamma$.] The
corrections appearing at order $O(Gm/rc^2)^2$ are due to the five
remaining diagrams of Fig.~\ref{fig11}. Three of them do not involve
any scalar field, and are thus proportional to $G_*^2 =
G_{A0}G_{B0}(1+\overline\gamma_{A0}/2) (1+\overline\gamma_{B0}/2)$. The
other two involve a scalar line between two material bodies, and yield
therefore contributions proportional to
$G_*^2(\alpha_A\alpha_B)_0=-G_*^2\overline\gamma_{AB}/
(2+\overline\gamma_{AB})$, where $G_*^2$ should be rewritten as above
in terms of effective gravitational constants. In the case of a single
material body $A$, it should be noted that the first two {\sf V}
diagrams of Fig.~\ref{fig11} do not contribute, since the same body
cannot be directly connected by a propagator. On the contrary, the
first {\sf T} diagram of this figure does contribute, because the two
blobs representing the same body $A$ are connected to a
scalar--scalar--graviton vertex, and not directly to each other. Hence
this diagram yields a contribution proportional to
$\overline\gamma_{AA}/ (2+\overline\gamma_{AA})$ in second-order
light-deflection and time-delay experiments. In conclusion, our
diagrammatic approach has allowed us to prove is a streamlined way that
these experiments do not depend on $\overline\beta^{\ A}_{BC}$ at the
2PK level, and more precisely that they can differ from general
relativity only by terms of the form indicated in Eq.~(\ref{eqC1}) in
the case of a single material body $A$.

Let us now give explicit expressions for the 2PN light deflection and
perihelion advance. Following Weinberg \cite{W72}, we use here
Schwarzschild-like coordinates ({\it i.e.}, an ``area radius'' $r$) and
write the integral giving the polar angle $\phi$ in the plane of the
trajectory as
\begin{equation}
\phi = \pm \int{{\cal J}\sqrt{{\cal AB}}\, dr/r^2\over
[{\cal E}^2/c^2 - (m_0^2c^2+{\cal J}^2/r^2){\cal B}]^{1\over 2}} \ ,
\label{eqC4}\end{equation}
where $\cal E$ denotes the conserved energy (including the rest mass
contribution) of a test particle of mass $m_0$, and
$\cal J$ its angular momentum. [Note that these conserved quantities
are related in a non-obvious way to the constants $E_{\rm W}$ and
$J_{\rm W}$ used by Weinberg in Eq.~(8.4.30) of \cite{W72}: $E_{\rm W}
\equiv (m_0 c^2/{\cal E})^2$; $J_{\rm W}\equiv{\cal J}c/{\cal E}$.] The
area radius $r$ should not be confused with the Just radius nor the
$g^*$-harmonic radius, both denoted also by $r$ in subsection IV--B.
To rewrite the metric (\ref{eq4.18}) in Schwarzschild coordinates
\begin{equation}
d\widetilde s^2/A^2(\varphi_0) =
-{\cal B}(r) c^2 dt^2 + {\cal A}(r) dr^2
+ r^2 (d\theta^2+\sin^2\theta d\phi^2)\ ,
\label{eqC5}\end{equation}
we need to express the area radius $r$ in terms of the isotropic one
$\rho$. We find easily
\begin{equation}
r = \rho\left[
1+{\mu_A\over \rho c^2}(1+\overline\gamma_{A0})
+{1\over4}\left({\mu_A\over \rho c^2}\right)^2
(1+3\overline\Gamma-4\overline\gamma_{A0}-2\overline\gamma_{A0}^2)
+O\left({1\over c^6}\right)
\right]\ ,
\label{eqC6}\end{equation}
where $\mu_A\equiv G_{A0}m_A$ is the Keplerian mass of the attracting
body $A$, and $\overline\gamma_{A0}$, $\overline\Gamma$ are the
deviations from general relativity in the spatial isotropic metric
(\ref{eq4.18}). Equations (\ref{eq4.19}) and (\ref{eq4.20}) give the
expressions of these coefficients in tensor--multi-scalar theories.
Note, however, that our present calculation is valid for any static
and spherically symmetric metric of the form (\ref{eq4.18}), not
necessarily the one predicted by tensor--scalar theories. The
replacement of (\ref{eqC6}) in (\ref{eq4.18}) yields
\begin{mathletters}
\label{eqC7}
\begin{eqnarray}
{\cal A}(r) & = & 1+2{\mu_A\over rc^2}(1+\overline\gamma_{A0})
+4\left({\mu_A\over rc^2}\right)^2\left(
1+{3\over 4}\overline\Gamma+{1\over 2}\overline\gamma_{A0}
+{1\over 4}\overline\gamma_{A0}^2
\right)
+O\left({1\over c^6}\right)\ ,
\label{eqC7a} \\
{\cal B}(r) & = & 1-2{\mu_A\over rc^2}
+2\left({\mu_A\over rc^2}\right)^2(\overline\beta^{\ 0}_{AA}
-\overline\gamma_{A0})
\nonumber \\
& & -{3\over 2}\left({\mu_A\over rc^2}\right)^3\left(
\overline{\rm B}+\overline \Gamma
-{8\over 3}(1+\overline\gamma_{A0})\overline\beta^{\ 0}_{AA}
-{2\over 3}(2-\overline\gamma_{A0})\overline\gamma_{A0}
\right)
+O\left({1\over c^8}\right)\ .
\label{eqC7b}\end{eqnarray}
\end{mathletters}
The polar angle (\ref{eqC4}) can now be obtained by a straightforward
integration. The case of light corresponds to $m_0=0$, and the
deflection angle $\Delta\phi$ is found to have the form
\begin{equation}
\Delta\phi = \Delta\phi_1+\Delta\phi_2+O(1/c^6)\ ,
\label{eqC8}\end{equation}
where
\begin{equation}
\Delta\phi_1 = {2\mu_A(2+\overline\gamma_{A0})\over\rho_0 c^2}
\left[1 - {\mu_A(2+\overline\gamma_{A0})\over \rho_0c^2}
\right]
\label{eqC9}\end{equation}
is the 1PK result up to a global correcting factor, and
\begin{equation}
\Delta\phi_2 = {\pi\over 4}
\left({\mu_A\over \rho_0 c^2}\right)^2
\left[15+3\overline\Gamma - 4\overline\beta^{\ 0}_{AA}
+8\overline\gamma_{A0}\right]
\label{eqC10}\end{equation}
is the actual 2PK contribution. Note that, after having performed the
integral (\ref{eqC4}) in Schwarzschild coordinates, we have expressed
the results in terms of $\rho_0$, the minimal distance between the
light ray and the center of body $A$, measured in {\it isotropic\/}
coordinates. [Note that $\rho_0$ differs from the impact parameter.]
The results are unchanged if one uses $g^*$-harmonic
coordinate to measure this minimal distance, since the transformation
(\ref{eq4.22}) introduces corrections only at order $O(1/c^6)$. By
contrast, the use of Schwarzschild coordinates transforms
(\ref{eqC9}) into $\Delta\phi_1 =
[2\mu_A(2+\overline\gamma_{A0})/r_0c^2]\times[1-(\mu_A/r_0c^2)]$,
while (\ref{eqC10}) remains unchanged. This second-order light
deflection (\ref{eqC8})--(\ref{eqC10}) agrees with previous
calculations in the literature \cite{ES80,FF80,RM82}. It can now be
particularized to the case of tensor--multi-scalar theories by using
the expression (\ref{eq4.19b}) for $\overline\Gamma$. We find that the
term ${4\over 3}\overline\beta^{\ 0}_{AA}$ in $\overline\Gamma$
cancels exactly the $-4\overline\beta^{\ 0}_{AA}$ contribution in
(\ref{eqC10}), and we get
\begin{equation}
\Delta\phi_2 = {\pi\over 16}\left({\mu_A(2+\overline\gamma_{A0})
\over \rho_0c^2}\right)^2
\left[15+{\overline\gamma_{AA}\over 2+\overline\gamma_{AA}}\right] \ .
\label{eqC11}\end{equation}
Together with Eq.~(\ref{eqC9}), this result confirms therefore the
conclusion (\ref{eqC1}) of our diagrammatic analysis: The first
order deviation from general relativity is proportional to
$\overline\gamma_{A0}$, while the second order contributions involve
both $\overline\gamma_{A0}$ and $\overline\gamma_{AA}$, but not
$\overline\beta^{\ 0}_{AA}$. Note also the appearance of the same
coefficient $\mu_A(2+\overline\gamma_{A0})\equiv2G_*m_A$ in
(\ref{eqC9}) and (\ref{eqC11}), which is in fact a mere rewriting of the
bare gravitational constant $G_*$ in terms of observable quantities.
The coefficient $\overline\gamma_{AA}/(2+\overline\gamma_{AA})\equiv -
(\alpha_A\alpha_A)_0$ entering (\ref{eqC11}) is due to the first {\sf T}
diagram of Fig.~\ref{fig11}, where both blobs represent body $A$.

Let us now take the second post-Newtonian limit of the above 2PK
results, as appropriate for interpreting future high-precision
experiments in the solar system. The coefficient $\overline\gamma_{A0}$
of the leading term in (\ref{eqC9}) should therefore be expanded as in
Eq.~(\ref{eq5.6}), together with the renormalization (\ref{eqB3b}) due
to the gravitational potential of external masses, while the 2PK
contribution (\ref{eqC10}) becomes
\begin{equation}
\Delta\phi_2^{\rm 2PN} = {\pi\over 4}\left({\mu_A\over \rho_0c^2}
\right)^2
\left[15+{31\over 2}\overline\gamma +4 \overline\gamma^2\right]\ .
\label{eqC12}\end{equation}

The explicit expression of the second-order time-delay has been
derived in \cite{RM83} for a general metric of the form
(\ref{eq4.18}). This result confirms also the diagrammatic analysis
discussed at the beginning of the present appendix: The parameter
$\overline\beta^{\ 0}_{AA}$ appears again in the combination
$3\overline\Gamma-4\overline\beta^{\ 0}_{AA}$, like in (\ref{eqC10})
above, and it vanishes therefore in the case of tensor--multi-scalar
theories, for which $\overline\Gamma = {4\over 3}\overline\beta^{\
0}_{AA}+O(\overline\gamma_{A0}, \overline\gamma_{AA})$.

The integral (\ref{eqC4}) can also be used to derive the periastron
advance of a test mass $m_0\ll m_A$. Let us introduce as in
Ref.~\cite{DS88} the notation $E\equiv({\cal E}-m_0c^2)/m_0$ for the
specific conserved energy minus rest mass, and $h\equiv{\cal
J}/m_0\mu_A$ for the reduced conserved angular momentum. A
straightforward integration yields the 2PK periastron shift per orbit:
\begin{eqnarray}
\Delta\phi & = & {6\pi\over h^2 c^2}\Biggl[
{3-\overline\beta^{\ 0}_{AA}+2\overline\gamma_{A0}\over 3}
\nonumber \\
& & +{1\over h^2c^2}\left(
{35\over 4}+{3\over 4}\overline{\rm B}+{3\over 2}\overline\Gamma
-9\overline\beta^{\ 0}_{AA} +{1\over 2}(\overline\beta^{\ 0}_{AA})^2
+8\overline\gamma_{A0}+2\overline\gamma_{A0}^2
-4\overline\beta^{\ 0}_{AA}\overline\gamma_{A0}
\right)
\nonumber \\
& & +{E\over c^2}\left(
{5\over 2}+{1\over 2}\overline\Gamma-{2\over 3}\overline\beta^{\ 0}_{AA}
+{4\over 3}\overline\gamma_{A0}
\right)\Biggr]
+O\left({1\over c^6}\right)\ .
\label{eqC13}\end{eqnarray}
This expression agrees with the general relativistic result derived in
\cite{H76,DS88}. Note that contrary to the light-deflection and
time-delay formulae, the periastron advance involves not only
$\overline\beta^{\ 0}_{AA}$ but also the 2PK parameter $\varepsilon^{\
\ 0}_{AAA}$, entering $\overline{\rm B} = {2\over 9}\varepsilon^{\ \
0}_{AAA} +O(\overline\beta, \overline\gamma)$, {\it cf.}
Eq.~(\ref{eq4.19a}). The 2PN limit of Eq.~(\ref{eqC13}) can be obtained
easily by using the expansions (\ref{eq4.23}) of the different
body-dependent parameters.

The result (\ref{eqC13}) is coordinate-independent since it is expressed
in terms of the conserved quantities $E$ and $h$. It can nevertheless
be helpful to rewrite it in a particular coordinate system. Both
isotropic and $g^*$-harmonic coordinates give the same result at the
2PK order. Let us denote by $a(1-e)$ the coordinate periastron radius,
and by $a(1+e)$ the coordinate apoastron radius (in isotropic or,
equivalently, $g^*$-harmonic coordinates). The conserved quantities can
then be rewritten as
\begin{mathletters}
\label{eqC14}
\begin{eqnarray}
E & = & -{\mu_A\over 2a} + {1\over 8c^2}\left({\mu_A\over a}\right)^2
(7+4\overline\gamma_{A0})
+O\left({1\over c^4}\right)\ ,
\label{eqC14a} \\
{1\over h^2} & = & {\mu_A\over a(1-e^2)}
-{4\over c^2}\left({\mu_A\over a(1-e^2)}\right)^2
\left(
1+{e^2\over 2}-{1\over 2}\overline\beta^{\ 0}_{AA}
+{3\over 4}\overline\gamma_{A0}
+{e^2\over 4}\overline\gamma_{A0}
\right)
+O\left({1\over c^4}\right)\ .
\label{eqC14b}\end{eqnarray}
\end{mathletters}
We recover in particular the standard 1PN formula, $\Delta\phi =
2\pi\mu_A (3-\overline\beta+2\overline\gamma)/a(1-e^2)c^2 +O(1/c^4)$,
in terms of the {\it semilatus rectum\/} $a(1-e^2)$. The 2PN result
takes the form displayed in Eq.~(\ref{eq5.9}) above.


\begin{figure}
\caption{Diagrammatic interpretation of the effective gravitational
constant $\widetilde G = G_* A_0^2 (1+\alpha_0^2)$.}
\label{fig1}
\end{figure}

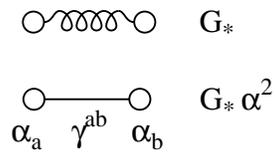
\begin{figure}
\caption{Diagrammatic interpretation of the contraction
$(\alpha\beta\alpha)_0$ involved in the Eddington parameter
$\overline\beta = {1\over 2} (\alpha\beta\alpha)_0/(1+\alpha_0^2)^2$.}
\label{fig2}
\end{figure}

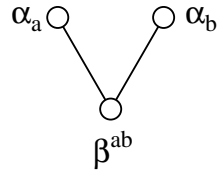
\begin{figure}
\caption{Diagrammatic notations for the material sources, the fields,
and their propagator.}
\label{fig3}
\end{figure}

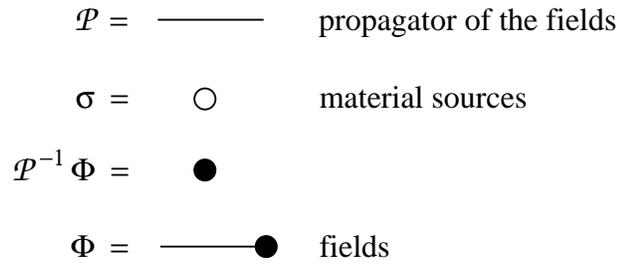
\begin{figure}
\caption{Diagrammatic expression of the $\Phi^i$-linear terms of the
total action (\protect\ref{eq3.6}), for $i=1,2,3,4$.}
\label{fig4}
\end{figure}

\begin{figure}
\caption{Equation (\protect\ref{eq3.2a}) satisfied by the field
$\overline\Phi[\sigma]$.}
\label{fig5}
\end{figure}

\begin{figure}
\caption{Diagrammatic expansion of the Fokker action
(\protect\ref{eq3.13}).}
\label{fig6}
\end{figure}

\begin{figure}
\caption{Expression of the diagrams of Fig.~\protect\ref{fig6} when the
graviton and scalar propagators are represented respectively as curly
and straight lines.}
\label{fig7}
\end{figure}

\begin{figure}
\caption{Constraints imposed by four different binary-pulsar data on the
2PN parameters $\varepsilon$, $\zeta$.}
\label{fig8}
\end{figure}

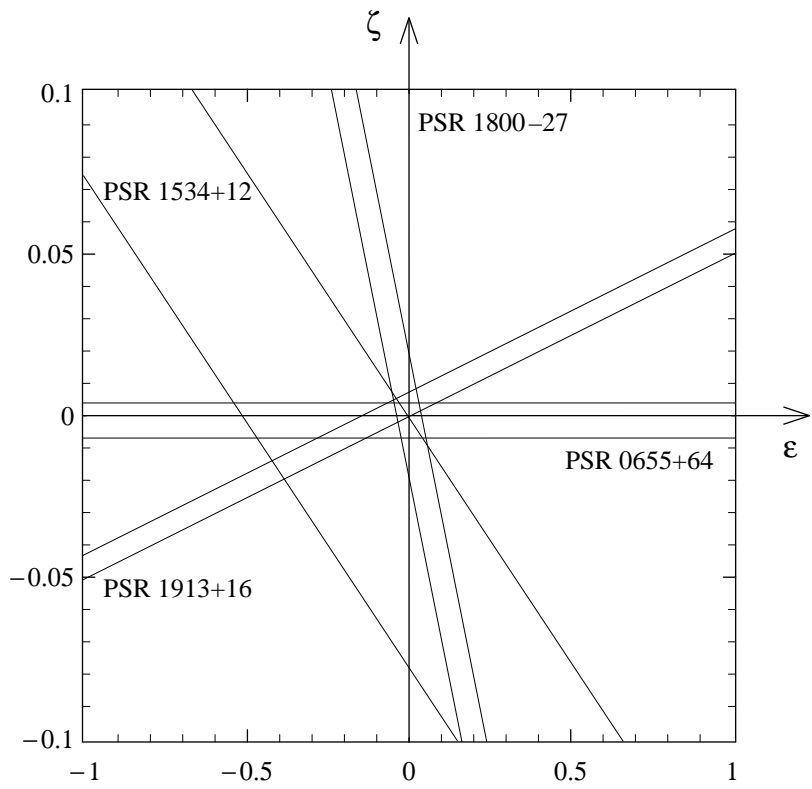
\begin{figure}
\caption{Region of the $\varepsilon$--$\zeta$ plane allowed at the
1$\sigma$ level by the four tests of Fig.~\protect\ref{fig8}.}
\label{fig9}
\end{figure}

\begin{figure}
\caption{Decomposition of the {\sf F} and {\sf H} diagrams of
Fig.~\protect\ref{fig6}, showing in bold the lines involving
derivatives in the vertex $\Phi\partial\Phi\partial\Phi$. Such a
decomposition is useful to compute the last two {\sf F} and {\sf H}
diagrams of Fig.~\protect\ref{fig7} without symmetrizing the 3-graviton
vertex they contain.}
\label{fig10}
\end{figure}

\begin{figure}
\caption{Diagrams describing the interaction of light (wavy lines) with
the gravitational field generated by material bodies, at orders
$O(Gm/rc^2)$ --1PK-- and $O(Gm/rc^2)^2$ --2PK--.}
\label{fig11}
\end{figure}

\begin{table}
\caption{Expression of Nordtvedt's mass parameters in
tensor--multi-scalar theories of gravity.}
\label{tab1}
\begin{tabular}{ll}
Nordtvedt's parameters&Tensor--multi-scalar theories\\
\tableline
$M(I)$&$\widetilde m_A$\\
$\Gamma_{ij}/M(I)_iM(I)_j$&$\widetilde G_{AB}$\\
$\gamma\Theta_{ij}/\Gamma_{ij}$&$1+\overline \gamma_{AB}$\\
$(2\beta-1)\Gamma_{ijk}/M(I)_iM(I)_jM(I)_k$&$(1+2\overline \beta^{\
A}_{BC})
\widetilde G_{AB} \widetilde G_{AC}$\\
$M(G)/M(I)$&$\widetilde G_{A0}/\widetilde G$\\
$M(\gamma)/M(I)$&$\widetilde G_{A0}(1+\overline\gamma_{A0})/\widetilde G
(1+\overline \gamma)$\\
$M(\beta)/M(I)$&$\widetilde G_{A0}(1+2\overline \beta^{\ 0}_{A0})
/\widetilde G (1+2\overline \beta)$\\
$M(\beta')/M(I)$&$\widetilde G_{A0}^2(1+2\overline
\beta^A_{00})/\widetilde G^2(1+2\overline \beta)$
\end{tabular}
\end{table}

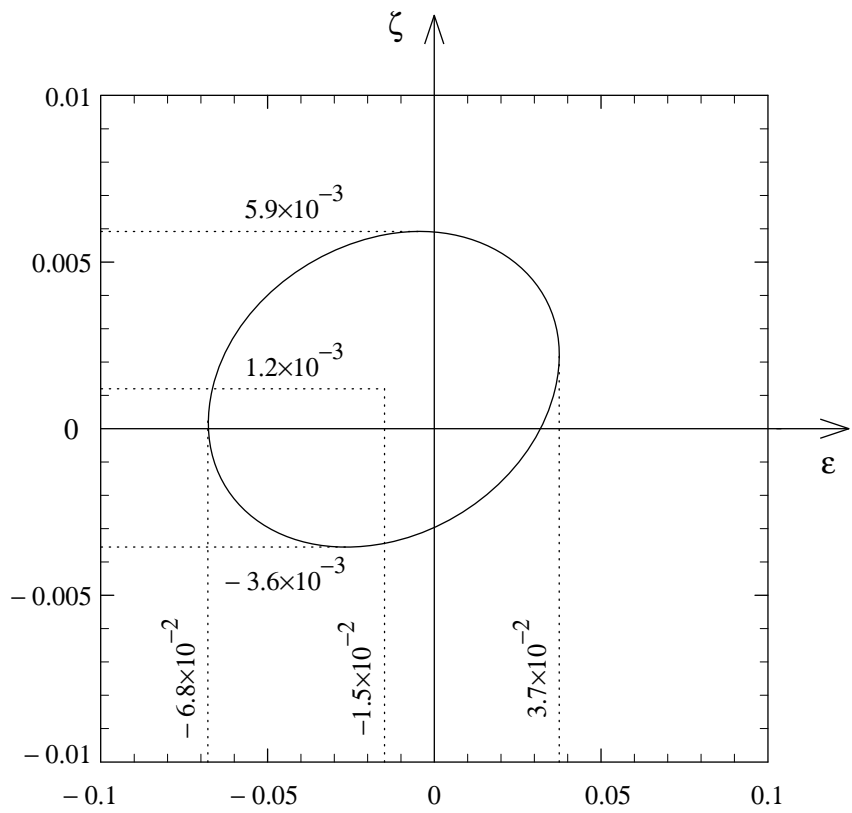
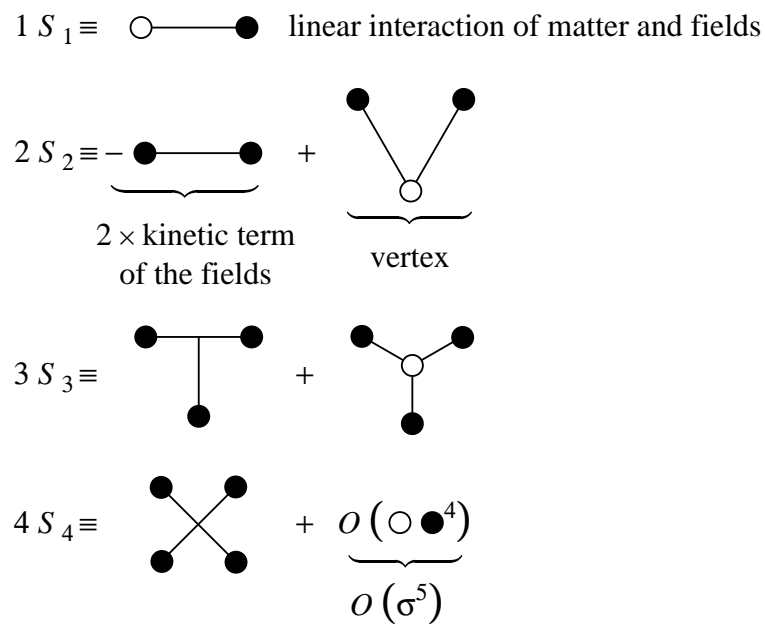
\end{document}